\documentclass[12pt]{article}
\usepackage{amsmath}
\usepackage{amssymb}
\begin{document}
\begin{center}
{\large\bf  {Massive 4D Abelian 2-Form  Theory: Nilpotent Symmetries  from the  (Anti-)Chiral Superfield Approach}}

\vskip 3.0cm

{\sf S. Kumar$^{(a)}$, B.Chauhan$^{(a)}$, A. Tripathi$^{(a)}$,  R. P. Malik$^{(a,b)}$}\\
$^{(a)}$ {\it Physics Department, Institute of Science,}\\
{\it Banaras Hindu University, Varanasi - 221 005, (U.P.), India}\\

\vskip 0.1cm

\vskip 0.1cm

$^{(b)}$ {\it DST Centre for Interdisciplinary Mathematical Sciences,}\\
{\it Institute of Science, Banaras Hindu University, Varanasi - 221 005, India}\\
{\small {\sf {e-mails: sunil.bhu93@gmail.com; bchauhan501@gmail.com; ankur1793@gmail.com; rpmalik1995@gmail.com}}}

\end{center}

\vskip 1cm

\noindent
{\bf Abstract:}
The off-shell nilpotent and absolutely anticommuting (anti-)BRST symmetries are obtained
 by using the (anti-)chiral superfield approach (ACSA) to 
 Becchi-Rouet-Stora-Tyutin (BRST) formalism for the four (3+1)-dimensional (4D) St$\ddot{u}$ckelberg-modified {\it massive} 
Abelian 2-form gauge theory. We perform exactly similar kind of exercise for the derivation of the off-shell 
nilpotent (anti-)co-BRST symmetry transformations, too. In the above derivations, the symmetry invariant restrictions 
on the superfields play very important and decisive roles. To prove the sanctity of the above nilpotent symmetries, 
we generalize  our 4D {\it ordinary} theory (defined on the 4D {\it flat} Minkowskian spacetime 
manifold) to its counterparts (4,1)-dimensional (anti-)chiral super sub-manifolds of the (4, 2)-dimensional 
supermanifold which is parameterized by the superspace coordinates $Z^{M} = (x^{\mu},\theta, \bar{\theta} ) $ 
 where $x^\mu ( \mu = 0,1,2,3 )$ are the {\it bosonic} coordinates and a pair of Grassmannian variables 
$(\theta, \bar{\theta})$ are {\it fermionic}: $\theta^{2} = \bar{\theta^{2}} = 0, 
\,\,\theta\,\bar{\theta} +\bar{\theta}\,\theta = 0$ in nature. One of the novel observations of
 our present endeavor is the derivation of the Curci-Ferrari (CF)-type restrictions from the requirement
 of the symmetry invariance of the coupled (but equivalent) Lagrangian
densities of our theory {\it within} the framework of ACSA to BRST formalism. We also exploit the standard techniques of ACSA to 
capture the off-shell nilpotency and absolute anticommutativity of the conserved (anti-)BRST as well as the conserved (anti-)co-BRST
 charges.  In a {\it subtle} manner, the proof of the absolute anticommutativity of the above conserved charges {\it also}   
 implies the existence of the appropriate  CF-type restrictions on our theory. This proof is {\it also} a {\it novel} result.

\vskip 0.5 cm
\noindent
{PACS numbers:   11.15.-q; 11.30.-j; 03.70.+k; 95.35.+d.}

\vskip 0.5 cm
\noindent
{\it {Keywords}}: {4D massive Abelian  2-form  gauge theory; (anti-)BRST 
symmetries; (anti-)co-BRST  symmetries; coupled Lagrangian densities;  
Curci--Ferrari type restrictions; conserved charges;   
 ACSA to BRST formalism; symmetry invariant restrictions;  nilpotency
 and absolute anticommutativity properties.}



\section{Introduction}
\label{intro} The superfield approach  to Becchi-Rouet-Stora-Tyutin (BRST) formalism [1-8] provides the geometrical
origin and interpretation for the nilpotency and absolute anticommutativity properties that are associated with
the {\it quantum} BRST and anti-BRST symmetries corresponding to a given {\it classical} local gauge symmetry 
transformation for a given $p$-form ($p$ = 1, 2, 3...) gauge theory which is generated by the first-class constraints 
[9, 10] that characterize {\it such} a theory. One of the key features of the usual superfield approach (USFA)
to BRST formalism [4-6], proposed by Bonora and Tonin (BT), is the derivation of (i) the 
{\it exact} off-shell nilpotent and absolutely anticommuting (anti-)BRST symmetry transformations, {\it and} (ii) 
the (anti-) BRST invariant CF-type restriction(s). The existence of the {\it latter} is the signature [11, 12] of the 
{\it quantum} version of a gauge theory  that is discussed and described within the framework of BRST formalism.
 The {\it other} characteristic feature of USFA (proposed by BT) is the observation that it leads {\it only} to the derivation of the nilpotent 
 (anti-)BRST symmetries for the gauge and associated (anti-)ghost fields of the given $p$-form {\it quantum} gauge theory.

The above USFA has been systemically generalized in our earlier works (see, e.g. [13-16]) 
which lead to the derivation of  proper (anti-)BRST symmetry transformations for the 
{\it matter} fields in addition to the gauge and (anti-)ghost fields of an {\it interacting}  $p$-form  (non-)Abelian 
gauge theory. The generalized version of the USFA has been christened as the augmented  
version of superfield approach (AVSA) to BRST formalism (see, e.g. [13-16]). In a very recent  
set of works (see, e.g. [17-20]), we have incorporated the diffeomorphism  transformation in the 
BT-superfield formalism  which has been called as the {\it modified} BT-superfield approach to
BRST formalism. One of the common features of the above superfield approaches [1-8, 13-20]  
is the observation that the fields of an {\it ordinary} D-dimensional gauge/diffeomorphism invariant theory 
have been generalized onto a suitably chosen (D, 2)-dimensional supermanifold 
which is parameterized by the super space coordinates $Z^M = (x^\mu, \theta, \bar\theta)$
where coordinates  $x^\mu( \mu = 0, 1, 2....D-1)$ define the {\it bosonic} D-dimensional {\it ordinary} coordinates of the
D-dimensional flat Minkowskian  sub-manifold of the above (D, 2)-dimensional 
supermanifold and the Grassmannian variables ($\theta, \bar\theta)$ 
satisfy: $\theta^2 =  \bar\theta^2 = 0, \theta\,\bar\theta + \bar\theta\,\theta = 0$
demonstrating that {\it they} are {\it fermionic} in nature. 
Furthermore, it has been observed that the {\it full} expansions of the superfields have been taken along {\it all} the possible
Grassmannian directions of the (D, 2)-dimensional supermanifold
in {\it all} the above superfield approaches to BRST formalism [1-8, 13-20].

In our very recent set of works (see, e.g. [21-23] for details), we have applied the simplified  version of the AVSA/USFA as well as the {\it modified} BT-superfield 
approaches  where the {\it ordinary} fields of a given {\it ordinary} D-dimensional {\it gauge} theory 
have been generalized onto  a {\it couple} of (D, 1)-dimensional (anti-)chiral super sub-manifolds of the {\it general} (D, 2)-dimensional 
supermanifold that has been considered  in [1-8, 13-20]. The purpose of our present investigation is to apply
the simplified version (see, e.g. [21-23]) of the superfield approach to BRST formalism where {\it only} the  (anti-)chiral super expansions are taken into account
for the (anti-)chiral superfields  [defined on the (D, 1)-dimensional (anti-)chiral super sub-manifolds] 
and apply the symmetry invariant  restrictions on {\it them} to obtain the (anti-)BRST 
as well as the (anti-)co-BRST  transformations for our 4D St$\ddot{u}$ckelberg-modified {\it massive} 
 Abelian 2-form theory\footnote{Our present 4D massive Abelian 2-form theory is {\it interesting} because there is some {\it physical}  
relevance  of {\it this} theory  in the context of dark matter, dark energy
and cosmological models of the universe (cf. Sec. 7 below for detail). It is precisely because of this reason that we have concentrated 
seriously on the proof of
its {\it quantum} symmetries as well as the CF-type restrictions within the framework of ACSA to BRST formalism.}
 which has been proven by us to be a {\it massive} model for the Hodge theory in physical (3+1)-dimensional (4D) spacetime [24].

In our present investigation, we briefly mention the most general 
forms of the Lagrangian densities [cf. Eqs. (10), (11) below]  by linearizing 
the kinetic and gauge-fixing terms for the fields  $(B_{\mu\nu}, \phi_\mu, \tilde\phi_\mu)$ by
 invoking the Nakanishi-Lautrup type auxiliary fields.  These coupled (but equivalent) Lagrangian
 densities are the generalizations of the {\it ordinary} Lagrangian density [cf. Eq. (6) below]
where the kinetic term for the $B_{\mu\nu}$ field and gauge-fixing terms 
for $(B_{\mu\nu},\phi_\mu, \tilde\phi_\mu)$ fields are {\it  not} 
linearized. We focus on  the (anti-)BRST and (anti-)co-BRST symmetries
  of these Lagrangian densities and obtain the Euler-Lagrange equations of motion (EL-EOMs)
 as well as the CF-type restrictions from them. The main results of
 our present investigation are the derivations of the  (anti-)BRST,  (anti-)co-BRST symmetries and CF-type restrictions  by using the 
ACSA  to BRST formalism. Furthermore, we express the coupled Lagrangian densities and (anti-)BRST as well as  
(anti-)co-BRST charges in terms of the (anti-)chiral superfields which are obtained  after the applications of the (anti-)BRST 
and (anti-)co-BRST invariant restrictions on the (anti-)chiral superfields. We prove the existence of the CF-type
 restrictions on our theory by demanding the  (anti-) BRST and (anti-)co-BRST symmetry  invariance of 
the {\it super} Lagrangian densities as well as  by proving the absolute anticommutativity of the 
(anti-)BRST as well as (anti-) co-BRST charges that are present in our theory.

The following motivating factors have spurred our curiosity in perusing our present investigation.
First of all, we have demonstrated the existence of the proper (anti-)BRST, (anti-)co-BRST symmetries and CF-type restrictions
in our earlier work [24] on the 4D {\it massive} Abelian 2-form gauge theory. To prove the sanctity of the above 
continuous symmetries and the CF-type restrictions, it is essential to verify them within the framework of 
ACSA to BRST formalism. Second, we have taken (anti-)ghost part of the coupled (but equivalent) Lagrangian densities 
of our 4D theory to be the {\it same} [24]. We provide the precise arguments for the above correct {\it choice} in our present
endeavor starting from the (dual-)gauge symmetry transformations (cf. Sec. 2). Finally, we know, from our earlier works
 [11, 12], that the existence of the CF-type restriction(s) is a decisive feature of a gauge theory (described 
within the framework of BRST formalism). We verify  {\it their} existence, using the   
ACSA to BRST formalism in our present endeavor.

Our present paper is organized as follows. In Sec. 2, we discuss the bare essentials of the 
continuous symmetry properties of the St$\ddot{u}$ckelberg-modified Lagrangian density for the 
{\it massive} 4D Abelian 2-form theory. Our Sec. 3 is devoted to the discussion of  the off-shell nilpotent (anti-)BRST
and (anti-)co-BRST symmetries for the coupled (but equivalent)
 Lagrangian densities. We derive the (anti-)BRST as well as the (anti-)co-BRST
symmetries within the framework of ACSA to BRST formalism and comment on their absolute 
anticommutativity property in Sec. 4.  In Sec. 5 of
our present endeavor, we show the existence of the CF-type restrictions  by proving the invariance of the 
{\it super} Lagrangian densities  within the framework of ACSA to BRST formalism.
 Our Sec. 6 deals with the proof of nilpotency and absolute anticommutativity
 of the (anti-)BRST and (anti-)co-BRST conserved charges within the 
framework of ACSA. We also demonstrate the existence of the CF-type 
 restrictions, in a {\it subtle} manner, in the proof of the absolute anticommutativity 
property of the  conserved and nilpotent charges.
Finally, in Sec. 7, we make some concluding remarks and point out a few future
directions. \\

In our Appendices   A and B, we perform some explicit computations which complement  the theoretical contents
of our sub-sections  5.2 and  6.1 in the main body of the text of our present investigation.\\

\noindent
{\it Convention and Notations}: We adopt the convention of the left derivative w.r.t. all the {\it fermionic} fields
of our theory. We take the 4D flat Minkowskian metric tensor $\eta_{\mu\nu}$ as: $\eta_{\mu\nu}$ = 
diag $(+ 1, -1, -1, -1)$ so that the dot product between two {\it non-null} 4D vectors $P_\mu$ and $Q_\mu$ is defined as: 
$P\cdot Q = \eta_{\mu\nu} P^\mu\,Q^\nu \equiv P_0\, Q_0 - P_i\,Q_i$ where the Greek indices 
$ \mu, \nu, \lambda ... = 0, 1,2,  3$ stand for the time and space directions and Latin indices $i, j, k... = 1, 2, 3$ correspond to 
the 3D space directions {\it only}. We denote the nilpotent (anti-)BRST and (anti-)co-BRST symmetry transformations (of {\it all} kinds)
by $s_{(a)b}$ and $s_{(a)d}$, respectively. The corresponding  conserved charges are represented by $Q_{(a)b}$ and $Q_{(a)d}$.
The 4D Levi-Civita tensor $\varepsilon_{\mu\nu\eta\kappa}$ is chosen such that  $\varepsilon_{0123} 
= +1 = - \varepsilon^{0123}$ and $\varepsilon_{\mu\nu\eta\kappa} \varepsilon^{\mu\nu\eta\kappa}= - \,4!$,
$\varepsilon_{\mu\nu\eta\kappa} \varepsilon^{\mu\nu\eta\rho}= - \,3! \,\delta^\rho_\kappa$,
 $\varepsilon_{\mu\nu\rho\sigma} \varepsilon^{\mu\nu\eta\kappa}= - \,2! \,(\delta^\eta_\rho \delta^\kappa_\sigma- 
\delta^\eta_\sigma \delta^\kappa_\rho)$, etc. 
We also adopt the convention: $(\delta B_{\mu\nu}/\delta B_{\rho\sigma}) = \frac{1}{2!}\,\big(\delta^\rho_\mu \,
\delta^\sigma_\nu - \delta^\rho_\nu \, \delta^\sigma_\mu \big)$, etc. for the tensorial dififferentiation/variation.

\section{Preliminaries: Lagrangian Formulation}

\label{sec:1}
In this section, we discuss  the infinitesimal  and continuous (dual-)gauge symmetry transformations of the St$\ddot u$ckelberg-modified 
Lagrangian densities  {\it before} their  generalizations to the off-shell nilpotent  and continuous (anti-)BRST and (anti-)co-BRST symmetry 
invariant coupled (but equivalent) Lagrangian densities for our present 4D {\it massive} Abelian
2-form theory. Our present section is divided
into {\it two } parts as discussed and described below:


\subsection {Infinitesimal Gauge Symmetry Transformations}

\noindent
We begin with the four $(3+1)$-dimensional (4D) Kalb-Ramond Lagrangian density 
${\cal L}_{(0)}$ for the free Abelian 2-form  {\it massive} theory (with the rest mass equal to $m$) as 
follows (see, e.g. [25] for details)
\begin{eqnarray}
{\cal L}_{(0)} = \frac{1}{12}\, H^{\mu\nu\eta} H_{\mu\nu\eta} - \frac{m^2}{4}\, B^{\mu\nu} B_{\mu\nu},
\end{eqnarray}
where the antisymmetric $(B_{\mu\nu} = - B_{\nu\mu})$ tensor field $B_{\mu\nu}$ is the 4D massive Abelian 2-form 
$\big[B^{(2)} = \frac{1}{2!}\,(dx^\mu \wedge dx^\nu) \, B_{\mu \nu} \big]$ field and the curvature (i.e. field strength) tensor 
$H_{\mu\nu\eta} = \partial_\mu B_{\nu\eta} + \partial_\nu B_{\eta\mu} + \partial_\eta B_{\mu\nu}$ is derived  
from the 3-form $ \big[H^{(3)} = d B^{(2)} \equiv \frac{1}{3!}\,(dx^\mu \wedge dx^\nu \wedge dx^\eta)\, H_{\mu\nu\eta} \big]$
 where $d  = d x^\mu\partial_\mu$ (with $d^2= 0$) is the exterior derivative of differential geometry.  
We note that the mass dimension of  $B_{\mu\nu}$ is $[M]$ in the natural units (where we take $\hbar = c = 1$) for our 4D theory. Due to the existence 
of mass term $(- \frac{m^2}{4}\, B_{\mu\nu}\, B^{\mu\nu})$, the gauge invariance is lost   
because the above Lagrangian density is endowed with the second-class constraints [26] in the terminology of 
Dirac's prescription for the classification scheme of constraints [9, 10]. We can find out the Euler-Lagrange equation of motion (EL-EOM) from
${\cal L}_{(0)}$ as: $\partial_\mu H^{\mu\nu\eta} + m^2\, B^{\nu\eta} = 0$. At this stage,
it is evident that, for $m^2\neq 0$, we have $\partial_\mu\,B^{\mu\nu} = 0 = \partial_\nu\,B^{\mu\nu}$.  
The latter conditions (i.e. $\partial_\mu\,B^{\mu\nu} = 0, \partial_\nu\,B^{\mu\nu} = 0$) emerge out
 because of  the totally  antisymmetric nature of $H^{\mu\nu\eta}$ (present in the original equation: 
$\partial_\mu H^{\mu\nu\eta} + m^2\,B^{\nu\eta} = 0$). Ultimately, we obtain the usual Klein-Gordan 
equation [i.e. $(\Box + m^2)\, B_{\mu\nu} = 0$] for the {\it massive} Abelian 2-form field $(B_{\mu\nu})$.
This observation, in a subtle way, implies that all the numerical factors in Eq. (1)
are {\it correct} with their proper signatures. 

By the application of  St{\"u}ckelberg's technique, it can be checked that, we have the following 
appropriate  transformation [24] of the antisymmetric tensor field $B_{\mu\nu}$:  
\begin{eqnarray}
B_{\mu\nu}  \longrightarrow   B_{\mu\nu} - \frac{1}{m}\, \Phi_{\mu\nu} - \frac{1}{2\,m}\, \varepsilon_{\mu\nu\eta\kappa}\, \tilde \Phi^{\eta\kappa} & \equiv &  B_{\mu\nu} - \frac{1}{m}\, \big(\partial_\mu \phi_\nu - \partial_\nu \phi_\mu + \varepsilon_{\mu\nu\eta\kappa}\, \partial^\eta \tilde \phi^\kappa\big)\nonumber\\
&\equiv & B_{\mu\nu} - \frac{1}{m}\, \Phi_{\mu\nu} - \frac{1}{m}\, {\cal F}_{\mu\nu}.
\end{eqnarray}
In the above, the Abelian 2-form $\Phi^{(2)} = \frac{1}{2!}\,(dx^\mu \wedge dx^\nu)\,\Phi_{\mu\nu} \equiv d \,\Phi^{(1)}$ 
(with vector 1-form  $\Phi^{(1)} = dx^\mu\, \phi_\mu \Rightarrow \Phi_{\mu\nu} = \partial_\mu \phi_\nu 
- \partial_\nu \phi_\mu$) is obtained from a vector field $\phi_\mu$. On the contrary, the   dual
antisymmetric tensor $\tilde \Phi_{\mu\nu} = \partial_\mu \tilde \phi_\nu - \partial_\nu \tilde \phi_\mu$ is constructed with the help of an
axial-vector field $\tilde \phi_\mu$ which is defined through  the axial-vector 1-form $\tilde \Phi^{(1)} = dx^\mu \,\tilde \phi_\mu$. In other words, 
the {\it axial} Abelian 2-form: $\tilde{\Phi}^{(2)} = d\,\tilde{\Phi}^{(1)} \equiv \big(\frac{d\,x^{\mu} \wedge d\,x^{\nu}}{2!}\big)\,\tilde{\Phi}_{\mu\nu}$ leads to the derivation: $\tilde{\Phi}_{\mu\nu} = \partial_{\mu}\,\tilde{\phi}_{\nu} - \partial_{\nu}\,\tilde{\phi}_{\mu}$.
We would like to add that we have ${\cal F}_{\mu\nu} = \frac {1}{2!}\varepsilon_{\mu\nu\eta\kappa}\, \tilde \Phi^{\eta\kappa}$
where $\tilde \Phi_{\mu\nu} = \partial_\mu \tilde\phi_\nu - \partial_\nu \tilde\phi_\mu $ is defined, as argued earlier,  from the {\it axial} 
Abelian 2-form $\tilde \Phi ^{(2)}$ to maintain the parity invariance in our Abelian 2-form massive gauge theory.  We shall comment on the
specific structure of the   antisymmetric tensor:
$\partial_\mu \phi_\nu - \partial_\nu \phi_\mu + \varepsilon_{\mu\nu\eta\kappa}\, \partial^\eta \tilde \phi^\kappa$
and its connection with the source-free Maxwell's theory 
in our Conclusions  section (cf. Sec. 7 below). Thus, we observe that, under the  above transformations (2), 
the Lagrangian density ${\cal L}_{(0)}$  transforms (i.e. ${\cal L}_{(0)}\rightarrow {\cal L}_{(1)}$) 
into the following form\footnote{It should be noted that the kinetic term 
(i.e. $\frac{1}{12}H_{\mu\nu\eta}H^{\mu\nu\eta}$) transforms, under (2), to {\it itself} plus extra terms [25]. 
The {\it latter} terms, however, turn out to be derivatives of {\it third} and {\it fourth} order. 
Hence, they have been ignored keeping in mind the renormalizibility of our theory [in four $(3+1)$-dimensions of spacetime].}:  
\begin{eqnarray}
{\cal L}_{(0)} \to {\cal L}_{(1)} &=& \frac{1}{12}\, H^{\mu\nu\eta} H_{\mu\nu\eta} - \frac{m^2}{4}\, B^{\mu\nu} B_{\mu\nu} 
- \frac{1}{4}\, \Phi^{\mu\nu} \Phi_{\mu\nu} + \frac{1}{4}\, \tilde \Phi^{\mu\nu} \tilde \Phi_{\mu\nu} \nonumber\\ 
&+& \frac{m}{2}\, B^{\mu\nu} \Phi_{\mu\nu} + \frac{m}{4}\, \varepsilon^{\mu\nu\eta\kappa} \, B_{\mu\nu} \tilde \Phi_{\eta\kappa}, 
\end{eqnarray}
The above St$\ddot{u}$ckelberg's modified Lagrangian density respects the following continuous 
and infinitesimal gauge symmetry transformations $(\delta_g)$
\begin{eqnarray} 
&&\delta_g B_{\mu\nu}  = - \,(\partial_\mu  \Lambda_\nu - \partial_\nu \Lambda_\mu), 
\quad \delta_g \phi_\mu = (\partial_\mu\,\Lambda -\, m\, \Lambda_\mu), \quad \delta_g \tilde \phi_\mu = 0,
\end{eqnarray}
where $\Lambda_\mu$ and $\Lambda$ are the vector and scalar gauge transformation parameters. 
Under the continuous gauge symmetry transformations $(4)$, the Lagrangian density ${\cal L}_{(1)}$ 
transforms to the following total spacetime derivative, namely;
\begin{eqnarray} 
\delta_g\,{\cal L}_{(1)} = \partial _\mu \Big[-\,m\,\varepsilon  ^{\mu\nu\eta\kappa} \Lambda_\nu\,\partial_\eta\,\tilde\phi_\kappa\Big]. 
\end{eqnarray}
As a consequence, the action integral $S = \int d^4 x\, {\cal L}_{(1)}$ remains invariant under (4)
for the physically well-defined fields which vanish off at infinity.

We would like to end this sub-section with the following remarks. First, we observe that the kinetic term
$\big(\frac{1}{12}H_{\mu\nu\eta}H^{\mu\nu\eta}\big)$ remains invariant under the infinitesimal gauge symmetry transformations
(4). To be precise, we note that the curvature (i.e. the field strength) tensor $H_{\mu\nu\eta}$, owing its origin to the
exterior derivative $d = d\,x^{\mu}\partial_{\mu}\;(d^2 = 0)$ of the de Rham cohomological operators [27-31], remains invariant 
(i.e. $\delta_g H_{\mu\nu\eta} = 0 $) under the infinitesimal {\it gauge} symmetry transformations (4). Second, 
we do note that the axial-vector field $\tilde\Phi_\mu$ and, hence $\tilde\Phi_{\mu\nu} = \partial_\mu\,\tilde\phi_\nu - \partial_\nu\,\tilde\phi _\mu  $   
as well as the kinetic term for {\it this} field, remain invariant (i.e. $\delta_g \tilde\phi_\mu  = 0, \delta_g \tilde\Phi_{\mu\nu} = 0  $)
under the {\it gauge} symmetry transformations (4). Third, we have {\it not} invoked any {\it new}
fields in the theory besides the St$\ddot{u}$ckelberg fields (e.g. vector field ${\cal \phi_\mu}$ and axial-vector field $\tilde\phi_\mu$)
due to the dimensionality of the spacetime and antisymmetric $(B_{\mu\nu} = - B_{\nu\mu})$ nature of the gauge field 
$(B_{\mu\nu})$. Finally, we note that, so far, our gauge symmetry transformations (4) are {\it classical}. However,
they can be elevated to their   {\it quantum} counterparts [i.e. nilpotent and anticommuting (anti-)BRST symmetries]  within the framework 
of BRST formalism (see, Sec. 3 below).


\subsection{Infinitesimal Dual-Gauge Symmetry Transformations}

To discuss the dual-gauge symmetry transformations, we have to add the {\it gauge-fixing} term to the 
St$\ddot{u}$ckelbrg-modified Lagrangian density ${\cal L}_{(1)}$ [cf. Eq. (3)] which respects the {\it classical}
gauge symmetry transformations quoted in Eq. (4). Furthermore, we have to {\it modify} the {\it kinetic term} as well 
as the {\it gauge-fixing} term by invoking  some {\it new } additional  fields. This has already been done 
systematically in our earlier work [24] where physical and mathematical 
arguments have been provided for the {\it same}. The {\it total} Lagrangian 
density with the  {\it modified} kinetic term, gauge-fixing term 
{\it and}  a mass term is as follows (see, e.g. [24] for details) 
\begin{eqnarray}
{\cal L}_{(1)} \to {\cal L}_{(2)}  &=&  -  \frac{1}{2}\bigg(\frac{1}{2}
\varepsilon_{\mu\nu\eta\kappa} \, \partial^\nu B^{\eta\kappa} - \frac{1}{2}\partial_\mu \tilde \varphi + m \,\tilde \phi_\mu\bigg)^2 
- \frac{m^2}{4} \,B^{\mu\nu}B_{\mu\nu}\nonumber\\
& - & \frac{1}{4}\, \Phi^{\mu\nu}\Phi_{\mu\nu} + \frac{m}{2}\, B^{\mu\nu}\Phi_{\mu\nu} + \frac{1}{4}\, \tilde\Phi^{\mu\nu}\tilde\Phi_{\mu\nu} 
+ \frac{m}{4}\, \varepsilon^{\mu\nu\eta\kappa} B_{\mu\nu} \tilde\Phi_{\eta\kappa}\nonumber\\
& + & \frac{1}{2}\left( \partial^\nu B_{\nu\mu}
 - \frac{1}{2}\,\partial_\mu \varphi + m\,\phi_\mu \right)^2 - \frac{1}{2} \left(\partial_\mu \phi^\mu + \frac{1}{2}\,{m}\,\varphi \right)^2\nonumber\\
& + & \frac{1}{2} \left(\partial_\mu \tilde \phi^\mu 
+ \frac{1}{2}m\,\tilde\varphi \right)^2,   
\end{eqnarray}
where the (pseudo-)scalar fields $(\tilde\varphi)\varphi$ are the {\it new} fields and the gauge-fixing term
 $(\partial^\nu B_{\nu\mu})$ for the antisymmetric tensor field owes its origin to the co-exterior derivative 
$\delta =  \pm \ast \, d\, \ast (\delta^2 = 0)$ of the de Rham cohomologial operators of differential geometry 
[26-30] because $\delta B^{(2)} \equiv (\partial^\nu B_{\nu\mu}) \, dx^\mu$.
 Here $\ast$ is the Hodge duality operator. It should be noted that the {\it last} two terms in Eq. (6) are nothing but the gauge-fixing terms for the 
vector and axial-vector fields $\phi_\mu$ and $\tilde\phi_\mu$, respectively. At this stage, it can be noted that
the {\it new} fields ($ \varphi, \tilde\varphi, \phi_\mu, \tilde\phi_\mu$) have mass dimension $[M]$ in the natural units
(i.e. $\hbar = c = 1$) for our present  theory.

It is obvious that the above {\it modified} Lagrangian density (6) would {\it not} have the {\it perfect} gauge symmetry 
transformations (4) because of  the additional terms. However, it is very interesting  to note that  
under the following infinitesimal and continuous transformations 
\begin{eqnarray} 
&&\delta_{dg} B_{\mu\nu}  = -\,\varepsilon _{\mu\nu\eta\kappa}\partial^\eta \Sigma ^\kappa, \qquad \delta_{dg} \tilde\phi_\mu = (\partial_\mu\,\Sigma -\, m\, \Sigma_\mu), \qquad \delta_{dg}  \phi_\mu = 0, \nonumber \\
&&\delta_{dg}\,\tilde\varphi = -\;\sigma, \qquad \delta_{dg}\,\varphi = 0, \qquad \delta_{dg}\big[\partial^\nu B_{\nu\mu} - \frac{1}{2}\,\partial_\mu \varphi + m\,\phi_\mu \big] = 0,    \nonumber \\
&&\delta_g B_{\mu\nu}  = - \,(\partial_\mu  \Lambda_\nu - \partial_\nu \Lambda_\mu), \qquad \delta_g \phi_\mu = (\partial_\mu\,\Lambda -\, m\, \Lambda_\mu), \qquad \delta_g \tilde \phi_\mu = 0, \nonumber \\
&&\delta_g\,\varphi = \tilde{\lambda}, \quad \delta_g\,\tilde\varphi = 0, \quad \delta_g \Big[\frac{1}{2}\varepsilon_{\mu\nu\eta\kappa} \, \partial^\nu B^{\eta\kappa} - \frac{1}{2}\partial_\mu \tilde \varphi + m \,\tilde \phi_\mu \Big] = 0,
\end{eqnarray}
we obtain the following transformations for the Lagrangian density (6):
\begin{eqnarray}
 \delta_{dg}\,{\cal L}_{(2)} &=& \partial_\mu\Big[\frac{1}{2}\,\tilde{\varphi}\,\partial^\mu\big((\partial\cdot\Sigma)
 + \frac{1}{2}\,\sigma + m\,\Sigma\big)\nonumber\\ 
& - & \big((\partial\cdot\Sigma) + \frac{1}{2}\,\sigma 
 m\,\Sigma\big)\Big(\frac{1}{2}\,\varepsilon^{\mu\nu\eta\kappa}\partial_{\nu}B_{\eta\kappa} + m\,{\tilde{\phi}}^\mu\Big)\nonumber\\ 
& - & \frac{1}{2}\,\tilde{\varphi}\,(\Box + m^2)\,\Sigma^\mu
  + \frac{m}{2}\,\varepsilon^{\mu\nu\eta\kappa}\Sigma_\nu\Phi_{\eta\kappa}\Big]\nonumber\\
& + & \Big(\frac{1}{2}\,\varepsilon^{\mu\nu\eta\kappa}
\partial_{\nu}B_{\eta\kappa} + m\,{\tilde{\phi}}^\mu\Big)(\Box + m^2)\,\Sigma_\mu \nonumber \\ 
&+& (\partial\cdot\tilde{\phi})\;(\Box + m^2)\,\Sigma - \frac{1}{4}\,\tilde{\varphi}\;(\Box + m^2)\,\sigma,\nonumber\\
\delta_{g}\,{\cal L}_{(2)} &=& -\,\partial_\mu\Big[\frac{1}{2}\,{\varphi}\,\partial^\mu\big((\partial\cdot\Lambda) 
- \frac{1}{2}\,\tilde{\lambda} + m\,\Lambda\big)\nonumber\\
& - & \big((\partial\cdot\Lambda) - \frac{1}{2}\,\tilde{\lambda} 
 m\,\Lambda\big)\Big(\partial_{\nu}B^{\nu\mu} + m\,{{\phi}}^\mu\Big)\nonumber\\
& - & \frac{1}{2}{\varphi}\,(\Box + m^2)\,\Lambda^\mu 
+ \frac{m}{2}\,\varepsilon^{\mu\nu\eta\kappa}\Lambda_\nu\tilde{\Phi}_{\eta\kappa}\Big]\nonumber\\
& - & \Big(\partial_{\nu}B^{\nu\mu} - m\,{{\phi}}^\mu\Big)(\Box + m^2)\,\Lambda_\mu\nonumber\\ 
& - & (\partial\cdot{\phi})\;(\Box + m^2)\,\Lambda -
 \frac{1}{4}\,{\varphi}\;(\Box + m^2)\,\tilde{\lambda}.   
\end{eqnarray}
We christen the above transformations  [cf. Eqs. (7), (4)] as the (dual-)gauge symmetry transformations because of 
the following arguments. First of all, we note that under the gauge symmetry transformations $(\delta_g)$, 
the {\it total} kinetic term remains invariant. On the other hand, it is the {\it total} gauge-fixing term that remains invariant 
under the dual-gauge symmetry transformations $(\delta_{dg})$. Second, as argued earlier, the kinetic term has its origin in the exterior  
derivative $d = dx^\mu\,\partial_\mu$ (with $d^2 = 0$) {\it but}  the gauge-fixing term owes its origin to the dual-exterior (i.e. co-exterior) 
derivative $\delta = \pm \ast\, d \,\ast  $ (with $\delta^2 = 0$). Thus, the nomenclature (dual-)gauge 
symmetry transformation $\delta_{(d)g}$ sounds  appropriate. Finally, even though there are {\it three} individual terms in the kinetic 
and gauge-fixing terms,  the origin of the additional terms like 
$[-\frac{1}{2}\partial_\mu \tilde{\varphi} + m\,\tilde{\phi}_\mu]$ 
and $[-\frac{1}{2}\partial_\mu \varphi + m\,\phi_\mu]$ is the $H^{(3)} = d\,B^{(2)}$ 
and $\delta\,B^{(2)} = (\partial^\nu B_{\nu\mu})\,dx^\mu$ which owe their origins 
to $d = d\,x^\mu \partial_\mu$ and $\delta = \pm \star d \star$, respectively. 
It is evident that the transformation parameters $(\Sigma_\mu, \Sigma, \sigma)$ are the axial-vector 
and pseudo-scalars (i.e. $\Sigma, \sigma)$ for the {\it full} dual-gauge symmetry transformations $(\delta_{dg})$.
 On the other hand, we have already noted that the Lorentz vector ($\Lambda_\mu$) and
Lorentz scalars ($\Lambda, \tilde\lambda$) are the transformation parameters for the {\it full}
 gauge symmetry transformations ($\delta_g$).

At this stage, we now comment on the transformations (8) which have been obtained after the applications of 
$\delta_{dg}$ and $\delta_{g}$. It is straightforward 
to note that under the following restrictions on the (dual-)gauge transformation parameters, namely;
\begin{eqnarray}
&&(\Box + m^2)\,\Sigma_\mu = 0, \quad\qquad (\Box + m^2)\,\Sigma = 0, \quad\qquad (\Box + m^2)\,\sigma = 0, \nonumber\\
&&(\Box + m^2)\,\Lambda_\mu = 0, \quad\qquad (\Box + m^2)\,\Lambda = 0, \qquad\quad (\Box + m^2)\,\tilde{\lambda} = 0,
\end{eqnarray} 
we achieve the {\it perfect} (dual-)gauge symmetry invariance of the Lagrangian density ${\cal L}_{(2)}$. It is very
 interesting to pinpoint that the mathematical structure of the restrictions in Eq. (9) is {\it exactly} the {\it same} on the (dual-)gauge 
symmetry transformation parameters. Thus, it is very clear that, within the framework of BRST 
approach, the (anti-)ghost part of the Lagrangian density would be exactly the {\it same} for the
 coupled (but equivalent) Lagrangian densities (cf. Sec. 3 below). The bosonic nature of the
 transformation parameters $(\Sigma_\mu, \Sigma, \sigma, \Lambda_\mu, \Lambda, \tilde{\lambda})$ 
implies that, at the {\it quantum} level, these parameters would be replaced by the {\it fermionic} (anti-)ghost fields
within the framework of BRST formalism.

We wrap up this sub-section with the following remarks. We note, as pointed out earlier,
that total gauge-fixing term remains invariant under the dual-gauge symmetry transformations 
$(\delta_{dg})$. Furthermore, the vector and scalar fields ($\phi_\mu$ and $\varphi$)  {\it do}
not transform at all under $\delta_{dg}$. Hence, the kinetic term ($-\, \frac {1}{4} \Phi_{\mu\nu}\, \Phi^{\mu\nu}$) for the vector field
$(\phi_\mu)$ {\it also} does {\it not} transform under the dual-gauge symmetry transformations. Whereas the gauge symmetry transformations $(\delta_g)$
exist at the {\it classical} level, we note that the dual-gauge symmetry transformations $(\delta_{dg})$ exist {\it only}
when the gauge-fixing term is incorporated for the purpose of {\it quantization}  (or the derivation  of the propagator)
 for the gauge field  $(B_{\mu\nu})$ of our {\it massive} 4D theory.  Finally, we observe that {\it new } fields $(\varphi, \tilde\varphi)$
 have been invoked for the discussion of the dual-gauge symmetries transformations $(\delta_{dg})$ in our theory [in contrast to the
gauge symmetry transformations $(\delta_{g})$ where {\it no} such fields have been invoked (see, e.g., the discussions after Eq. (5))].


\section {Coupled Lagrangian Densities: Off-Shell Nilpotent (Anti-)BRST and (Anti-)co-BRST Symmetry Transformations}

\vskip 0.4cm

In this section, we concisely mention the off-shell nilpotent symmetries and the
CF-type restrictions  for the most generalized version of the 
Lagrangian density (6) where the Nakanishi-Lautrup type auxiliary fields are invoked for the linearizations   of the 
kinetic and gauge-fixing terms for the $B_{\mu\nu}$ field {\it and}  
gauge-fixing terms for the fields $\phi_\mu$ and $\tilde\phi_\mu$.
The central theme and purpose of this section is to mention {\it all} the appropriate  equations in the 4D ordinary spacetime 
which are important in the context of superfield approach to BRST formalism (cf. Secs. 4, 5, 6 below). 
We begin with the following coupled (but equivalent) (dual-)BRST invariant 
Lagrangian densities (see, e.g. [24] for details)
\begin{eqnarray}
{\cal L}_{(B,\cal B)} &=& \frac{1}{2}{\cal B}_\mu \;{\cal B}^\mu - {\cal B}^\mu \left(\frac{1}{2}
\varepsilon_{\mu\nu\eta\kappa} \, \partial^\nu B^{\eta\kappa} - \frac{1}{2}\,\partial_\mu \tilde \varphi + m \;\tilde \phi_\mu\right) 
- \frac{m^2}{4} \,B^{\mu\nu}B_{\mu\nu}\nonumber\\
& - & \frac{1}{4}\, \Phi^{\mu\nu}\Phi_{\mu\nu} + \frac{m}{2}\, B^{\mu\nu}\Phi_{\mu\nu} + \frac{1}{4}\, \tilde\Phi^{\mu\nu}\tilde\Phi_{\mu\nu}
+ \frac{m}{4}\, \varepsilon^{\mu\nu\eta\kappa} B_{\mu\nu} \tilde\Phi_{\eta\kappa} 
- \frac{1}{2}\,B^\mu B_\mu  \nonumber\\
&+& B^{\mu}\left( \partial^\nu B_{\nu\mu} - \frac{1}{2}\, \partial_\mu \varphi + m \;\phi_\mu \right) + \frac{1}{2}\, B^2 
+ B \left(\partial_\mu \phi^\mu + \frac{m}{2} \,\varphi \right) \nonumber\\
&-& \frac{1}{2}\,{\cal B}^2 - {\cal B} \left(\partial_\mu \tilde \phi^\mu + \frac{m}{2} \, \tilde\varphi \right)
+  \big(\partial_\mu \bar C - m \;\bar C_\mu \big) \big(\partial^\mu C - m \;C^\mu \big) \nonumber\\
&-& \big(\partial_\mu \bar C_\nu - \partial_\nu \bar C_\mu \big) \big(\partial^\mu C^\nu \big) 
- \frac{1}{2}\,\partial_\mu \bar \beta \,\partial^\mu \beta + \frac{1}{2}\, m^2\, \bar \beta \beta \nonumber\\
&-& \frac{1}{2}\left(\partial_\mu \bar C^\mu +  m \, \bar C + \frac{\rho}{4} \right) \lambda 
- \frac{1}{2}\left(\partial_\mu C^\mu +  m \, C - \frac{\lambda}{4} \right) \rho, 
\end{eqnarray}
\begin{eqnarray}
{\cal L}_{(\bar B,\bar {\cal B})} &=& \frac{1}{2} \bar {\cal B}_\mu \bar {\cal B}^\mu + \bar {\cal B}^\mu \left(\frac{1}{2}
\varepsilon_{\mu\nu\eta\kappa} \, \partial^\nu B^{\eta\kappa} + \frac{1}{2}\,\partial_\mu \tilde \varphi + m\; \tilde \phi_\mu\right) 
- \frac{m^2}{4} \,B^{\mu\nu}B_{\mu\nu}\nonumber\\
& - & \frac{1}{4}\, \Phi^{\mu\nu}\Phi_{\mu\nu}  + \frac{m}{2}\, B^{\mu\nu}\Phi_{\mu\nu} + \frac{1}{4}\, \tilde\Phi^{\mu\nu}\tilde\Phi_{\mu\nu}
+ \frac{m}{4}\, \varepsilon^{\mu\nu\eta\kappa} B_{\mu\nu} \tilde\Phi_{\eta\kappa} 
- \frac{1}{2}\, \bar B^\mu \bar B_\mu  \nonumber\\
&-& \bar B^{\mu}\left( \partial^\nu B_{\nu\mu} + \frac{1}{2}\, \partial_\mu \varphi + m\, \phi_\mu \right) 
+ \frac{1}{2}\, \bar B^2 - \bar B \left(\partial_\mu \phi^\mu - \frac{m}{2} \,\varphi \right) \nonumber\\
&-& \frac{1}{2}\,\bar {\cal B}^2 + \bar {\cal B} \left(\partial_\mu \tilde \phi^\mu - \frac{m}{2} \, \tilde\varphi \right)
+  \big(\partial_\mu \bar C - m\, \bar C_\mu \big) \big(\partial^\mu C - m\, C^\mu \big) \nonumber\\
&-& \big(\partial_\mu \bar C_\nu - \partial_\nu \bar C_\mu \big) \big(\partial^\mu C^\nu \big)
- \frac{1}{2}\,\partial_\mu \bar \beta \,\partial^\mu \beta + \frac{1}{2}\, m^2\, \bar \beta \beta \nonumber\\
&-& \frac{1}{2}\left(\partial_\mu \bar C^\mu +  m \, \bar C + \frac{\rho}{4} \right) \lambda 
- \frac{1}{2}\left(\partial_\mu C^\mu +  m \, C - \frac{\lambda}{4} \right) \rho, 
\end{eqnarray}
where the auxiliary fields $({\cal B}_\mu, B_\mu,  B,  {\cal B},\bar{\cal B}_\mu, \bar B_\mu,  \bar B,  \bar {\cal B})$ 
are nothing but the {\it bosonic} Nakanishi-Lautrup type auxiliary fields. The {\it fermionic} 
(anti-)ghost fields\footnote{These {\it fermionic} (anti-)ghost fields are the generalizations 
of the {\it bosonic} (dual-)gauge symmetry transformation parameters ($\Sigma_\mu, \Sigma, \sigma, \Lambda_\mu, \Lambda, \tilde\lambda $) 
which have been mentioned at the fag end of Sec. 2 of our present endeavor.} are:
$(\bar C_\mu) C_\mu$, $(\bar C)C$,  $(\rho) \lambda$ and {\it bosonic}
(anti-)ghost fields are $(\bar \beta) \beta$. Because of the stage-one reducibility in our theory, we 
have the ghost-for-ghost {\it bosonic} fields $(\bar \beta) \beta$.
It should be noted that the ghost-part of the Lagrangian densities (10) and (11) are {\it same}. We have provided 
some arguments regarding it  (cf. Sec. 2) in the language of the (dual-)gauge symmetry transformations and the 
restrictions on the transformation gauge parameters [cf. Eq. (9)] for their invariance. We also note here that the 
fields $(\rho)\lambda$ are  auxiliary fields but they are {\it fermionic} in nature and they carry the ghost number 
(-1)+1, respectively. In addition, the ghost numbers 
for the {\it fermionic} (anti-)ghost fields $(\bar C_\mu)C_\mu$ and $(\bar C)C$ 
are (-1)+1 and that of the bosonic (anti-)ghost fields $(\bar\beta)\beta$ are (-2)+2, respectively.

We observe that the following, nilpotent ($s_{(a)b}^2 = 0$) (anti-)BRST 
 transformations ($s_{(a)b}$) 
\begin{eqnarray}
&&  s_{ab} B_{\mu\nu} = - (\partial_\mu \bar C_\nu - \partial_\nu \bar C_\mu), \quad 
s_{ab} \bar C_\mu  = - \partial_\mu \bar \beta, \quad s_{ab}  C_\mu =  \bar B_\mu, \quad s_{ab} \beta = - \lambda, \nonumber\\
&&  \;\;  s_{ab} \bar C = - m\, \bar \beta, 
\;\; s_{ab}  C = \bar B, \;\;s_{ab} B = -\, m \,\rho, \;\; s_{ab} \varphi = \rho, s_{ab} B_\mu =  - \partial_\mu \rho,  
\;\; \nonumber\\
&&s_{ab} \phi_\mu = \partial_\mu \bar C - m\, \bar C_\mu,\;\;  s_{ab} [\bar B, \rho, \lambda, \bar \beta, \bar B_\mu, {\cal B}_\mu, {\bar{\cal B}}_\mu, 
\tilde\phi_\mu, \tilde \varphi, {\cal B}, \bar {\cal B},  H_{\mu\nu\kappa}] = 0, 
\end{eqnarray}
\begin{eqnarray}
&&  s_b B_{\mu\nu} = - (\partial_\mu C_\nu - \partial_\nu C_\mu), \quad 
s_b C_\mu  = - \partial_\mu \beta, \quad s_b \bar C_\mu = B_\mu, \quad s_b \bar \beta = - \rho, \nonumber\\
&& s_b \phi_\mu = \partial_\mu C - m\, C_\mu, \;\; s_b C = - \,m\,\beta, 
\;\; s_b \bar C =  B, \;\; s_b \bar B = -\, m\, \lambda, \; s_b \varphi = \lambda,\nonumber\\
&& s_b \bar B_\mu =  - \partial_\mu \lambda,  
\quad s_b [B, \rho, \lambda, \beta, B_\mu, {\cal B}_\mu, {\bar {\cal B}}_\mu, \tilde\phi_\mu,
\tilde \varphi, {\cal B}, \bar {\cal B}, H_{\mu\nu\kappa}] = 0, \quad
\end{eqnarray}
leave the action integrals $S_1 = \int d^4 x \;{\cal L}_{(B,{\cal B})}$ and $S_2
 = \int\, d^4 x\; {\cal L}_{(\bar B,\bar {\cal B})}$ invariant because the Lagrangian densities
transform to the total spacetime derivatives under the $s_{(a)b}$ as follows [24]:
\begin{eqnarray}
s_{ab} {\cal L}_{(\bar B, \bar {\cal B})} &=& - \partial_\mu \bigg[m\, \varepsilon^{\mu\nu\eta\kappa}\, 
\tilde \phi_\nu \big(\partial_\eta \bar C_\kappa \big) 
- \bar B_\nu \big(\partial^\mu \bar C^\nu - \partial^\nu \bar C^\mu  \big) + \frac{1}{2}\, \bar B^\mu\, \rho \nonumber\\
&+& \bar B \big(\partial^\mu \bar C - m\, \bar C^\mu\big) - \frac{1}{2}\, \big(\partial^\mu \bar \beta \big)\,\lambda \bigg],
\end{eqnarray}
\begin{eqnarray}
s_b {\cal L}_{(B, {\cal B})} &=& - \partial_\mu \bigg[m\, \varepsilon^{\mu\nu\eta\kappa}\, \tilde \phi_\nu \big(\partial_\eta C_\kappa \big) 
+ B_\nu \,\big(\partial^\mu C^\nu - \partial^\nu C^\mu  \big) + \frac{1}{2}\, B^\mu\, \lambda \nonumber\\
&-& B \,\big(\partial^\mu C - m \,C^\mu\big) - \frac{1}{2}\, \big(\partial^\mu \beta \big)\,\rho \bigg].
\end{eqnarray}
We point out that the above action integrals remain invariant due to Gauss's divergence theorem (as {\it all}
the physical fields vanish off at infinity).
It is very interesting to note that the above coupled (but equivalent) Lagrangian densities  
${\cal L}_{(\bar B,\bar {\cal B})}$ and ${\cal L}_{(\bar B,\bar {\cal B})}$ {\it also} respect another set of
off-shell nilpotent ($s_{(a)d}^2 = 0$) symmetries which are known as  the  (anti-)co-BRST
[or (anti-)dual BRST] symmetries [$s_{(a)d}$]. This is due to the fact that we observe the 
following transformations for the coupled (but equivalent) Lagrangian densities:
\begin{eqnarray}
s_{ad} {\cal L}_{(\bar B, {\cal{\bar B}})} &=& - \partial_\mu \, 
\bigg[ m\, \varepsilon^{\mu\nu\eta\kappa} \phi_\nu \,\big(\partial_\eta C_\kappa \big) 
+ \bar {\cal B}_\nu \big(\partial^\mu C^\nu-\partial^\nu C^\mu \big) + \frac{1}{2}\,\bar {\cal B}^\mu\, \lambda  \nonumber\\
&-& \bar {\cal B} \,\big(\partial^\mu C - m \,C^\mu \big) + \frac{1}{2}\, \big(\partial^\mu \beta \big)\,\rho \bigg], 
\end{eqnarray}
\begin{eqnarray}
s_d {\cal L}_{(B, {\cal B})} &=& - \partial_\mu \, 
\bigg[ m\, \varepsilon^{\mu\nu\eta\kappa} \phi_\nu \big(\partial_\eta \bar C_\kappa \big) 
- {\cal B}_\nu \, \big(\partial^\mu {\bar C}^\nu-\partial^\nu \bar C^\mu \big) + \frac{1}{2}\, {\cal B}^\mu\, \rho  \nonumber\\
&+&  {\cal B}\, \big(\partial^\mu \bar C - m \;\bar C^\mu \big) - \frac{1}{2}\, \big(\partial^\mu \bar\beta \big)\,\lambda \bigg]. 
\end{eqnarray}
Hence, the action integrals $S_1 = \int d^4 x \;{\cal L}_{(B, {\cal B})}$ and $S_2
 = \int d^4 x\; {\cal L}_{(\bar B,\bar {\cal B})}$ remain invariant under the following
infinitesimal and continuous (anti-)co-BRST symmetry transformations [$s_{(a)d}$]: 
\begin{eqnarray}
s_{ad} B_{\mu\nu} = - \,\varepsilon_{\mu\nu\eta\kappa} \partial^\eta  C^\kappa, \quad 
s_{ad} \bar C_\mu = {\bar{\cal  B}}_\mu, \quad\; s_{ad} C_\mu  =  \partial_\mu  \beta, 
\quad\; s_{ad} \tilde \varphi = -\,\lambda,\nonumber\\
s_{ad} \tilde\phi_\mu = \partial_\mu  C - m\,  C_\mu,
 \quad  s_{ad}   C =  m\, \beta, 
 \qquad s_{ad}  {\cal B} =  m \,\lambda,  \quad s_{ad} {\cal{ B}}_\mu =   \partial_\mu \lambda, \nonumber\\
s_{ad}  \bar\beta =  \rho,  \, s_{ad} {\bar C} =  \bar {\cal  B},\,
 s_{ad} [\partial^\nu B_{\nu\mu}, B_\mu, {\bar{\cal B}}_\mu,
 \bar B_\mu, \bar {\cal B}, B, \bar B, \varphi, \phi_\mu, \rho, \lambda, \beta] = 0,
 \end{eqnarray}
\begin{eqnarray}
s_d B_{\mu\nu} = - \varepsilon_{\mu\nu\eta\kappa} \partial^\eta \bar C^\kappa, \qquad 
s_d  C_\mu = {\cal B}_\mu, \qquad s_d \bar C_\mu  = - \partial_\mu \bar \beta, 
\qquad s_d  \beta = - \lambda, \nonumber\\
s_d \tilde\phi_\mu = \partial_\mu \bar C - m\, \bar C_\mu,
\qquad s_d  C =  {\cal B}, \qquad\;  \quad s_d \bar  C = - \,m\,\bar \beta, 
 \qquad s_d \bar {\cal B} =  m\, \rho, \nonumber\\
s_d {\bar {\cal B}}_\mu =   \partial_\mu \rho, \; s_d \tilde \varphi = -\,\rho, \;\;
 s_d [\partial^\nu B_{\nu\mu}, B_\mu, {\cal{ B}}_\mu,
 \bar B_\mu, {\cal B}, B, \bar B, \varphi, \phi_\mu, \rho, \lambda, \bar\beta] = 0. 
 \end{eqnarray}
It is straightforward to note that the (anti-)BRST and (anti-)co-BRST symmetry transformations 
are off-shell nilpotent $(s_{(a)b}^2 = 0, s_{(a)d}^2 = 0)$ of order two and, hence, are {\it fermionic}
in nature.  We note that the {\it total} gauge-fixing term  for the massive gauge field 
$B_{\mu\nu}$ remains invariant under $s_{(a)d}$. This observation should be contrasted  with the (anti-)BRST
symmetry transformations where the {\it total} kinetic term for the massive gauge field $B_{\mu\nu}$ is found to 
be invariant.  Furthermore, it is worth pointing out that 
the BRST as well as anti-co-BRST transformations {\it increase} the ghost number by {\it one} and anti-BRST 
as well as the co-BRST symmetry transformations {\it decrease} the 
 ghost number by {\it one} when  they {\it operate} on an ordinary field.

A few comments are in order as far as the absolute anticommutativity $({\{s_b, s_{ab}}\} = 0, {\{s_d, s_{ad}}\} = 0)$ of 
the (anti-)BRST $(s_{(a)b})$ and (anti-)co-BRST $(s_{(a)d})$  symmetry transformations  are concerned. It can be checked that the following 
anticommutators, namely;
\begin{eqnarray}
&& \{s_b, \,s_{ab}\}\,B_{\mu\nu} = - \partial_\mu\big(B_\nu + \bar B_\nu \big) + \partial_\nu \big(B_\mu + \bar B_\mu \big),  \nonumber\\
&& \{s_b, \,s_{ab}\}\,\phi_\mu = \partial_\mu\big(B + \bar B \big) - m\,\big(B_\mu + \bar B_\mu \big), \nonumber \\
&&{\{s_d, s_{ad}\}}\,B_{\mu\nu} = -\,\varepsilon_{\mu\nu\eta\kappa}\partial^\eta({\cal B}^{\kappa} + \bar {\cal B}^\kappa),\nonumber\\
&&{\{s_d, s_{ad}\}}\, \tilde\phi_\mu = \partial_\mu({\cal B} + \bar {\cal B}) - m\,({\cal B}_\mu + {\bar {\cal B}}_\mu), 
\end{eqnarray}
are equal to zero {\it only} when the following  {\it physically}  allowed CF-type restrictions [24] are imposed
from {\it outside}, namely;
\begin{eqnarray}
&&B_\mu + \bar{B}_\mu + \partial_\mu \varphi = 0, \qquad \qquad B + \bar{B} +m\,\varphi = 0, \nonumber \\
&&{\cal B}_\mu + \bar{\cal B}_\mu + \partial_\mu \tilde{\varphi} = 0, \qquad \qquad \; {\cal B} + \bar{\cal B} +m\,\tilde{\varphi} = 0. 
\end{eqnarray}     
We can explicitly check that the above CF-type restrictions are (anti-)BRST as well as (anti-)co-BRST {\it invariant}. 
To corroborate the above statement, we point out the following precise observations:
\begin{eqnarray}
&&s_b \big[B_\mu + \bar{B}_\mu + \partial_\mu \varphi\big] = 0, \qquad \;\, s_d \big[{\cal B}_\mu 
+ \bar{\cal B}_\mu + \partial_\mu \tilde{\varphi}\big] = 0, \nonumber \\
&&s_{ab}\big[B_\mu + \bar{B}_\mu + \partial_\mu \varphi\big] = 0,     
\qquad    s_{ad}\big[{\cal B}_\mu + \bar{\cal B}_\mu + \partial_\mu \tilde{\varphi}\big] = 0, \nonumber \\
&&s_d\big[B + \bar{B} +m\,\varphi\big] = 0,  \qquad \quad\, s_b\big[{\cal B} + \bar{\cal B} +m\,\tilde{\varphi}\big] = 0, \nonumber \\
&&s_{ad}\big[B + \bar{B} +m\,\varphi\big] = 0, \qquad \;\;\;   s_{ab}\big[{\cal B} + \bar{\cal B} +m\,\tilde{\varphi}\big] = 0.
\end{eqnarray}  
Thus, for a model of the Hodge theory (i.e. 4D {\it massive} Abelian 2-form gauge theory), the CF-type restrictions (21) are 
{\it physical} constraints on the theory because the restrictions (21) are (anti-)BRST as well as
 (anti-)co-BRST invariant, {\it together}. It would be worthwhile to note that some of the pertinent
 equations of motion from the coupled Lagrangian densities   
${\cal L}_{(B, {\cal B})}$ and ${\cal L}_{(\bar B, \bar {\cal B})}$ are:
\begin{eqnarray}
&& {\cal B}_\mu  = \left(\frac{1}{2}
\varepsilon_{\mu\nu\eta\kappa} \, \partial^\nu B^{\eta\kappa} - \frac{1}{2}\,\partial_\mu \tilde \varphi + m \;\tilde \phi_\mu\right), \quad B = - \left(\partial_\mu \phi^\mu + \frac{m}{2} \,\varphi \right), 
 \nonumber\\
&&B_{\mu} = \left(\partial^\nu B_{\nu\mu} - \frac{1}{2}\, \partial_\mu \varphi + m \;\phi_\mu \right), \qquad\quad 
{\cal B} = - \left(\partial_\mu \tilde \phi^\mu + \frac{m}{2}\, \tilde\varphi \right), \nonumber\\
&& \bar {\cal B}_\mu  = - \left(\frac{1}{2}
\varepsilon_{\mu\nu\eta\kappa} \, \partial^\nu B^{\eta\kappa} + \frac{1}{2}\,\partial_\mu \tilde \varphi + m \;\tilde \phi_\mu\right), \qquad \bar B =  \left(\partial_\mu \phi^\mu - \frac{m}{2} \,\varphi \right),
 \nonumber\\
&& \bar B_{\mu} = - \left(\partial^\nu B_{\nu\mu} + \frac{1}{2}\, \partial_\mu \varphi + m \;\phi_\mu \right),  \qquad\quad\;\; 
\bar {\cal B} =  \left(\partial_\mu \tilde \phi^\mu - \frac{m}{2}\, \tilde\varphi \right).
\end{eqnarray}
We note that the (anti-)BRST and (anti-)co-BRST invariant restrictions (21) can be
derived from the above equations (23). Thus, in a subtle manner, we provide  the derivation of 
the CF-type restrictions (21)   from the coupled (but equivalent) Lagrangian  densities 
${\cal L}_{(B, {\cal B})}$ and ${\cal L}_{(\bar B, \bar {\cal B})}$ in the sense that the appropriate EL-EOM
 lead to {\it their} existence on our theory.

In the context of the existence of the above CF-type restrictions (21), we note the following transformations
for the  (anti-)BRST {\it and} (anti-)co-BRST invariant  Lagrangian densities ${\cal L}_{(B, {\cal B})}$ and ${\cal L}_{(\bar B, \bar {\cal B})}$: 
\begin{eqnarray}
s_{ab} {\cal L}_{(B, {\cal B})} &=& - \partial_\mu \bigg[m\, \varepsilon^{\mu\nu\eta\kappa}\, \tilde \phi_\nu \big(\partial_\eta \bar C_\kappa \big)
+ \Big(\partial_\nu B^{\nu\mu} + \frac{1}{2}\, \bar B^\mu + m \,\phi^\mu\Big) \rho \nonumber\\
&+& B_\nu \big(\partial^\mu \bar C^\nu - \partial^\nu \bar C^\mu  \big)  
- B \big(\partial^\mu \bar C - m\; \bar C^\mu\big) - \frac{1}{2}\, \big(\partial^\mu \bar \beta \big)\,\lambda \bigg] \nonumber\\
&+& \frac{1}{2}\, \big[B_\mu + \bar B_\mu + \partial_\mu \varphi \big] \big(\partial^\mu \rho \big) 
+ \partial_\mu\big[B_\nu + \bar B_\nu + \partial_\nu \varphi  \big] \big(\partial^\mu \bar C^\nu - \partial^\nu \bar C^\mu  \big) \nonumber\\
&+& m \big[B_\mu + \bar B_\mu + \partial_\mu \varphi \big] \big(\partial^\mu \bar C - m\, \bar C^\mu  \big)
- \frac{m}{2}\, \big[B + \bar B + m\; \varphi \big] \rho \nonumber\\
&-& \partial_\mu \big[B + \bar B + m\; \varphi \big] \big(\partial^\mu \bar C - m \,\bar C^\mu  \big),
\end{eqnarray} 
\begin{eqnarray}
s_b {\cal L}_{(\bar B, \bar {\cal B})} &=& - \partial_\mu \bigg[m\, \varepsilon^{\mu\nu\eta\kappa}\, \tilde \phi_\nu \big(\partial_\eta C_\kappa \big)
- \Big(\partial_\nu B^{\nu\mu} - \frac{1}{2}\, B^\mu + m\, \phi^\mu\Big) \lambda \nonumber\\
&-& \bar B_\nu \big(\partial^\mu C^\nu - \partial^\nu C^\mu  \big)  
+ \bar B \big(\partial^\mu C - m \, C^\mu\big) - \frac{1}{2}\, \big(\partial^\mu \beta \big)\,\rho \bigg] \nonumber\\
&+& \frac{1}{2}\, \big[B_\mu + \bar B_\mu + \partial_\mu \varphi \big] \big(\partial^\mu \lambda \big) 
- \partial_\mu\big[B_\nu + \bar B_\nu + \partial_\nu \varphi  \big] \big(\partial^\mu  C^\nu - \partial^\nu  C^\mu  \big) \nonumber\\
&-& m \big[B_\mu + \bar B_\mu + \partial_\mu \varphi \big] \big(\partial^\mu  C - m \, C^\mu  \big)
- \frac{m}{2}\, \big[B + \bar B + m \,\varphi \big] \lambda \nonumber\\
&+& \partial_\mu \big[B + \bar B + m \;\varphi \big] \big(\partial^\mu  C - m \, C^\mu  \big),
\end{eqnarray}
\begin{eqnarray}
s_d {\cal L}_{(\bar B, \bar {\cal B})} &=& - \partial_\mu \bigg[m\, \varepsilon^{\mu\nu\eta\kappa}\,  \phi_\nu \big(\partial_\eta \bar C_\kappa \big)
- \Big(\frac{1}{2}\,\varepsilon^{\mu\nu\eta\kappa}\,\partial_\nu B_{\eta\kappa} - \frac{1}{2}\, {\cal B}^\mu + m \,\tilde \phi^\mu\Big) \rho \nonumber\\
&+& \bar {\cal B}_\nu \big(\partial^\mu \bar C^\nu - \partial^\nu \bar C^\mu  \big)  
- \bar {\cal B} \big(\partial^\mu \bar C - m\, \bar C^\mu\big) - \frac{1}{2}\, \big(\partial^\mu \bar \beta \big)\,\lambda \bigg] \nonumber\\
&+& \frac{1}{2}\, \big[{\cal B}_\mu + \bar {\cal B}_\mu + \partial_\mu \tilde \varphi \big] \big(\partial^\mu \rho \big) 
+ \partial_\mu\big[{\cal B}_\nu + \bar {\cal B}_\nu + \partial_\nu \tilde \varphi  \big] \big(\partial^\mu \bar C^\nu 
- \partial^\nu \bar C^\mu  \big) \nonumber\\
&+& m \big[{\cal B}_\mu + \bar {\cal B}_\mu + \partial_\mu \tilde \varphi \big] \big(\partial^\mu \bar C - m\, \bar C^\mu  \big)
- \frac{m}{2}\, \big[{\cal B} + \bar {\cal B} + m \,\tilde \varphi \big] \rho \nonumber\\
&-& \partial_\mu \big[{\cal B} + \bar {\cal B} + m \;\tilde \varphi \big] \big(\partial^\mu \bar C - m \,\bar C^\mu  \big),
\end{eqnarray}
\begin{eqnarray}
s_{ad} {\cal L}_{( B,  {\cal B})} &=& - \partial_\mu \bigg[m\, \varepsilon^{\mu\nu\eta\kappa}\, \phi_\nu \big(\partial_\eta C_\kappa \big)
+ \Big(\frac{1}{2}\,\varepsilon^{\mu\nu\eta\kappa}\,\partial_\nu B_{\eta\kappa} + \frac{1}{2}\, \bar {\cal B}^\mu 
+ m \,\tilde \phi^\mu\Big) \lambda \nonumber\\
&-&  {\cal B}_\nu \big(\partial^\mu C^\nu - \partial^\nu C^\mu  \big)  
+ {\cal B} \big(\partial^\mu C - m \, C^\mu\big) + \frac{1}{2}\, \big(\partial^\mu \beta \big)\,\rho \bigg] \nonumber\\
&+& \frac{1}{2}\, \big[{\cal B}_\mu + \bar {\cal B}_\mu + \partial_\mu \tilde \varphi \big] \big(\partial^\mu \lambda \big) 
- \partial_\mu\big[{\cal B}_\nu + \bar {\cal B}_\nu + \partial_\nu \tilde \varphi  \big] \big(\partial^\mu  C^\nu - \partial^\nu  C^\mu  \big) \nonumber\\
&-& m \big[{\cal B}_\mu + \bar {\cal B}_\mu + \partial_\mu \tilde \varphi \big] \big(\partial^\mu  C - m \, C^\mu  \big)
- \frac{m}{2}\, \big[{\cal B} + \bar {\cal B} + m \,\tilde \varphi \big] \lambda \nonumber\\
&+& \partial_\mu \big[{\cal B} + \bar {\cal B} + m\; \tilde \varphi \big] \big(\partial^\mu  C - m\,  C^\mu  \big).
\end{eqnarray}
It is crystal clear, at this stage, that {\it both} the Lagrangian densities ${\cal L}_{(B, \cal B)}$ and
${\cal L}_{(\bar B, \bar{\cal B})}$ are {\it equivalent} in the sense that {\it both} of them respect the 
(anti-)BRST as well as  (anti-)co-BRST symmetry transformations {\it together} provided our whole theory is 
considered on the {\it submanifold} of space of fields which is defined by the field equations 
(21). The {\it latter} are nothing but the (anti-) BRST and (anti-)co-BRST invariant [cf. Eq. (22)] CF-type
restrictions on our theory. We observe that, besides the {\it perfect} symmetry invariance(s) [cf. Eqs. (14), 
(15), (16), (17)], we have the following, namely;
\begin{eqnarray}
s_{ab} {\cal L}_{(B, {\cal B})} &=& - \partial_\mu \bigg[m\, \varepsilon^{\mu\nu\eta\kappa}\, \tilde \phi_\nu \big(\partial_\eta \bar C_\kappa \big)
+ \Big(\partial_\nu B^{\nu\mu} + \frac{1}{2}\, \bar B^\mu + m\; \phi^\mu\Big) \rho \nonumber\\
&+&  \big(\partial^\mu \bar C^\nu - \partial^\nu \bar C^\mu  \big)\;B_\nu  
- B\; \big(\partial^\mu \bar C - m\; \bar C^\mu\big) - \frac{1}{2}\, \big(\partial^\mu \bar \beta \big)\,\lambda \bigg],\nonumber\\ 
s_b {\cal L}_{(\bar B, \bar {\cal B})} &=& - \partial_\mu \bigg[m\, \varepsilon^{\mu\nu\eta\kappa}\, \tilde \phi_\nu \big(\partial_\eta C_\kappa \big)
- \Big(\partial_\nu B^{\nu\mu} - \frac{1}{2}\, B^\mu + m \;\phi^\mu\Big) \lambda \nonumber\\
&-& \big(\partial^\mu C^\nu - \partial^\nu C^\mu  \big)\;\bar B_\nu   
+ \bar B \;\big(\partial^\mu C - m \; C^\mu\big) - \frac{1}{2}\, \big(\partial^\mu \beta \big)\,\rho \bigg],\nonumber\\ 
s_d {\cal L}_{(\bar B, \bar {\cal B})} &=& - \partial_\mu \bigg[m\, \varepsilon^{\mu\nu\eta\kappa}\,  \phi_\nu \big(\partial_\eta \bar C_\kappa \big)
- \Big(\frac{1}{2}\,\varepsilon^{\mu\nu\eta\kappa}\,\partial_\nu B_{\eta\kappa} - \frac{1}{2}\, {\cal B}^\mu + m\; \tilde \phi^\mu\Big) \rho \nonumber\\
&+& \big(\partial^\mu \bar C^\nu - \partial^\nu \bar C^\mu  \big)\;\bar {\cal B}_\nu   
- \bar {\cal B} \;\big(\partial^\mu \bar C - m \;\bar C^\mu\big) - \frac{1}{2}\, \big(\partial^\mu \bar \beta \big)\,\lambda \bigg],\nonumber\\
s_{ad} {\cal L}_{( B,  {\cal B})} &=& - \partial_\mu \bigg[m\, \varepsilon^{\mu\nu\eta\kappa}\, \phi_\nu \big(\partial_\eta C_\kappa \big)
+ \Big(\frac{1}{2}\,\varepsilon^{\mu\nu\eta\kappa}\,\partial_\nu B_{\eta\kappa} + \frac{1}{2}\, \bar {\cal B}^\mu 
+ m \tilde \phi^\mu\Big) \lambda \nonumber\\
&-&  \big(\partial^\mu C^\nu - \partial^\nu C^\mu  \big)\;{\cal B}_\nu   
+ {\cal B} \;\big(\partial^\mu C - m \; C^\mu\big) + \frac{1}{2}\, \big(\partial^\mu \beta \big)\,\rho \bigg], 
\end{eqnarray}
which are also {\it perfect} symmetry transformations on a {\it submanifold} of the space of fields where
the (anti-)BRST and (anti-)co-BRST invariant CF-type restrictions are {\it satisfied}. Hence, the {\it latter} are {\it physical} 
constraints on our theory.

We end this section with the remark that the most appropriate generalized versions of the Lagrangian density
(6) (that are nothing but ${\cal L}_{(\bar B, \bar{\cal B})}$ and ${\cal L}_{(B, {\cal B})}$) respect {\it both}
types of off-shell nilpotent symmetries [i.e. (anti-)BRST and (anti-)co-BRST symmetries] provided the whole theory is
restricted to be defined on the submanifold of space of fields where the CF-type restrictions (21) are {\it respected}.
In fact, on {\it this} submanifold, the (anti-)BRST and (anti-)co-BRST symmetries are found to be absolutely 
anticommutating, too.  Hence, the submanifold of the space of fields, defined by the  
(anti-)BRST and (anti-)co-BRST invariant [cf. Eq. (22)] field equations (21), are {\it physical} subspace of the {\it quantum} fields
where the proper off-shell nilpotent symmetries and corresponding {\it proper} (i.e. coupled and equivalent)
Lagrangian densities  ${\cal L}_{(\bar B, \bar{\cal B})}$ and ${\cal L}_{(B, {\cal B})}$) are defined.


\section { Off-Shell Nilpotent (Anti-)BRST and (Anti-)co-BRST Symmetries: ACSA to BRST Formalism}

In this section, we exploit the basic tenets of ACSA to BRST formalism [21-23]
 to derive the (anti-)BRST and (anti-)co-BRST symmetries  for the coupled (but equivalent)
Lagrangian densities ${\cal L}_{(\bar B, \bar{\cal B})}$ and ${\cal L}_{(B, {\cal B})}$)
[(cf. Eqs. (10),(11))] of our theory by invoking the symmetry  invariant restrictions  
on the (anti-)chiral superfields.  Our present section is divided into {\it two} sub-sections as described in 4.1 and 4.2.


\subsection{(Anti-)BRST Symmetries: ACSA}

First of all, we derive the BRST symmetries [cf. Eq. (13)]. For this purpose, we generalize the {\it ordinary} 
fields of the Lagrangian density  ${\cal L}_{(B, {\cal B})}$) onto their counterpart {\it anti-chiral}
superfields on the (4, 1)-dimensional (anti-) chiral super sub-manifold as:
\begin{eqnarray*}
&&B_{\mu\nu} (x) \longrightarrow \;\;{\tilde {B}_{\mu\nu}} (x, \bar\theta) \;= B_{\mu\nu} (x) + \bar\theta\, R_{\mu\nu} (x),\nonumber\\
&&C_\mu (x) \;\longrightarrow \;\;{\tilde {\cal F}}_\mu (x, \bar\theta) = C_\mu (x) + \bar\theta\, B^{(1)}_\mu (x),\nonumber\\ 
&&\bar C_\mu \;(x)\longrightarrow \;\;{\tilde {\bar {\cal F}}}_\mu (x, \bar\theta) = \bar C_\mu (x) + \bar\theta\, B^{(2)}_\mu (x), \nonumber\\
&&\beta (x)\;\;\longrightarrow \;\; {\tilde \beta} (x, \bar\theta ) \; = \beta (x) + \bar\theta\, f_1 (x),\nonumber\\ 
 &&\bar\beta (x)\;\;\longrightarrow \;\; {\tilde {\bar \beta}} (x, \bar\theta) = \bar\beta (x) + \bar \theta\, f_2 (x), \nonumber\\ 
&&\varphi (x)\;\;\longrightarrow \;\; {\tilde \Phi} (x,\bar\theta) = \varphi (x) + \bar\theta\, f_3 (x),\nonumber\\
&&\phi_\mu(x)\;\longrightarrow \;\; {\tilde {\Phi}}_\mu (x,  \bar\theta) = \phi_\mu(x) +\bar\theta \, R_\mu (x),\nonumber\\
&&C(x) \;\;\longrightarrow \;\; {\tilde {\cal F}}(x, \bar\theta) = C(x) + \bar\theta \, B_1 (x), \nonumber\\
&&\bar C (x)\;\;\longrightarrow \;\; {\tilde {\bar {\cal F}}} (x,  \bar\theta) = \bar C (x) +  \bar\theta \, B_2 (x),\nonumber\\
&&B_\mu (x)\; \longrightarrow \;\; \tilde {B}_\mu (x, \bar\theta) = B_\mu (x)  + \bar\theta\,f^{(1)}_\mu (x),\nonumber\\
&&\bar B_\mu (x)\; \longrightarrow \;\; \tilde {\bar B}_\mu (x, \bar\theta) = \bar B_\mu (x)  + \bar\theta\,f^{(2)}_\mu(x),\nonumber\\
\end{eqnarray*}
\begin{eqnarray}
&&B(x) \;\;\longrightarrow \;\; \tilde {B} (x, \bar\theta) = B (x)  + \bar\theta\,f_4 (x),\nonumber\\
&&\bar B(x) \;\;\longrightarrow \;\; \tilde {\bar B} (x, \bar\theta) = \bar B (x)  + \bar\theta\,f_5 (x),\nonumber\\
&&{\cal B}_\mu (x)\; \longrightarrow \;\; \tilde {{\cal B}}_\mu (x, \bar\theta) = {\cal B}_\mu (x)  + \bar\theta\,f^{(3)}_\mu(x),\nonumber\\
&&\bar {\cal B}_\mu (x)\; \longrightarrow \;\; {\tilde {\bar {\cal B}}}_{\mu} (x, \bar\theta) = \bar {\cal B}_\mu (x)   + \bar\theta \, f^{(4)}_\mu(x),\nonumber\\
&&\tilde\varphi (x)\;\;\longrightarrow \;\; {\it {\tilde{\Phi}}} (x,\bar\theta) = \tilde\varphi (x) + \bar\theta\, f_6 (x),\nonumber\\
&&\tilde\phi_\mu(x)\;\longrightarrow \;\; {\it {\tilde {\Phi}}}_\mu (x,  \bar\theta) = \tilde\phi_\mu(x) +\bar\theta \, R_\mu^{(1)} (x),\nonumber\\
&&\lambda(x)\;\; \longrightarrow \;\;\tilde\lambda (x, \bar\theta) = \lambda(x)  + \bar\theta\,B_3 (x),\nonumber\\
&&\rho(x)\;\; \longrightarrow \;\; \tilde\rho  (x, \bar\theta) = \rho(x)  + \bar\theta\,B_4 (x),\nonumber\\
&&{\cal B}(x)\;\; \longrightarrow \;\;\tilde{\cal B} (x, \bar\theta) = {\cal B}(x) + \bar\theta f_7(x),\nonumber\\
&&{\bar{\cal B}}(x)\;\; \longrightarrow \;\;{\tilde{\bar {\cal B}}} (x, \bar\theta) = {\bar{\cal B}}(x) + \bar\theta f_8(x).
\end{eqnarray}
In the above {\it anti-chiral} super expansions, it is worthwhile as well as pertinent to point out that the secondary fields 
$(R_{\mu\nu}, f_1, f_2, f_3, R_\mu, f^{(1)}_\mu, f^{(2)}_\mu, f_4,$ $f_5, f^{(3)}_\mu, f^{(4)}_\mu, f_{6}, R_\mu^{(1)}, f_{7}, f_{8})$
are {\it fermionic} and $({B_\mu^{(1)}}, B_\mu^{(2)}, B_1, B_2, B_3, B_4)$ are {\it bosonic} 
in nature due to the {\it fermionic} $(\bar\theta^2 = 0)$ nature
of the Grassmannian variable $\bar\theta$ that characterizes the anti-chiral super 
sub-manifold [along with the bosonic coordinates
$x^{\mu} (\mu = 0, 1,2,3)$].

We note that the following non-trivial quantities are BRST invariant in view of the symmetry  transformations (13), namely;
\begin{eqnarray}
&&s_b(\lambda\,\varphi) = 0,\quad\qquad s_b(\bar{\beta}\,\lambda - \varphi\,\rho) = 0,\quad\qquad s_b(C_\mu\partial^\mu\rho 
- \partial_\mu\bar{\beta}\,\partial^\mu\beta) = 0,  \nonumber \\
&&s_b(C_\mu\partial^\mu{C} + \phi_\mu\partial^\mu\beta) = 0, \qquad s_b(\bar{B}_\mu + \partial_\mu\varphi) = 0, \qquad s_b(\bar{B} + m\,\varphi) = 0, \nonumber \\
&&s_b \big[\bar{C}_\mu(m\,C^\mu - \partial^\mu{C}) + B^\mu \,\phi_\mu\big] = 0, \qquad s_b\big[\bar{C}(m\,C_\mu 
- \partial_\mu{C}) + B\,\phi_\mu\big] = 0, \nonumber \\
&&s_b\big[m\,B_{\mu\nu} - (\partial_\mu\phi_\nu - \partial_\nu\phi_\mu)\big] = 0, \quad\qquad s_b\big[m\,\bar{\beta}\,\beta + \rho\,C\big] = 0.  
\end{eqnarray}
In addition, we have some {\it trivial} BRST invariant quantities as: $s_b[ B, B_\mu, \rho, \lambda, \beta,
 {{\cal B}_\mu}, {\bar {\cal B}_\mu},\\ \tilde\varphi, \tilde\phi_\mu,$ ${\cal B}, {\bar{\cal B}}, H_{\mu\nu\lambda}] = 0$ [cf. Eq. (13)]. 
It is because of the {\it latter} observation that we have the following {\it trivial} super-expansions and equalities
for the appropriate {\it anti-chiral} superfields  
\begin{eqnarray}
&&\tilde B ^{(b)} (x, \bar\theta) = B(x),\qquad \tilde\rho ^{(b)} (x, \bar\theta) = \rho (x),\qquad \tilde\lambda ^{(b)} (x, \bar\theta) = \lambda (x),\nonumber\\
&&{{\tilde{\bar {\cal B}}}}_\mu ^{(b)} (x, \bar\theta) = \bar {\cal B}_\mu (x),\quad {\it {\tilde{\Phi}}} ^{(b)} (x, \bar\theta) =  \tilde\varphi (x),\quad
 {{\it {\tilde{\Phi}}}}_\mu ^{(b)} (x, \bar\theta)  =  \tilde {\phi}_\mu (x),\nonumber\\
&&{\tilde{\cal B}}_\mu ^{(b)} (x, \bar\theta) = {\cal B}_\mu (x),\quad {\tilde{\bar{\cal B}}} ^{(b)} (x, \bar\theta)
 = {\bar{\cal B}} (x), \quad \tilde B_\mu^{(b)}(x, \bar\theta) = B_\mu(x),\nonumber\\
&&\tilde\beta ^{(b)} (x, \bar\theta) =  \beta (x),\qquad{\tilde {\cal B}} ^{(b)} (x, \bar\theta)
 = {\cal B} (x),\quad\;\tilde H_{\mu\nu\lambda} ^{(b)} (x, \bar\theta) = H_{\mu\nu\lambda} (x),
\end{eqnarray}
where the superscript $(b)$ denotes the superfields that have been obtained after the applications of the 
BRST invariant {\it trivial} restrictions. The above equation (31) {\it also} implies that we have some of the 
secondary fields {\it trivially} equal to zero: $B_3 = B_4 = 0, f_1 = f_4 =  f_6 = f_7 =  f_8 =  f_\mu ^{(1)} = 
f_\mu ^{(3)} =f_\mu ^{(4)}= R_\mu^{(1)}  = 0.$  The above expansions (31) can  now be utilized in the 
{\it non-trivial} BRST-invariant equalities  as  follows:
\begin{eqnarray}
&& \tilde \lambda ^{(b)}(x, \bar\theta)\; \tilde \Phi  (x, \bar\theta) = \lambda (x)\,\varphi (x),\quad {\tilde {\bar{B}}}_\mu (x, \bar\theta) + \partial_\mu\tilde {\Phi}(x, \bar\theta) = \bar{B}_\mu (x) + \partial_\mu\varphi (x),\nonumber \\
&&\tilde {\bar\beta} (x, \bar\theta)\; \tilde\lambda ^{(b)}(x, \bar\theta)- \tilde \Phi  (x, \bar\theta)\;\tilde \rho ^{(b)}  (x, \bar\theta) = \bar{\beta} (x)\,\lambda (x) - \varphi (x)\,\rho (x),\nonumber\\
&& m\,\tilde {\bar\beta} (x, \bar\theta)\;\tilde\beta^{(b)} (x, \bar\theta) + \tilde \rho ^{(b)}  (x, \bar\theta)\;
{\tilde{\cal F}} (x, \bar\theta) = m\,\bar{\beta} (x)\,\beta (x) + \rho (x)\,C (x), \nonumber \\
&& {\tilde {\cal F}}_\mu (x, \bar\theta)\;\partial^\mu {\tilde{\cal F}} (x, \bar\theta) +  
{{\tilde \Phi}}_\mu (x, \bar\theta) \; \partial^\mu\tilde\beta^{(b)} (x, \bar\theta)    
= C_\mu (x)\partial^\mu{C} (x)\nonumber\\
&& + \phi_\mu (x)\;\partial^\mu\beta (x),\quad {\tilde {\bar{B}}}(x, \bar\theta)  + m\,\tilde {\Phi}(x, \bar\theta) 
= \bar{B} (x) + m\,\varphi (x), \nonumber\\
&& {\tilde {\bar {\cal F}}}_\mu(x, \bar\theta)\; \big[m\,{\tilde {\cal F}}^\mu (x, \bar\theta) -
 \partial^\mu {\tilde {\cal F}}(x, \bar\theta)\big] + {\tilde {{B}}}^{\mu (b)} (x, \bar\theta)\;{\tilde \Phi}_\mu (x, \bar\theta)\nonumber\\
  && = \bar{C}_\mu (x)\,\big[(m\,C^\mu (x) - \partial^\mu{C} (x)\big] + B_\mu (x)\;\phi^\mu (x),\nonumber \\
&& {\tilde {\bar {\cal F}}}(x, \bar\theta)\big [m\,{\tilde {\cal F}}_\mu (x, \bar\theta) - \partial_\mu {\tilde {\cal F}}(x, \bar\theta)\big ] 
 +  {\tilde {{B}}}^{(b)} (x, \bar\theta)\;{\tilde \Phi}_\mu (x, \bar\theta)
  = \bar{C}  (x)\big[ m\,C_\mu (x)\nonumber \\ 
 && - \partial_\mu{C} (x)\big] + B(x)\,\phi_\mu (x), \,\qquad m\,{\tilde B}_{\mu\nu} (x, \bar\theta)
- \big[\partial _\mu \tilde {\Phi}_\nu (x, \bar\theta) - \partial _\nu \tilde {\Phi}_\mu (x, \bar\theta) \big] \nonumber\\
&&= m\,B_{\mu\nu} (x) - \big[\partial_\mu\phi_\nu (x) - \partial_\nu \phi_\mu (x)\big], 
\quad\tilde {\cal F}_\mu (x, \bar\theta)\;\partial^\mu \tilde \rho ^{(b)} (x, \bar\theta)\nonumber\\
&& - \partial_\mu \tilde {\bar\beta} (x, \bar\theta)\;\partial ^{\mu}\tilde\beta^{(b)} (x, \bar\theta)=
 C_\mu (x)\;\partial^\mu\rho (x) - \partial_\mu\bar{\beta} (x)\,\partial^\mu\beta (x).
\end{eqnarray}
Here we have utilized the {\it basic} tenets of ACSA to BRST formalism and imposed the condition   that
the BRST invariant {\it (physical)} quantities should be {\it independent} of the fermionic ($\bar\theta ^2 = 0$) Grassmannian variable $\bar\theta$.  
The above restrictions on the superfields lead to the derivation of all the {\it rest} of the {\it secondary} fields of the super expansion (29)
in terms of the basic and auxiliary fields of the Lagrangian density ${\cal L} _{(B, {\cal B})}$ as:  
\begin{eqnarray}
&&R_{\mu\nu} (x) = - (\partial_\mu  C_\nu (x) - \partial_\nu  C_\mu (x)), \; B_\mu ^{(1)} (x) = -\,\,\partial_\mu\, \beta (x),\;
B_\mu ^{(2)} (x) = B_\mu(x),\nonumber\\
&&f_{2} (x) =  -\,\rho (x),\;\;\;R_\mu (x) = \partial_\mu  C (x) - m\,  C_\mu (x),\;\;\; B_1 (x) = -\,m\,\beta (x),\nonumber\\
&&f_7 (x) = -\,m\;\lambda (x),\;f_\mu^{(2)} (x) = -\,\partial _\mu \lambda (x),\;\, f_3 (x) =  \lambda (x),\; B_2 (x) = B(x). 
\end{eqnarray}
Plugging in these inputs  into  the {\it anti-chiral} super expansion (29), we obtain the coefficients of $\bar\theta$
as the {\it non-trivial} BRST symmetry transformations  (13) as illustrated below:  
\begin{eqnarray*}
B_{\mu\nu} (x) \longrightarrow {\tilde {B}_{\mu\nu}} ^{(b)} (x, \bar\theta) & = & B_{\mu\nu} (x) +
 \bar\theta\,[- (\partial_\mu  C_\nu (x) - \partial_\nu  C_\mu (x))] \nonumber\\
 &\equiv & B_{\mu\nu} (x) + \bar\theta\,[s_b\,B_{\mu\nu} (x)],\nonumber\\
C_\mu (x) \;\,\longrightarrow \;\;{\tilde {\cal F}}_\mu ^{(b)} (x, \bar\theta) & = & C_\mu (x) + \bar\theta\,(-\,\,\partial_\mu\, \beta (x))\equiv
C_\mu (x) + \bar\theta\,[s_b\, C_\mu(x)], \nonumber\\ 
\bar C_\mu (x)\longrightarrow \;{\tilde {\bar {\cal F}}}_\mu ^{(b)} (x, \bar\theta) & = & \bar C_\mu (x) + \bar\theta\,(B_\mu(x)) \equiv
\bar C_\mu (x) + \bar\theta\,  [s_b\, \bar C_\mu(x)], \nonumber\\
\beta (x)\longrightarrow \;\; {\tilde \beta}^{(b)} (x, \bar\theta ) \; & = & \beta (x) + \bar\theta\, (0) \equiv  \beta (x)
 + \bar\theta\, [s_b\,\beta(x)], \nonumber\\ 
\bar\beta (x)\longrightarrow \;\; {\tilde {\bar \beta}} ^{(b)} (x, \bar\theta) & = & \bar\beta (x) 
+ \bar \theta\, \,(-\,\rho(x))\equiv \bar\beta (x) + \bar \theta\, \,[s_b\, \bar\beta (x)], \nonumber\\ 
\varphi (x)\;\;\longrightarrow \;\; {\tilde \Phi}^{(b)} (x,\bar\theta) & = &\varphi (x) + \bar\theta\, (\lambda (x))\equiv \varphi (x)
 + \bar\theta\, [s_b\,\varphi (x)] ,\nonumber\\
 \phi_\mu(x)\quad\longrightarrow  {\tilde {\Phi}}_\mu^{(b)} (x,  \bar\theta) & = &\phi_\mu(x) +\bar\theta \, [\partial_\mu  C(x) - m\, C_\mu (x)]\nonumber\\
 &\equiv & \phi_\mu(x) +\bar\theta \,[s_b\phi_\mu(x)],\nonumber\\ 
 C(x) \;\;\longrightarrow \;\; {\tilde {\cal F}}^{(b)} (x, \bar\theta) & = & C(x) + \bar\theta \, (-\, m\, \beta (x))\equiv 
C(x) + \bar\theta \, [s_b \, C (x)], \nonumber\\
\bar C (x)\;\;\longrightarrow \;\; {\tilde {\bar {\cal F}}}^{(b)} (x,  \bar\theta) & = &\bar C (x) +  \bar\theta \, (B (x))
\equiv \bar C (x) +  \bar\theta \,[s_b \bar C (x)],\nonumber\\
 \end{eqnarray*}
 \begin{eqnarray}
B_\mu (x)\;\;\; \longrightarrow  \tilde {B}_\mu ^{(b)} (x, \bar\theta) & = & B_\mu (x) 
 + \bar\theta\, (0)\equiv B_\mu (x)  + \bar\theta\,[s_b B_\mu (x)],\nonumber\\
\bar B_\mu (x)\; \longrightarrow  \tilde {\bar B}_\mu ^{(b)} (x, \bar\theta) & = & \bar B_\mu (x)  + \bar\theta\,(-\partial _\mu \lambda(x))
\equiv \bar B_\mu (x)  + \bar\theta\,[s_b \, \bar B_\mu (x)],\nonumber\\
B(x) \;\;\longrightarrow \;\; \tilde {B} ^{(b)} (x, \bar\theta) & = & B (x)  + \bar\theta\,(0)\equiv B (x)  + \bar\theta\, [s_b\, B (x)] ,\nonumber\\
\bar B(x) \;\;\longrightarrow \;\; \tilde {\bar B} ^{(b)} (x, \bar\theta)& = &\bar B (x)  + \bar\theta\,(-\,m\lambda(x))\equiv 
\bar B (x)  + \bar\theta\,[s_b\, \bar B (x)],\nonumber\\
{\cal B}_\mu (x)\; \longrightarrow  \tilde {{\cal B}}_\mu ^{(b)} (x, \bar\theta) & = &{\cal B}_\mu (x)  + \bar\theta\, (0)\equiv
{\cal B}_\mu (x)  + \bar\theta\, [s_b \, {\cal B}_\mu (x)],\nonumber\\
\bar {\cal B}_\mu (x)\; \longrightarrow  {\tilde {\bar {\cal B}}}_{\mu} ^{(b)} (x, \bar\theta)& = &\bar {\cal B}_\mu (x)   + \bar\theta \,  (0)
\equiv \bar {\cal B}_\mu (x)  + \bar\theta \,[s_b \,\bar {\cal B}_\mu (x)],\nonumber\\
\tilde\varphi (x)\;\;\;\longrightarrow  {\it {\tilde\Phi}}^{(b)} (x,\bar\theta)& = & \tilde\varphi (x) + \bar\theta\,  (0)\equiv 
\tilde\varphi (x) + \bar\theta \;[s_b \,\tilde\varphi (x)],\nonumber\\
\tilde\phi_\mu(x)\;\;\;\longrightarrow  {\it {\tilde\Phi}}_\mu ^{(b)} (x,  \bar\theta) & = & \tilde\phi_\mu(x) +\bar\theta \,  (0)\equiv 
\tilde\phi_\mu(x) +\bar\theta \; [s_b \,\tilde\phi_\mu (x)],\nonumber\\
\lambda(x)\;\; \longrightarrow \;\;\tilde\lambda ^{(b)} (x, \bar\theta)& = & \lambda(x)  + \bar\theta\,(0)\equiv 
\lambda(x)  + \bar\theta \,[s_b \,\lambda (x)] ,\nonumber\\
\rho(x)\;\; \longrightarrow \;\; \tilde\rho  ^{(b)} (x, \bar\theta) & = &\rho(x)  + \bar\theta\, (0)\equiv 
\rho(x)  + \bar\theta\, [s_b \,  \rho(x)],\nonumber\\
{\cal B}(x)\;\; \longrightarrow \;\; {\tilde{\cal B}}  ^{(b)} (x, \bar\theta) & = &{\cal B} (x)  + \bar\theta\, (0)\equiv 
{\cal B}(x)  + \bar\theta\, [s_b \,  {\cal B}(x)],\nonumber\\
{\bar {\cal B}}(x)\;\; \longrightarrow \;\; {\tilde{\bar {\cal B}}}  ^{(b)} (x, \bar\theta) & = &{\bar {\cal B}} (x)  + \bar\theta\, (0)\equiv 
{\bar {\cal B}} (x)  + \bar\theta\, [s_b \,  {\bar {\cal B}}(x)].
\end{eqnarray}
In the above, the superscript $(b)$ denotes the fact that the super expansions
have been obtained after exploiting the {\it basic} tenets of ACSA 
{\it and} coefficients of $\bar\theta$ lead to the determination  of the
BRST symmetry transformations (13). It is elementary to note that 
$\partial_{\bar\theta}\,\Omega^{(b)}(x, \bar\theta) = s_b\,\omega (x)$
where $\Omega^{(b)}(x, \bar\theta)$ is the generic superfield obtained after the 
application of the BRST invariant restrictions (31) and (32) and $\omega (x)$ is the {\it ordinary} (4D)
generic field of the Lagrangian density ${\cal L}_{(B, {\cal B})}$.
Thus, we note that the nilpotency ($s_b^2 = 0$) property of $s_b$ is deeply connected with the nilpotency
$(\partial_{\bar\theta}^2 = 0)$ of the translational generators ($\partial_{\bar\theta}$)
along the $\bar\theta$-direction of the (4, 1)-dimensional {\it anti-chiral} 
super sub-manifold of the (4, 2)-dimensional {\it general} supermanifold (on which our 4D
{\it ordinary} massive Abelian 2-form  theory is generalized).

We now concentrate on the derivation of the off-shell nilpotent anti-BRST symmetry transformations (12) for ${\cal L}_{(\bar B,{\bar{\cal B})}}$
within the framework of ACSA to BRST formalism. It may be mentioned here that the anti-BRST transformations (12) are {\it perfect}
symmetry transformations  for ${\cal L}_{(\bar B, \bar{\cal B})}$ [cf. Eq. (14)]. Keeping this objective  in our mind, first of all, we note that
the following interesting and useful quantities are anti-BRST invariant in view of the {\it quantum} nilpotent, infinitesimal and continuous
 anti-BRST 
symmetry transformations (12), namely; 
\begin{eqnarray}
&&s_{ab}[\rho\,\varphi] = 0,\quad s_{ab}[{\beta}\,\rho - \varphi\,\lambda] = 0,\quad s_{ab}[\bar C_\mu\partial^\mu\lambda
 - \partial_\mu{\beta}\,\partial^\mu\bar\beta] = 0, \nonumber \\
&&s_{ab}[\bar C_\mu\partial^\mu{\bar C} + \phi_\mu\partial^\mu\bar\beta] = 0, \quad s_{ab}[{B}_\mu + \partial_\mu\varphi]
 = 0, \quad s_{ab}[{B} + m\,\varphi] = 0, \nonumber \\
&&s_{ab}[{C}_\mu(m\,\bar C^\mu - \partial^\mu\bar{C}) + \bar B^\mu\,\phi_\mu] = 0, \quad s_{ab}[{C}(m\,\bar C_\mu
 - \partial_\mu{\bar C}) + \bar B\,\phi_\mu] = 0, \nonumber \\
&&s_{ab}[m\,B_{\mu\nu} - (\partial_\mu\phi_\nu - \partial_\nu\phi_\mu)] = 0,\quad s_{ab}[m\,\bar{\beta}\,\beta + \lambda\,\bar C] = 0.  
\end{eqnarray}
According to the {\it basic} tenets of ACSA to BRST formalism, the above quantities should be 
independent of the Grassmannian variable $\theta$ of the (4, 1)-dimensional {\it chiral} super sub-manifold 
on which the ordinary {\it fields} of  ${\cal L}_{(\bar B,{\bar{\cal B})}}$ are generalized in the following manner
\begin{eqnarray}
&&B_{\mu\nu} (x) \longrightarrow \;\;{\tilde {B}_{\mu\nu}} (x, \theta) \;= B_{\mu\nu} (x) + \theta\, \bar R_{\mu\nu} (x),\nonumber\\
&&C_\mu (x) \;\longrightarrow \;\;{\tilde {\cal F}}_\mu (x, \theta) = C_\mu (x) + \theta\, {\bar B}^{(1)}_\mu (x),\nonumber\\ 
&&\bar C_\mu \;(x)\longrightarrow \;\;{\tilde {\bar {\cal F}}}_\mu (x, \theta) = \bar C_\mu (x) + \theta\, {\bar B}^{(2)}_\mu (x), \nonumber\\
&&\beta (x)\;\;\longrightarrow \;\; {\tilde \beta} (x, \theta ) \; = \beta (x) + \theta\, \bar f_1 (x), \nonumber\\ 
&&\bar\beta (x)\;\;\longrightarrow \;\; {\tilde {\bar \beta}} (x, \theta) = \bar\beta (x) + \theta\, \bar f_2 (x), \nonumber\\ 
&&\varphi (x)\;\;\longrightarrow \;\; {\tilde \Phi} (x, \theta) = \varphi (x) + \theta\, \bar f_3 (x),\nonumber\\
&&\phi_\mu(x)\;\longrightarrow \;\; {\tilde {\bf \Phi}}_\mu (x,  \theta) = \phi_\mu(x) + \theta  \bar R_\mu (x),\nonumber\\
&&C(x) \;\;\longrightarrow \;\; {\tilde {\cal F}}(x, \theta) = C(x) + \theta \,\bar  B_1 (x), \nonumber\\
&&\bar C (x)\;\;\longrightarrow \;\; {\tilde {\bar {\cal F}}} (x,  \theta) = \bar C (x) +  \theta\,  \bar B_2 (x),\nonumber\\
&&B_\mu (x)\; \longrightarrow \;\; \tilde {B}_\mu (x, \theta) = B_\mu (x)  + \theta\,\bar f_\mu^{(1)} (x),\nonumber\\
&&\bar B_\mu (x)\; \longrightarrow \;\; \tilde {\bar B}_\mu (x, \theta) = \bar B_\mu (x)  + \theta\,\bar f_\mu^{(2)} (x),\nonumber\\
&&B(x) \;\;\longrightarrow \;\; \tilde {B} (x, \theta) = B (x)  + \theta\,\bar f_4 (x),\nonumber\\  
&&\bar B(x) \;\;\longrightarrow \;\; \tilde {\bar B} (x, \theta) = \bar B (x)  + \theta\,\bar f_5 (x),\nonumber\\
&&{\cal B}_\mu (x)\; \longrightarrow \;\; \tilde {{\cal B}}_\mu (x, \theta) = {\cal B}_\mu (x)  + \theta\,\bar f_\mu^{(3)} (x),\nonumber\\
&&\bar {\cal B}_\mu (x)\; \longrightarrow \;\; {\tilde {\bar {\cal B}}}_{\mu} (x, \theta) = \bar {\cal B}_\mu (x)   + \theta \, \bar f_\mu^{(4)} (x),\nonumber\\
&&\tilde\varphi (x)\;\;\longrightarrow \;\; {\it {\tilde\Phi}} (x,\theta) = \tilde\varphi (x) + \theta\, \bar f_{6} (x),\nonumber\\
&&\tilde\phi_\mu(x)\;\longrightarrow \;\; {\it {\tilde\Phi}}_\mu (x,  \theta) = \tilde\phi_\mu(x) + \theta \, \bar R_\mu^{(1)} (x),\nonumber\\
&&\lambda(x)\; \;\longrightarrow \;\; \tilde\lambda (x, \theta) = \lambda(x)  + \theta\,\bar B_3 (x),\nonumber\\
&&\rho(x)\;\; \longrightarrow \;\; \tilde\rho  (x, \theta) = \rho(x)  + \theta\,\bar B_4 (x), \nonumber\\
&&{\cal B}(x)\;\; \longrightarrow \;\;\tilde{\cal B} (x, \theta) = {\cal B}(x) + \theta \bar f_{7}(x),\nonumber\\
&&{\bar{\cal B}}(x)\;\; \longrightarrow \;\;{\tilde{\bar {\cal B}}} (x, \theta) = {\bar{\cal B}}(x) + \theta \,\bar f_{8}(x),
\end{eqnarray}
where the {\it fermionic} $(\theta^2 = 0)$ Grassmannian variable $\theta$ [along with the
{\it bosonic} coordinates $x^{\mu} (\mu = 0, 1,2,3)$] characterize  the {\it chiral} 
super sub-manifold. We note that the secondary fields $(\bar R_{\mu\nu}, \bar f_1, \bar f_2, \bar f_3, 
\bar R_\mu, \bar f_\mu^{(1)}, \bar f_\mu^{(2)}, \bar f_4, \bar f_5,  
\bar f_{\mu}^{(3)}, \bar f_\mu^{(4)}, \bar f_{6}, \bar R_\mu ^{(1)}, \bar f_{7}, \bar f_{8})$
are {\it fermionic} in nature and the other secondary fields $(\bar {B_\mu^{(1)}}, \bar  B_\mu^{(2)}, \bar B_1, \bar B_2, \bar B_3, \bar B_4)$
are {\it bosonic}. These secondary fields are to be determined in terms of the {\it basic} and {\it auxiliary} 
fields of the Lagrangian density ${\cal L}_{(\bar B,{\bar{\cal B})}}$. At this stage, as  argued earlier, the 
following restrictions on the specific  combination of the {\it chiral} superfields, namely;    
\begin{eqnarray*}
&& \tilde \rho ^{(ab)}(x, \theta)\; \tilde \Phi  (x, \theta) = \rho (x)\,\varphi (x),\;\;{\tilde {{B}}}_\mu (x, \theta)
 + \partial_\mu\tilde {\Phi}(x, \theta) = {B}_\mu (x) + \partial_\mu\varphi (x),\nonumber \\
  &&\tilde {{\beta}} (x, \theta)\; \tilde\rho ^{(ab)}(x, \theta)- \tilde \Phi  (x, \theta)\;\tilde \lambda ^{(ab)}  (x, \theta)
 = {\beta} (x)\,\rho (x) - \varphi (x)\,\lambda (x),\nonumber\\
&& m\,\tilde {\beta}(x, \theta)\;\tilde {\bar\beta} ^{(ab)} (x, \theta) + \tilde \lambda ^{(ab)}  (x, \theta)\;
{\tilde{\bar {\cal F}}} (x, \theta) = m\,\bar{\beta} (x)\,\beta (x) + \lambda (x)\,\bar C (x), \nonumber \\
&& {\tilde {\bar {\cal F}}_\mu }(x, \theta)\;\partial^\mu {\tilde{\bar{\cal F}}} (x, \theta) +  
{{\tilde \Phi}}_\mu (x, \theta) \; \partial ^{\mu}{\tilde{\bar{\beta}}}^{(ab)} (x, \theta)   
= \bar C_\mu (x)\partial^\mu{\bar C} (x),\nonumber \\
&& + \phi_\mu (x)\partial^\mu\bar\beta (x), \;{\tilde {\cal F}}_\mu(x, \theta)\; \big[m\,{\tilde {\bar {\cal F}}}^{\mu }(x, \theta)
 - \partial^\mu {\tilde {\bar {\cal F}}}(x, \theta)\big] + 
 {\tilde {{\bar B}}}^{\mu (ab)} (x, \theta)\;{\tilde \Phi}_\mu (x, \theta),\nonumber \\
\end{eqnarray*}
 \begin{eqnarray}
  && = {C}_\mu (x)\,[m\,\bar C^\mu (x) - \partial^\mu{\bar C} (x)] + \bar B^\mu(x)\phi_\mu (x),\;\;{\tilde { {\cal F}}}(x, \theta)\big [m\,{\tilde {\bar {\cal F}}}_{\mu }(x, \theta)\nonumber\\
&& - \partial_\mu {\tilde{\bar  {\cal F}}}^{}(x, \theta)\big ] +  {\tilde {{\bar B}}}^{ } (x, \theta)\;{\tilde \Phi}_\mu ^{}(x, \theta) 
= {C} (x) \big[ m\,\bar C_\mu (x) 
- \partial_\mu{\bar C} (x)\big] + \bar B (x)\,\phi_\mu (x),\nonumber\\
&& m\,{\tilde B}_{\mu\nu}(x, \theta) - \big[\partial _\mu \tilde {\Phi}_\nu (x, \theta) - \partial _\nu \tilde {\Phi}_\mu (x, \theta) \big] 
= m\,B_{\mu\nu} (x) - [\partial_\mu\phi_\nu (x)\nonumber\\
&& - \partial_\nu \phi_\mu (x)], \quad {\tilde{\bar {\cal F }}}_\mu (x, \theta)\;\partial^\mu \tilde \lambda ^{(ab)} (x, \theta) -
 \partial_\mu \tilde {\beta} (x, \theta)\;\partial ^{\mu}{\tilde{\bar{\beta}}}^{(ab)} (x, \theta) = \nonumber\\
&& \bar C_\mu (x)\partial^\mu\lambda (x) - \partial_\mu{\beta} (x)\,\partial^\mu\bar\beta (x), \; {\tilde {{B}}} (x, \theta) 
 + m\,\tilde {\Phi}(x, \theta) = {B} (x) + m\,\varphi (x),
\end{eqnarray}
are to be {\it satisfied} due to the basic tenets of ACSA to BRST formalism. The above equalities [cf. Eq. (37)] are
 nothing but the {\it generalizations} of the useful and interesting  anti-BRST invariant quantities (35)
onto the {\it chiral} super sub-manifold [of the {\it general} (4, 2)-dimensional supermanifold] 
 with the following {\it inputs} [due to the {\it trivial} anti-BRST invariant quantities:
 $s_{ab} [\bar B, \rho, \lambda, \bar \beta, \bar B_\mu, {\cal B}_\mu, {\bar{\cal B}}_\mu, 
\tilde\phi_\mu, \tilde \varphi, {\cal B}, \bar {\cal B},  H_{\mu\nu\kappa}] = 0$ 
that are useful and important for our purpose], namely;
\begin{eqnarray}
&&\tilde {\bar B} ^{(ab)} (x, \theta) = \bar B(x),\qquad \tilde\rho ^{(ab)} (x, \theta) 
= \rho (x),\qquad \tilde\lambda ^{(ab)} (x, \theta) = \lambda (x),\nonumber\\
&&{\tilde {\bar {\cal B}}}_\mu ^{(ab)} (x, \theta) = \bar {\cal B}_\mu (x),\qquad {\it {\tilde\Phi}} ^{(ab)} (x, \theta) 
=  \tilde\varphi (x),\qquad {{\it {\tilde\Phi}}_\mu} ^{(ab)} (x, \theta)  =  \tilde {\phi}_\mu (x),\nonumber\\
&&{\tilde{\cal B}}_\mu ^{(ab)} (x, \theta) = {\cal B}_\mu (x),\qquad {\tilde{\bar{\cal B}}} ^{(ab)} (x, \theta) 
= {\bar{\cal B}}(x),\qquad {\tilde {\cal B}} ^{(ab)} (x, \theta) = {\cal B} (x),\nonumber\\
&& \,\tilde H_{\mu\nu\eta} ^{(ab)} (x, \theta) = H_{\mu\nu\eta} (x), \quad \bar B_\mu^{(ab)}(x, \theta)
 = \bar B_\mu(x),\quad \tilde{\bar\beta} ^{(ab)} (x, \theta) =  \bar\beta (x).
\end{eqnarray}
The above equalities/restrictions [i.e. Eq. (38)] {\it also} imply that the secondary fields: 
$\bar f_2 = \bar f_3 = \bar f_5 = \bar f_6 = \bar f_7 = \bar f_8 = \bar B_3 = \bar B_4 = \bar f_\mu ^{(2)} =
\bar f_\mu ^{(3)} = \bar f_\mu ^{(4)} = \bar R_\mu ^{(1)} =0$ in the {\it chiral} super eaxpansions of the {\it chiral} superfilelds in (36).
In other words, the coefficients of $\theta$ in the chiral super expansions (36) (that correspond
to the anti-BRST symmetry transformations ($s_{ab}$) are {\it trivially} zero for all the {\it ordinary}
fields that are present on the r.h.s. of (38).

The anti-BRST invariant restrictions (37) lead to the following precise expressions for the {\it secondary} fields 
in terms of the {\it basic} and {\it auxiliary} fields of Lagrangian density ${\cal L}_{(\bar B,{\bar{\cal B})}}$:
\begin{eqnarray}
&&\bar R_{\mu\nu} (x) = - [\partial_\mu \bar  C_\nu (x) - \partial_\nu  \bar C_\mu (x)], \;\; \bar B_\mu ^{(2)} (x) 
= -\,\partial_\mu\, \bar\beta (x),\;\;\bar B_\mu ^{1} (x) = \bar B_\mu (x),\nonumber\\
&&\bar f_{1} =  -\,\lambda (x),\;\;\;\bar R_\mu (x) = \partial_\mu  \bar C (x) - m\,  \bar C_\mu (x),\;\;\; \bar B_2 (x) 
= -\,m\;\bar\beta (x),\nonumber\\
&&\bar B_1 (x) = \bar B(x),\;\,\bar f_4 (x) = -\,m\;\rho(x),\;\,\bar f^{(1)}_\mu (x) 
= -\partial _\mu \rho (x),\;\, \bar f_3 (x) =  \rho (x). 
\end{eqnarray}
The above derivations are straightforward as there are no complicated tricks involved in their deductions.
Substitutions of the above precise values of the secondary fields into the {\it chiral}
super expansions (36) lead to the determination of anti-BRST symmetries $(s_{ab})$ [cf. Eq. (12)] as the 
coefficients of $\theta$ as illustrated in the following super expansions:
\begin{eqnarray*}
B_{\mu\nu} (x) \longrightarrow {\tilde {B}_{\mu\nu}}^{(ab)} (x, \theta) &= &B_{\mu\nu} (x) +
 \theta\,[- (\partial_\mu  \bar C_\nu (x) - \partial_\nu  \bar C_\mu (x))]\nonumber\\
 &\equiv & \; B_{\mu\nu} (x) + \theta\,[s_{ab}\,B_{\mu\nu} (x)],\nonumber\\
C_\mu (x) \;\,\longrightarrow \;\;{\tilde {\cal F}}_\mu^{(ab)} (x, \theta) & = & C_\mu (x) + \theta\,(\,\bar B_\mu(x))\equiv
C_\mu (x) + \theta\,[s_{ab}\, C_\mu(x)], \nonumber\\ 
\bar C_\mu (x)\longrightarrow \;{\tilde {\bar {\cal F}}}_\mu ^{(ab)} (x, \theta) & = & \bar C_\mu (x) + \theta\,(-\,\partial_\mu{\bar {\beta}}(x)) \equiv
\bar C_\mu (x) + \theta\,  [s_{ab}\, \bar C_\mu(x)], \nonumber\\
\beta (x)\longrightarrow \;\; {\tilde \beta} ^{(ab)}(x, \theta ) \; & = & \beta (x) + \theta\, (-\,\lambda (x)) \equiv  \beta (x) 
+ \theta\, [s_{ab}\,\beta(x)], \nonumber\\ 
\end{eqnarray*}
\begin{eqnarray}
\bar\beta (x)\longrightarrow \;\; {\tilde {\bar \beta}} ^{(ab)} (x, \theta) & = & \bar\beta (x)
 +  \theta\, \,(0)\equiv \beta (x) + \theta\, \,[s_{ab}\, \bar\beta (x)], \nonumber\\ 
\varphi (x)\;\;\longrightarrow \;\; {\tilde \Phi}^{(ab)} (x,\theta) & = &\varphi (x) + \theta\, (\rho (x))\equiv \varphi (x)
 + \theta\, [s_{ab}\,\varphi (x)] ,\nonumber\\
\phi_\mu(x)\;\;\longrightarrow  {\tilde {\Phi}}_\mu ^{(ab)}(x,  \theta) & = &\phi_\mu(x) +\theta \, [\partial_\mu  \bar C (x) - m\, \bar C_\mu (x)]\nonumber\\
&\equiv & \phi_\mu(x) +\theta \,[s_{ab}\phi_\mu(x)],\nonumber\\
C(x) \;\;\longrightarrow \;\; {\tilde {\cal F}}^{(ab)} (x, \theta) & = & C(x) + \theta \, (\bar B (x))\equiv 
C(x) + \theta \; [s_{ab} \, C (x)], \nonumber\\
\bar C (x)\;\;\longrightarrow \;\; {\tilde {\bar {\cal F}}}^{(ab)} (x,  \theta) & = &\bar C (x) +  \theta \, (-\,m\, \bar\beta(x))
\equiv \bar C (x) +  \theta \,[s_{ab} \bar C (x)],\nonumber\\
B_\mu (x)\;\;\; \longrightarrow  \tilde {B}_\mu ^{(ab)} (x, \theta) & = & B_\mu (x)  + \theta\, ( - \, \partial_\mu\,\rho)\equiv B_\mu (x)
  + \theta\,[s_{ab} B_\mu (x)],\nonumber\\
 \bar B_\mu (x)\; \longrightarrow  \tilde {\bar B}_\mu ^{(ab)} (x, \theta) & = & \bar B_\mu (x)  + \theta\,(0)
\equiv \bar B_\mu (x)  + \theta\,[s_{ab} \, \bar B_\mu (x)],\nonumber\\
B(x) \;\;\longrightarrow \;\; \tilde {B} ^{(ab)} (x, \theta) & = & B (x)  + \theta\,(-\,m\rho (x))\equiv B (x)  + \theta\, [s_{ab}\, B (x)] ,\nonumber\\
\bar B(x) \;\;\longrightarrow \;\; \tilde {\bar B} ^{(ab)} (x, \theta)& = &\bar B (x)  + \theta\,(0)\equiv 
\bar B (x)  + \theta\,[s_{ab}\, \bar B (x)],\nonumber\\
{\cal B}_\mu (x)\; \longrightarrow  \tilde {{\cal B}}_\mu ^{(ab)} (x, \theta) & = &{\cal B}_\mu (x)  + \theta\, (0)\equiv
{\cal B}_\mu (x)  + \theta\, [s_{ab} \, {\cal B}_\mu (x)],\nonumber\\
\bar {\cal B}_\mu (x)\; \longrightarrow  {\tilde {\bar {\cal B}}}_{\mu} ^{(ab)} (x, \theta)& = &\bar {\cal B}_\mu (x)   + \theta \,  (0)
\equiv \bar {\cal B}_\mu (x)  + \theta \,[s_{ab} \,\bar {\cal B}_\mu (x)],\nonumber\\
\tilde\varphi (x)\;\;\;\longrightarrow  {\it {\tilde\Phi}}^{(ab)} (x,\theta) & = & \tilde\varphi (x) + \theta\,  (0)\equiv 
\tilde\varphi (x) + \theta \,[s_{ab} \,\tilde\varphi (x)],\nonumber\\
\tilde\phi_\mu(x)\;\;\longrightarrow  {\it {\tilde\Phi}}_\mu^{(ab)} (x,  \theta) & = & \tilde\phi_\mu(x) +\theta \,  (0)\equiv 
\tilde\phi_\mu(x) +\theta \, [s_{ab} \,\tilde\phi_\mu(x) ],\nonumber\\
\lambda(x)\;\; \longrightarrow \;\;\tilde\lambda ^{(ab)} (x, \theta)& = & \lambda(x)  + \theta\,(0)\equiv 
\lambda(x)  + \theta \,[s_{ab} \,\lambda(x)] ,\nonumber\\
\rho(x)\;\; \longrightarrow \;\; \tilde\rho  ^{(ab)} (x, \theta) & = &\rho(x)  + \theta\, (0)\equiv 
\rho(x)  + \theta\, [s_{ab} \,  \rho(x)], \nonumber \\
{\cal B}(x)\;\; \longrightarrow \;\;\tilde{\cal B}^{(ab)} (x, \theta) &=& {\cal B}(x) + \theta\, (0) \equiv {\cal B}(x)
 + \theta\, [s_{ab} \,  {\cal B} (x)] ,\nonumber\\
{\bar{\cal B}}(x)\;\; \longrightarrow \;\;{\tilde{\bar {\cal B}}}^{(ab)} (x, \theta) &=& {\bar{\cal B}}(x) + \theta \,(0)\equiv 
{\bar{\cal B}}(x) + \theta \,[s_{ab} \, {{\bar {\cal B}}} (x)].
\end{eqnarray}
It is self-evident that we have an interesting relationship: $\partial_\theta\, \Omega ^{(ab)}(x, \theta) = s_{ab}\,\omega (x)$ where the generic superfield 
$\Omega ^{(ab)}(x, \theta)$  represents nothing but the {\it chiral} superfields present on the l.h.s. 
of Eqs. (38) as well as (40) {\it and} $\omega (x)$ denotes   nothing but the generic 4D field which stands  for the {\it ordinary} basic and auxiliary 
fields of the 4D {\it ordinary} Lagrangian density ${\cal L}_{(\bar B,{\bar{\cal B})}}$ (that respects anti-BRST symmetry transformations 
(12) in a {\it perfect} manner [cf. Eq. (14)]).  In other words, the translation of the superfields 
(obtained after the application of  anti-BRST invariant restrictions) along the  {\it chiral} $\theta$-direction of the {\it chiral} (4, 1)-dimensional super sub-manifold generates the anti-BRST symmetry transformations $(s_{ab})$ in the {\it ordinary} 4D space [cf. Eq. (12)]. We also observe 
that the nilpotency $(s_{ab}^2 = 0)$ of $s_{ab}$ and the nilpotency $(\partial_\theta ^2 = 0)$ of the translational 
generator $(\partial _\theta)$ (along the $\theta$-direction of the {\it chiral} super sub-manifold) are deeply related to each-other.


\subsection{(Anti-)co-BRST Symmetries: ACSA}

We focus now on the derivation of the (anti-)co-BRST symmetry transformations $s_{(a)d}$ by applying the ACSA to BRST formalism.
For this purpose, we use the (anti-)chiral super expansions (29) and (36) for the sake of brevity\footnote{We can
choose a (4, 4)-dimensional supermanifold for the generalization of our ordinary 4D theory where the Grassmannian
variables can be chosen to be $(\theta_1, \bar\theta_1, \theta_2, \bar\theta_2)$ with the {\it fermionic} properties: 
$\theta_1^2 = \theta_2^2 = 0, \,\bar{\theta}_1^2 =  \bar\theta_2^2 = 0, \, \theta_1\,\bar{\theta}_1 + \bar{\theta}_1\,\theta_1 = 0, \, 
\theta_2\,\bar{\theta}_2 + \bar{\theta}_2\,\theta_2 = 0$, etc. The pair $(\theta_1, \bar\theta_1)$ and corresponding derivatives 
$(\partial_{\theta1}, \partial_{\bar\theta1})$ can be associated with the BRST and anti-BRST symmetries [keeping the pair  
$(\theta_2, \bar\theta_2)$ intact].
On the other hand, the pair $(\theta_2, \bar\theta_2)$ and 
corresponding derivatives   $(\partial_{\theta 2}, \partial_{\bar\theta 2})$ could be associated with the 
co-BRST and anti-co-BRST symmetries [keeping the pair  
$(\theta_1, \bar\theta_1)$ intact]. This is required because the (anti-)BRST and (anti-)co-BRST symmetries are
{\it independent} of each-other as {\it are} the exterior and co-exterior derivatives of differential geometry. However, for the sake of brevity, 
we have considered {\it only} a single  pair  [i.e. $(\theta, \bar\theta)$]  of Grassmannian variables $\theta$ and 
$\bar\theta$ for the discussions of the off-shell nilpotent (anti-)BRST and (anti-)co-BRST symmetries 
{\it together} in our present endeavor.}. First of all, we derive the co-BRST symmetry transformations
$(s_d)$ by taking into account the {\it chiral} super expansions (36). The basic tenets of ACSA to BRST formalism 
states that the following useful and interesting co-BRST invariant quantities, namely;
\begin{eqnarray} 
&&s_{d}\big[\rho\,\tilde{\varphi}\big] = 0,\qquad s_{d}\big[{\beta}\,\rho + \tilde{\varphi}\,\lambda\big] = 0,
\qquad s_{d}\big[\bar{C}_\mu\partial^\mu\lambda - \partial_\mu{\beta}\,\partial^\mu\bar{\beta}\big] = 0,  \nonumber \\
&&s_{d}\big[\bar{C}_\mu\partial^\mu\bar{C} + \tilde{\phi}_\mu\partial^\mu\bar{\beta}\big] = 0, \quad s_{d}\big[\bar{\cal B}_\mu 
+ \partial_\mu\tilde{\varphi}\big] = 0, \quad s_{d}\big[\bar{\cal B} + m\,\tilde{\varphi}\big] = 0, \nonumber \\
&&s_{d}\big[{C}_\mu \;(m\,\bar{C}^\mu - \partial^\mu\bar{C}) + {\cal B}^\mu\tilde{\phi}_\mu\big] = 0, \quad s_{d}\big[{C}\;(m\,\bar{C}_\mu 
- \partial_\mu\bar{C}) + {\cal B}\;\tilde{\phi}_\mu\big] = 0, \nonumber \\
&&s_{d}\big[m\,B_{\mu\nu} - \varepsilon_{\mu\nu\eta\kappa}\partial^\eta\tilde{\phi}^\kappa\big] = 0,
\quad s_{d}\big[m\,\bar{\beta}\,\beta + \lambda\,\bar{C}\big] = 0, 
\end{eqnarray}
should be independent of the Grassmannian variables $ \theta $ when they are generalized onto the (4, 1)-dimensional
{\it chiral} super sub-manifold [of the {\it general} (4, 2)-dimensional supermanifold]. In other words, we demand the 
following conditions on the {\it chiral} superfields  
\begin{eqnarray}
&& \tilde \rho ^{(d)}(x, \theta)\; {\it {\tilde\Phi}}  (x, \theta) = \rho (x)\,\tilde\varphi (x),\;\;{\tilde {{\bar{\cal B}}}}_\mu (x, \theta)
 + \partial_\mu {\it {\tilde {\Phi}}}(x, \theta) = {\bar {\cal B}}_\mu (x) + 
\partial_\mu \tilde\varphi (x),\nonumber \\
&&\tilde {\beta} (x, \theta)\; \tilde\rho ^{(d)}(x, \theta) + {\it {\tilde \Phi}}  (x, \theta)\;\tilde \lambda ^{(d)}  (x, \theta) 
= {\beta} (x)\,\rho (x) + \tilde\varphi (x)\,\lambda (x),\nonumber\\
&&  + {{\cal B}}^\mu (x)\;\tilde\phi_\mu (x),
\;\;{\tilde {{\cal F}}}(x, \theta)\big [m\,{\tilde {\bar {\cal F}}}_{\mu}(x, \theta) - \partial_\mu {\tilde{\bar  {\cal F}}}(x, \theta)\big] 
+  {\tilde {{\cal B}}}^{(d)} (x, \theta)\;{\it {\tilde \Phi}}_\mu (x, \theta)\nonumber\\
&& = {C}\big[ m\,\bar C_\mu (x) - \partial_\mu{\bar C} (x) \big]
 + {\cal B} (x)\,\tilde \phi_\mu (x),\;\;{\tilde {{\bar{\cal B}}}}(x, \theta)  + m\,{\it {\tilde \Phi}} (x, \theta) \nonumber \\
&& = {\bar {\cal B}} (x) + m\,\tilde\varphi (x), \;\;{\bar {\cal F }}_\mu (x, \theta)\;\partial^\mu \tilde \lambda ^{(d)} (x, \theta) 
- \partial_\mu \tilde {\beta} (x, \theta)\;\partial ^{\mu} \;\tilde{\bar\beta} ^{(d)} (x, \theta)\nonumber\\
&&= \bar C_\mu (x)\partial^\mu\lambda (x) - \partial_\mu{\beta} (x)\,\partial^\mu\bar\beta (x), \;m\,\tilde {\beta} 
(x, \theta)\;\tilde{\bar{\beta}}^{(d)} (x, \theta) + \tilde \lambda ^{(d)}  (x, \theta)\;
{\tilde{\bar {\cal F}}} (x, \theta)\nonumber\\
&& = m\,\bar{\beta} (x)\,\beta (x) + \lambda (x)\,\bar C (x), \;\; {\tilde {\bar {\cal F}}_\mu }(x, \theta)\;\partial^\mu {\tilde{\bar{\cal F}}} (x, \theta) +  
{\it {\tilde \Phi}}_\mu ^{(d)}(x, \theta) \; \partial^\mu \;{\tilde{\bar\beta}}^{(d)} (x, \theta) \nonumber\\   
&&= \bar C_\mu (x)\;\partial^\mu{\bar C} (x) + \tilde\phi_\mu (x)\;\partial^\mu\bar\beta (x),\; m\,
{\tilde {B}}_{\mu\nu} (x, \theta) - \varepsilon_{\mu\nu\eta\kappa} \partial^\eta {\it {\tilde{\Phi}}}^\kappa (x, \theta) =\nonumber\\
&& m\,{B}_{\mu\nu} (x) - \varepsilon _{\mu\nu\eta\kappa} \partial^\eta \tilde\phi^\kappa (x),\;\;
 {\tilde {\cal F}}_\mu(x, \theta)\; \big[m\,{\tilde {\bar {\cal F}}}^{\mu}(x, \theta)
 - \partial^\mu {\tilde {\bar {\cal F}}}(x, \theta)\big]\nonumber\\
&& + {\tilde {{\cal B}}}^{\mu (d)} (x, \theta)\;{\it {\tilde\Phi}}_\mu ^{(d)}(x, \theta)= {C}_\mu (x)\,[m\,\bar C^\mu (x) - \partial^\mu{\bar C} (x)], 
\end{eqnarray}
where the superfields  with the superscript $(d)$ have been derived from our earlier observation: 
$s_d [\partial^\nu B_{\nu\mu}, B_\mu, {\bar B}_\mu, B, \bar B,  {\cal B}_\mu, {\cal B},  \phi_\mu, \bar\beta, \varphi, \rho, \lambda] = 0$
[cf. Eq. (19)] which imply  the following {\it trivial} restrictions on the {\it chiral} superfields:
\begin{eqnarray}
&&\partial^\nu \tilde B_{\nu\mu}^{(d)}  (x, \theta) = \partial^\nu B_{\nu\mu} (x),\quad  \tilde\rho ^{(d)} (x, \theta)
 =\rho (x), \quad \tilde\lambda ^{(d)} (x, \theta)= \lambda (x),\nonumber\\
&& \tilde {\cal B}_\mu ^{(d)} (x, \theta) = {\cal B}_\mu (x),\quad
 \Phi_\mu ^{(d)} (x, \theta)  = \phi_\mu (x),\quad \tilde {\cal B}^{(d)} (x, \theta) = {\cal B}(x) ,\nonumber\\
&& \tilde B ^{(d)} (x, \theta)  = B (x),\quad \tilde {\bar B} ^{(d)} (x, \theta) = \bar B (x), \quad \tilde B_\mu^{(d)} (x, \theta) = B_\mu (x),\nonumber\\
&& \tilde {\bar\beta} ^{(d)} (x, \theta) = \bar\beta (x),\quad \quad \tilde\Phi ^{(d)} (x, \theta)  = \varphi  (x),\quad 
{\tilde {\bar B}}_\mu^{(d)}(x, \theta) = \bar B_\mu(x).
 \end{eqnarray}
A close look at the {\it chiral} super expansions (36) demonstrates that the secondary 
fields $(\bar R_\mu, \bar B_3,  \bar B _4, \bar f_2,  \bar f_3, \bar f_4, \bar f_5, \bar f_7,\bar f_\mu ^{(1)}, \bar f_\mu ^{(2)}, \bar f_\mu ^{(3)})$ are {\it all trivially} equal to zero due to the straightforward co-BRST invariant restrictions in (43).

The restrictions (42) on the {\it chiral} superfields, along with the inputs from (43), 
lead to the determination of  the secondary fields [in terms of the {\it basic} and {\it auxiliary}
 fields of ${\cal L}_{(B, {\cal B})}$] of the {\it chiral} expansions in (36) as:
\begin{eqnarray}
&&  \bar R_{\mu\nu} = - \varepsilon_{\mu\nu\eta\kappa} \partial^\eta \bar C^\kappa, \;
\bar B_\mu^{(1)} = {\cal B}_\mu, \; \bar B_\mu^{(2)}  = - \partial_\mu \bar \beta, 
\; \bar f_1 = - \lambda,\; \bar f^{(4)}_\mu =   \partial_\mu \rho, \nonumber\\
&& \bar R_{\mu}^{(1)} = \partial_\mu \bar C - m\, \bar C_\mu,
\;\, \bar B_1 =  {\cal B}, \quad \bar B_2 = - \,m\,\bar \beta, 
 \;\, \bar f_8 =  m\, \rho,\; \bar f_6 = -\,\rho.
\end{eqnarray}
Substitutions of these precise values into the {\it chiral} expansions (36) lead to the following in terms
of the super expansions along $\theta$, namely;
\begin{eqnarray}
B_{\mu\nu} (x) \longrightarrow {\tilde {B}_{\mu\nu}}^{(d)} (x, \theta) &= &B_{\mu\nu} (x) +
 \theta\,[- \varepsilon_{\mu\nu\eta\kappa} \partial^\eta \bar C^\kappa (x)]\nonumber\\
&\equiv & B_{\mu\nu} (x) + \theta\,[s_d\,B_{\mu\nu} (x)],\nonumber\\
C_\mu (x) \;\,\longrightarrow \;\;{\tilde {\cal F}}_\mu^{(d)} (x, \theta) & = & C_\mu (x) + \theta\,(\,{\cal B}_\mu (x))\equiv
C_\mu (x) + \theta\;[s_d\; C_\mu(x)], \nonumber\\ 
\bar C_\mu (x)\longrightarrow \;{\tilde {\bar {\cal F}}}_\mu ^{(d)} (x, \theta) & = & \bar C_\mu (x) + \theta\,(-\,\partial_\mu{\bar {\beta}}(x)) \equiv
\bar C_\mu (x) + \theta\,  [s_d\, \bar C_\mu(x)], \nonumber\\
\beta (x)\longrightarrow \;\; {\tilde \beta} ^{(d)}(x, \theta ) \; & = & \beta (x) + \theta\, (-\,\lambda (x)) \equiv  \beta (x) 
+ \theta\, [s_d\,\beta(x)], \nonumber\\ 
\bar\beta (x)\longrightarrow \;\; {\tilde {\bar \beta}} ^{(d)} (x, \theta) & = & \bar\beta (x) +  \theta\, \,(0)\equiv \beta (x)
 +  \theta\, \,[s_d\, \bar\beta (x)], \nonumber\\
\varphi (x)\;\;\longrightarrow \;\; {\tilde \Phi}^{(d)} (x,\theta) & = &\varphi (x) + \theta\, ((0))\equiv \varphi (x) + \theta\, [s_d\,\varphi (x)] ,\nonumber\\
\phi_\mu(x)\;\;\longrightarrow  {\tilde { \Phi}}_\mu ^{(d)}(x,  \theta) & = &\phi_\mu(x) +\theta \, (0)\equiv 
\phi_\mu(x) +\theta \,[s_d\phi_\mu (x)] ,\nonumber\\
C(x) \;\;\longrightarrow \;\; {\tilde {\cal F}}^{(d)} (x, \theta) & = & C(x) + \theta \, ({\cal B} (x)) \equiv 
C(x) + \theta \; [s_d \, C(x)], \nonumber\\
\bar C (x)\;\;\longrightarrow \;\; {\tilde {\bar {\cal F}}}^{(d)} (x,  \theta) & = &\bar C (x) +  \theta \, (-\,m\, \bar\beta(x))
\equiv \bar C (x) +  \theta \,[s_d \bar C(x)],\nonumber\\
B_\mu (x)\;\;\; \longrightarrow  \tilde {B}_\mu ^{(d)} (x, \theta) & = & B_\mu (x)  + \theta\, (0)\equiv B_\mu (x)  + \theta\,[s_d B_\mu(x)],\nonumber\\
\bar B_\mu (x)\; \longrightarrow  \tilde {\bar B}_\mu ^{(d)} (x, \theta) & = & \bar B_\mu (x)  + \theta\,(0)
\equiv \bar B_\mu (x)  + \theta\;[s_d \, \bar B_\mu (x)],\nonumber\\ 
B(x) \;\;\longrightarrow \;\; \tilde {B} ^{(d)} (x, \theta) & = & B (x)  + \theta\, (0)\equiv B (x)  + \theta\, [s_d\, B (x)] ,\nonumber\\
\bar B(x) \;\;\longrightarrow \;\; \tilde {\bar B} ^{(d)} (x, \theta)& = &\bar B (x)  + \theta\,(0)\equiv 
\bar B (x)  + \theta\;[s_d\, \bar B (x)],\nonumber\\
{\cal B}_\mu (x)\; \longrightarrow  \tilde {{\cal B}}_\mu ^{(d)} (x, \theta) & = &{\cal B}_\mu (x)  + \theta\, (0)\equiv
{\cal B}_\mu (x)  + \theta\; [s_d \, {\cal B}_\mu (x)],\nonumber\\
\bar {\cal B}_\mu (x)\; \longrightarrow  {\tilde {\bar {\cal B}}}_{\mu} ^{(d)} (x, \theta)& = &\bar {\cal B}_\mu (x)   + \theta \;  (\partial_\mu\,\rho(x))
\equiv \bar {\cal B}_\mu (x)  + \theta \;[s_d \,\bar {\cal B}_\mu (x)],\nonumber\\
\tilde\varphi (x)\;\;\;\longrightarrow  {\it {\tilde\Phi}}^{(d)} (x,\theta)& = & \tilde\varphi (x) + \theta\;  (-\,\rho(x))\equiv 
\tilde\varphi (x) + \theta \;[s_d \,\tilde\varphi (x)]\nonumber\\
\tilde\phi_\mu(x)\;\;\longrightarrow  {\it {\tilde {\Phi}}}_\mu^{(d)} (x,  \theta) & = & \tilde\phi_\mu(x)
 +\theta \,  [\partial_\mu \bar C(x) - m\, \bar C_\mu(x)]\nonumber\\
 &\equiv & \tilde\phi_\mu(x) +\theta \; [s_d \,\tilde\phi_\mu(x)],\nonumber\\
\lambda(x)\;\; \longrightarrow \;\;\tilde\lambda ^{(d)} (x, \theta)& = & \lambda(x)  + \theta\,(0)\equiv 
\lambda(x)  + \theta \,[s_d \,\lambda(x)] ,\nonumber\\
\rho(x)\;\; \longrightarrow \;\; \tilde\rho  ^{(d)} (x, \theta) & = &\rho(x)  + \theta\, (0)\equiv 
\rho(x)  + \theta\, [s_d \,  \rho(x)], \nonumber \\
{\cal B}(x)\;\; \longrightarrow \;\;\tilde{\cal B}^{(d)} (x, \theta) &=& {\cal B}(x) + \theta\, (0) \equiv {\cal B}(x) 
+ \theta\, [s_d \,  {\cal B} (x)] ,\nonumber\\
{\bar{\cal B}}(x)\;\; \longrightarrow \;\;{\tilde{\bar {\cal B}}}^{(d)} (x, \theta) &=& {\bar{\cal B}}(x) + \theta \,(m\,\rho(x))\equiv 
{\bar{\cal B}}(x) + \theta \,[s_d \, {{\bar {\cal B}}} (x)],
\end{eqnarray}
where the superscript $(d)$ denotes the {\it chiral} super expansions of the superfields that lead to the 
derivation of  the co-BRST  transformations (19) as the coefficients of  $\theta$. 
It is crystal clear that $\partial_{\theta}\,\Omega^{(d)}(x, \theta) = s_d\, \omega (x)$ where  
the generic superfield $\Omega^{(d)}(x, \theta)$ stands for {\it all} the {\it chiral} superfields 
that are present on the l.h.s. of Eqs. (43) as well as  (45) and the ordinary generic field $\omega(x)$
corresponds to {\it all} the basic and auxiliary  fields of the Lagrangian density ${\cal L}_{(B, {\cal B})}$.
We also note that the off-shell nilpotency  $(s_d^2 = 0)$ of  $s_d$ is deeply connected with the 
 nilpotency $(\partial_{\theta}^2 = 0)$  of the translational generator $({\partial_{\theta}})$
  along the $\theta$-direction of the (4, 1)-dimensional {\it chiral} super sub-manifold 
[of the {\it general} (4, 2)-dimensional supermanifold].

We devote time on the derivation of the anti-co-BRST symmetry transformations $s_{ad}$ by applying the basic 
tenets  of ACSA to BRST formalism. In this connection, first of all, we observe  that the following anti-co-BRST
invariant quantities of interest, namely;
\begin{eqnarray}
&&s_{ad}\big [\rho\,\bar{\beta}\big] = 0,\qquad s_{ad}\big[\bar{\beta}\,\lambda + {\tilde{\varphi}}\,\rho \big] = 0,\qquad s_{ad} \big[C_\mu\partial^\mu\rho - \partial_\mu\bar{\beta}\,\partial^\mu\beta \big] = 0,  \nonumber \\
&&s_{ad} \big[C_\mu\partial^\mu{C} + \tilde{\phi}_\mu\partial^\mu\beta \big] = 0, \quad s_{ad} \big[{\cal B}_\mu + \partial_\mu\tilde{\varphi} \big] = 0, \quad s_{ad} \big[{\cal B} + m\,\tilde{\varphi} \big] = 0, \nonumber \\
&&s_{ad} \big[\bar{C}_\mu\;(m\,C^\mu - \partial^\mu{C}) + \bar{{\cal B}}^\mu\tilde{\phi}_\mu \big] = 0, \quad s_{ad} \big[\bar{C}\;(m\,C_\mu - \partial_\mu{C}) + \bar{{\cal B}}\tilde{\phi}_\mu \big] = 0, \nonumber \\
&&s_{ad} \big[m\,B_{\mu\nu} - \varepsilon_{\mu\nu\eta\kappa}\partial^\eta{\tilde{\phi}}^\kappa \big] = 0,\quad s_{ad} \big[m\,\bar{\beta}\,\beta + \rho\,C \big] = 0, 
\end{eqnarray}
can be generalized onto the suitably chosen (4, 1)-dimensional {\it anti-chiral} super sub-manifold 
where we have to take into account the anti-chiral superfield expansions (29). However,  before we perform that,
we note that (due to the trivial anti-co-BRST symmetry invariance $s_{ad} [\partial^\nu B_{\nu\mu}, B_\mu, {\bar B}_\mu, B, \bar B,  \bar {\cal B}_\mu, 
\bar {\cal B},  \phi_\mu, \beta, \varphi, \rho, \lambda] = 0$), we have the following
\begin{eqnarray}
&&\partial^\nu \tilde B_{\nu\mu}^{(ad)}  (x, \bar\theta) = \partial^\nu B_{\nu\mu} (x),\quad  \tilde\rho ^{(ad)} (x, \bar\theta) =\rho (x), \quad \tilde\lambda ^{(ad)} (x, \bar\theta)= \lambda (x),\nonumber\\
&& {\tilde {\bar{\cal B}}}_\mu ^{(ad)} (x, \bar\theta) = {\bar{\cal B}}_\mu (x),\quad
 \tilde\Phi_\mu ^{(ad)} (x, \bar\theta)  = \phi_\mu (x),\quad {\tilde {\bar {\cal B}}}^{(ad)} (x, \bar\theta) = {\bar{\cal B}}(x) ,\nonumber\\
&& \tilde B^{(ad)} (x, \bar\theta)  = B (x),\quad \tilde {\bar B} ^{(ad)} (x, \bar\theta) = \bar B (x), \quad \tilde B_\mu^{(ad)} (x, \bar\theta)
 = B_\mu(x), \nonumber \\ 
&&\tilde \beta ^{(ad)} (x, \bar\theta) = \beta (x),\quad  \tilde\Phi ^{(ad)} (x, \bar\theta)  = \varphi  (x), \quad 
{\tilde {\bar B}}_\mu^{(ad)}(x, \bar\theta) = \bar B_\mu(x),
\end{eqnarray}
where the superscript $(ad)$ denotes the anti-chiral superfields which have been obtained due to the  {\it trivial} anti-co-BRST
invariance [cf. Eq. (18)]. Another way of saying the fact that the secondary fields 
$(R_\mu,  B_3,   B _4,  f_1,   f_3,  f_4, f_5,  f_8, f_\mu ^{(1)}, f_\mu ^{(2)},  f_\mu ^{(3)})$ are {\it trivially} equal to
zero so that we can have $s_{ad} [\partial^\nu B_{\nu\mu}, B_\mu,  {\bar B}_\mu, B, \bar B,  \bar {\cal B}_\mu, 
\bar {\cal B},  \phi_\mu, \beta, \\ \varphi, \rho, \lambda] = 0$ as the coefficients of $\bar\theta$
in the super expansions that have been listed in Eq. (29) and represented in the super expansions (47).

Using the trivial equalities (47), we have the following {\it generalizations} of the anti-co-BRST invariant
quantities (46) [located inside the square brackets] in terms of the anti-chiral superfields, namely;
\begin{eqnarray*}
&& \tilde \rho ^{(ad)}(x, \bar\theta)\; {\tilde {\bar{\beta }}} (x, \bar\theta) = \rho (x)\,\bar\beta (x),\quad {\tilde {\cal B}}_\mu (x, \bar\theta) 
+ \partial_\mu{\it{\tilde\Phi}}(x, \bar\theta) = {\cal B}_\mu (x) + \partial_\mu\tilde\varphi (x),\nonumber \\
&&\tilde {\bar\beta} (x, \bar\theta)\; \tilde\lambda ^{(ad)}(x, \bar\theta) + {\it {\tilde \Phi}}  (x, \bar\theta)\;\tilde \rho ^{(ad)}  
(x, \bar\theta) = \bar{\beta} (x)\,\lambda (x) + \tilde\varphi (x)\,\rho (x),\nonumber\\
&& m\,\tilde {\bar\beta} (x, \bar\theta)\;\tilde\beta^{(ad)} (x, \bar\theta) + \tilde \rho ^{(ad)}  (x, \bar\theta)\;
{\tilde{\cal F}}(x, \bar\theta) = m\,\bar{\beta} (x)\,\beta (x) + \rho (x)\,C (x), \nonumber \\
&& {\tilde {\cal F}}_\mu (x, \bar\theta)\;\partial^\mu {\tilde{\cal F}} (x, \bar\theta) +  
{\it {\tilde \Phi}}_\mu (x, \bar\theta) \; \partial^\mu\tilde\beta^{(ad)} (x, \bar\theta)    
= C_\mu (x)\partial^\mu{C} (x) + \tilde\phi_\mu (x)\partial^\mu\beta (x),\nonumber \\
&& {\tilde {\cal B}}(x, \bar\theta)  + m\,{\it \tilde {\Phi}}(x, \bar\theta) = {\cal B} (x) + m\,\tilde\varphi (x), \quad
{\tilde {\bar {\cal F}}}_\mu(x, \bar\theta)\; \big[m\,{\tilde {\cal F}}^{\mu} (x, \bar\theta) - \partial^\mu {\tilde {\cal F}}(x, \bar\theta)\big]\nonumber\\
 && +\; {\tilde {\bar {\cal B}}}^{\mu} (x, \bar\theta)\;{\it {\tilde \Phi}}_\mu (x, \bar\theta)
  = \bar{C}_\mu (x)\big[m\,C^\mu (x) - \partial^\mu{C} (x)\big] + {\bar {\cal B}}_\mu (x)\;\tilde\phi^\mu (x),\nonumber \\
 \end{eqnarray*}
\begin{eqnarray}
&& {\tilde {\bar {\cal F}}}(x, \bar\theta)\big [m\,{\tilde {\cal F}}_\mu (x, \bar\theta) - \partial_\mu {\tilde {\cal F}}(x, \bar\theta)\big] 
 +\;  {\tilde {\bar {\cal B}}}^{(ad)} (x, \bar\theta)\;{\it{\tilde \Phi}}_\mu (x, \bar\theta)
  = \bar{C}  (x)\big[ m\,C_\mu (x)\nonumber\\
 && - \partial_\mu{C} (x)\big] + \; {\bar {\cal B}}(x)\,\tilde\phi_\mu (x), \,\qquad
 m\,{\tilde B}_{\mu\nu} (x, \bar\theta) - 
\varepsilon_{\mu\nu\eta\kappa}\; \partial^\eta {\it \tilde{\Phi}}^{\kappa}(x, \bar\theta)  = m\,B_{\mu\nu} (x) \nonumber\\
&&- \varepsilon_{\mu\nu\eta\kappa}\; \partial^\eta {\it \tilde{\Phi}}^\kappa (x),\;\; \tilde{\cal F}_\mu  
(x, \bar\theta)\;\partial^\mu \tilde \rho ^{(ad)} (x, \bar\theta)
 - \partial_\mu \tilde {\bar\beta} (x, \bar\theta)\;\partial ^{\mu}\tilde\beta^{(ad)} (x, \theta)\nonumber\\
&& = C_\mu (x)\;\partial^\mu\rho (x) - \partial_\mu\bar{\beta} (x)\,\partial^\mu\beta (x). 
\end{eqnarray}
 At this stage, we substitute the anti-chiral 
super expansions (29) into the above equalities which lead to the determination  of {\it secondary} fields in terms of the {\it basic} and 
{\it auxiliary} fields of the Lagrangian density ${\cal L}_{(\bar B, {\bar {\cal B}})}$ as:
\begin{eqnarray}
&&   R_{\mu\nu} = - \varepsilon_{\mu\nu\eta\kappa} \partial^\eta  C^\kappa, \quad 
 B_\mu^{(2)} = \bar{\cal B}_\mu, \quad  B_\mu^{(1)}  =  \partial_\mu  \beta, 
\quad  f_6 = - \lambda,\quad \bar f^{(3)}_\mu =   \partial_\mu \rho, \nonumber\\
&& R_\mu^{(1)} = \partial_\mu  C - m\,  C_\mu,
\quad  B_2 =  {\bar {\cal B}}, \quad  \quad  B_1 = \,m\, \beta, 
 \quad  f_7 =  m\, \lambda,\quad  f_2 = \rho.
\end{eqnarray}
The substitutions of (49) {\it and} observations in  (47) enable us to write the {\it anti-chiral} super expansions 
(29), in terms of the anti-co-BRST symmetry transformations (18), as:
\begin{eqnarray*}
B_{\mu\nu} (x) \longrightarrow {\tilde {B}_{\mu\nu}}^{(ad)} (x, \bar\theta) &= &B_{\mu\nu} (x) +
 \bar\theta\,[- \varepsilon_{\mu\nu\eta\kappa} \partial^\eta \bar C^\kappa (x)]\nonumber\\
&\equiv & B_{\mu\nu} (x) + \bar\theta\,[s_{ad}\,B_{\mu\nu} (x)],\nonumber\\
C_\mu (x) \;\,\longrightarrow \;\;{\tilde {\cal F}}_\mu^{(ad)} (x, \bar\theta) & = & C_\mu (x) + \bar\theta\,(\partial_\mu \beta (x))\equiv
C_\mu (x) + \bar\theta\,[s_{ad}\, C_\mu(x)], \nonumber\\ 
\bar C_\mu (x)\longrightarrow \;{\tilde {\bar {\cal F}}}_\mu ^{(ad)} (x, \bar\theta) & = & \bar C_\mu (x) + \bar\theta\,(\bar{\cal B}_\mu (x)) \equiv
\bar C_\mu (x) + \bar\theta\,  [s_{ad}\, \bar C_\mu(x)], \nonumber\\
\beta (x)\longrightarrow \;\; {\tilde \beta} ^{(ad)}(x, \bar\theta ) \; & = & \beta (x) + \bar\theta\, (0) \equiv  \beta (x)
 + \bar\theta\, [s_{ad}\,\beta(x)], \nonumber\\ 
\bar\beta (x)\longrightarrow \;\; {\tilde {\bar \beta}} ^{(ad)} (x, \bar\theta) & = & \bar\beta (x) +  \bar\theta\, \,(\rho(x))\equiv \bar\beta (x) + \bar\theta\, \,[s_{ad}\, \bar\beta (x)], \nonumber\\ 
\varphi (x)\;\;\longrightarrow \;\; {\tilde \Phi}^{(ad)} (x,\bar\theta) & = &\varphi (x) + \bar\theta\, ((0))\equiv \varphi (x)
 + \bar\theta\, [s_{ad}\,\varphi (x)] ,\nonumber\\
\phi_\mu(x)\;\;\longrightarrow  {\tilde { \Phi}}_\mu ^{(ad)}(x,  \bar\theta) & = &\phi_\mu(x) +\bar\theta \, (0)\equiv 
\phi_\mu(x) +\bar\theta \,[s_{ad}\phi_\mu(x)] ,\nonumber\\
C(x) \;\;\longrightarrow \;\; {\tilde {\cal F}}^{(ad)} (x, \bar\theta) & = & C(x) + \bar\theta \, (m\, \beta(x)) \equiv 
C(x) + \bar\theta \; [s_{ad} \, C (x)], \nonumber\\
\bar C (x)\;\;\longrightarrow \;\; {\tilde {\bar {\cal F}}}^{(ad)} (x,  \bar\theta) & = &\bar C (x) +  \bar\theta \, (\bar{\cal B}(x))
\equiv \bar C (x) +  \bar\theta \,[s_{ad} \bar C (x)],\nonumber\\
B_\mu (x)\;\;\; \longrightarrow  \tilde {B}_\mu ^{(ad)} (x, \bar\theta) & = & B_\mu (x)  + \bar\theta\, (0)\equiv B_\mu (x) 
 + \bar\theta\;[s_{ad} B_\mu (x)],\nonumber\\
\bar B_\mu (x)\; \longrightarrow  \tilde {\bar B}_\mu ^{(ad)} (x, \bar\theta) & = & \bar B_\mu (x)  + \bar\theta\,(0)
\equiv \bar B_\mu (x)  + \bar\theta\;[s_{ad} \, \bar B_\mu (x)],\nonumber\\
B(x) \;\;\longrightarrow \;\; \tilde {B} ^{(ad)} (x, \bar\theta) & = & B (x)  + \bar\theta\, (0)\equiv B (x)  + \bar\theta\; [s_{ad}\, B (x)] ,\nonumber\\
\bar B(x) \;\;\longrightarrow \;\; \tilde {\bar B} ^{(ad)} (x, \bar\theta)& = &\bar B (x)  + \bar\theta\,(0)\equiv 
\bar B (x)  + \bar\theta\,[s_{ad}\, \bar B (x)],\nonumber\\
{\cal B}_\mu (x)\; \longrightarrow  \tilde {{\cal B}}_\mu ^{(ad)} (x,\bar\theta) & = &{\cal B}_\mu (x)  + \bar\theta\; (\partial_\mu \lambda (x))\equiv
{\cal B}_\mu (x)  + \bar\theta\; [s_{ad} \, {\cal B}_\mu (x)],\nonumber\\
\bar {\cal B}_\mu (x)\; \longrightarrow  {\tilde {\bar {\cal B}}}_{\mu} ^{(ad)} (x, \bar\theta)& = &\bar {\cal B}_\mu (x)   + \bar\theta \;  (0)
\equiv \bar {\cal B}_\mu (x)  + \bar\theta\; [s_{ad} \,\bar {\cal B}_\mu (x)],\nonumber\\
\tilde\varphi (x)\;\;\;\longrightarrow  {\it {\tilde \Phi}}^{(ad)} (x,\bar\theta)& = & \tilde\varphi (x) + \bar\theta\;  (-\,\lambda)\equiv 
\tilde\varphi (x) + \bar\theta\; [s_{ad} \,\tilde\varphi (x)], \nonumber\\
\tilde\phi_\mu(x)\;\;\longrightarrow  {\it {\tilde {\Phi}}}_\mu^{(ad)} (x,  \bar\theta) & = & \tilde\phi_\mu(x)
 +\bar\theta \,  [\partial_\mu  C(x) - m\,  C_\mu(x)]\nonumber\\
&\equiv & 
\tilde\phi_\mu(x) +\bar\theta \; [s_{ad} \,\tilde\phi_\mu(x)],\nonumber\\
\end{eqnarray*}
\begin{eqnarray}
\lambda(x)\;\; \longrightarrow \;\;\tilde\lambda ^{(ad)} (x, \bar\theta)& = & \lambda(x)  + \bar\theta\,(0)\equiv 
\lambda(x)  + \bar\theta \,[s_{ad} \,\lambda(x)] ,\nonumber\\
\rho(x)\;\; \longrightarrow \;\; \tilde\rho  ^{(ad)} (x, \bar\theta) & = &\rho(x)  + \bar\theta\, (0)\equiv 
\rho(x)  + \bar\theta\, [s_{ad} \,  \rho(x)], \nonumber \\
{\cal B}(x)\;\; \longrightarrow \;\;\tilde{\cal B}^{(ad)} (x, \bar\theta) & = & {\cal B}(x) + \bar\theta\, (m \, \lambda) \equiv {\cal B}(x) 
+ \bar\theta\, [s_{ad} \,  {\cal B} (x)] ,\nonumber\\
{\bar{\cal B}}(x)\;\; \longrightarrow \;\;{\tilde{\bar {\cal B}}}^{(ad)} (x, \bar\theta) & = & {\bar{\cal B}}(x) + \bar\theta \,(0)\equiv 
{\bar{\cal B}}(x) + \bar\theta \,[s_{ad} \, {{\bar {\cal B}}} (x)].
\end{eqnarray}
The above {\it final} expansions explicitly show that we have already derived the anti-co-BRST 
symmetry transformations $(s_{ad})$ as the coefficients of $\bar\theta$. In other words, we note that
 $\partial_{\bar\theta} \,\Omega^{(ad)} (x, \bar\theta) = s_{ad}\, \omega (x)$ where $\Omega^{(ad)} (x, \bar\theta)$
 is the generic superfield that stands for all the { \it anti-chiral} superfields which are present on the l.h.s. of Eqs. (47) as well as (50) and $\omega (x)$
 is the generic field which corresponds to the {\it ordinary} basic and {\it auxiliary} fields of ${\cal L}_{(\bar B,{\bar{\cal B})}}$ that are the {\it first}
 terms on the r.h.s. of Eqs. (47) as well as (50). This observation also implies that we have interconnection between the nilpotency 
 ($s_{ad}^2 = 0$) of $s_{ad}$ and nilpotency ($\partial_{\bar\theta}^2 = 0$) of the translational generator 
$\partial_{\bar\theta}$ along $\bar\theta$-direction of the anti-chiral super sub-manifold as:
$ s_{ad}^2 = 0\Leftrightarrow\partial_{\bar\theta}^2 = 0$.


\section {Invariance of the Lagrangian Densities:\\ ACSA to BRST Formalism}

In this section, we establish the existence of the CF-type restrictions [cf. Eq. (21)] within the framework   ACSA by
capturing the symmetry invariance of the Lagrangian densities  ${\cal L}_{(B,{{\cal B})}}$ and 
${\cal L}_{(\bar B, \bar{\cal B})}$. This exercise {\it also} proves the {\it equivalence} of the {\it coupled}
Lagrangian densities ${\cal L}_{( B, {\cal B})}$ and ${\cal L}_{(\bar B, {\bar{\cal B})}}$ w.r.t. the (anti-)BRST
and (anti-)co-BRST symmetry transformations in the space of fields
where the CF-type restrictions [cf. Eq. (21)] are satisfied. Our present section is divided into two parts.
In sub-section 5.1, we discuss the (anti-) BRST invariance and the CF-type restrictions (associated with the nilpotent (anti-)BRST symmetries).
Our sub-section 5.2  is devoted to the discussion of the (anti-)co-BRST invariance and derivation  of the CF-type restrictions 
(associated with {\it these} nilpotent symmetries).


\subsection {(Anti-)BRST Invariance and CF-Type Restrictions}

In this sub-section, we discuss the (anti-)BRST invariance and derivation of the proper 
CF-type restrictions within the framework of ACSA to BRST formalism. Toward this objective  in mind, we perform  the following generalization
of the {\it ordinary} Lagrangian density:  ${\cal L}_{(B,\cal B)} \longrightarrow  \tilde {\cal L}^{(ac)}_{(B,\cal B)}(x, \bar\theta),$ namely;
\begin{eqnarray*}
\tilde {\cal L}^{(ac)}_{(B, \cal B)} (x, \bar\theta)  
 & = & \frac{1}{2}{\cal B}_\mu (x) \; {\cal B}^\mu (x)
 -  {\cal B}^\mu (x) \Big(\frac{1}{2}\varepsilon_{\mu\nu\eta\kappa} \, \partial^\nu 
{\tilde B}^{\eta \kappa (b)} (x, \bar\theta)\nonumber\\
 &- & \frac{1}{2}\,\partial_\mu \tilde \varphi (x) + m \tilde \phi_\mu (x)\Big)
- \frac{m^2}{4} \,{\tilde {B}}^{\mu\nu{(b)}}(x, \bar\theta)\,{\tilde {B}}_{\mu\nu}^{(b)}(x, \bar\theta)\nonumber\\
& - &\frac{1}{2}\,\partial^\mu {\tilde \Phi}^{\nu (b)} (x, \bar\theta)
\,\Big(\partial_\mu {\tilde \Phi}_\nu ^{(b)} (x, \bar\theta) - \partial_\nu {\tilde \Phi}_\mu ^{(b)} (x, \bar\theta)\Big)  \nonumber\\
& + & m\,\tilde B^{\mu\nu (b)} (x, \bar\theta)\,\partial_\mu {\tilde\Phi}_\nu^{(b)} (x, \bar\theta)
 +  \frac{1}{4}\, {\tilde\Phi}^{\mu\nu}(x)\,
{\tilde\Phi}_{\mu\nu}(x)\nonumber\\
\end{eqnarray*}
 \begin{eqnarray}
 &+ &  \frac{m}{2}\, \varepsilon^{\mu\nu\eta\kappa} {\tilde B}^{(b)}_{\mu\nu}(x, \bar\theta) \partial_\eta {\tilde\phi}_\kappa (x) 
 - \frac{1}{2}\,B^\mu (x)\, B_\mu (x)\nonumber\\
&  + & B^{\mu} (x)\;\left(\partial^\nu {\tilde B}_{\nu\mu}^{(b)} (x, \bar\theta) - \frac{1}{2}\, 
\partial_\mu \tilde\Phi ^{(b)} (x, \bar\theta)   
+ m {\tilde\Phi}_\mu ^{(b)} (x, \bar\theta) \right)   \nonumber\\
&+& \frac{1}{2}\, B (x)\, B (x) + B (x) \,\Big(\partial_\mu {\tilde\Phi}^{\mu (b)} (x, \bar\theta) + \frac{m}{2} \,\tilde\Phi ^{(b)} (x, \bar\theta) \Big)\nonumber\\ 
&-& \frac{1}{2}\,{\cal B} (x)\,{\cal B} (x) 
- {\cal B}(x)  \Big(\partial_\mu \tilde \phi^\mu (x)  + \frac{m}{2} \, \tilde\varphi (x)  \Big)\nonumber\\
& + & \Big(\partial_\mu {\tilde{\bar {\cal F}}}^{(b)} (x, \bar\theta)
- m \,{\tilde {\bar {\cal F}}}_\mu ^{(b)} (x, \bar\theta)\,\Big)\nonumber\\
& &\Big(\partial^\mu \,{\tilde{\cal F}}^{(b)} (x, \bar\theta) 
 - m \,{\tilde {\cal F}}^{\mu (b)} (x, \bar\theta) \Big)\nonumber\\
& - & \Big(\partial_\mu {\tilde {\bar {\cal F}}}_\nu ^{(b)} (x, \bar\theta) - \partial_\nu {\tilde {\bar {\cal F}}}_\mu ^{(b)} (x, \bar\theta) \Big)\Big(\partial^\mu {\tilde {\cal F}}^{\nu(b)} (x, \bar\theta)  \Big)\nonumber\\  
&-& \frac{1}{2}\,\partial_\mu {\tilde {\bar\beta}}^{(b)} (x, \bar\theta) \,\partial^\mu \beta (x) + 
\frac{1}{2}\, m^2\, \tilde{\bar\beta}^{(b)} (x, \bar\theta)\, \beta (x) \nonumber\\
&-& \frac{1}{2}\left(\partial_\mu {\tilde {\bar {\cal F}}}^{\mu (b)} (x, \bar\theta) +  m \, {\tilde{\bar {\cal F}}}^{(b)} (x, \bar\theta) + \frac{1}{4}\,\rho   (x) \right) \lambda  (x) \nonumber\\ 
&-&\frac{1}{2}\left(\partial_\mu {\tilde {\cal F}}^{\mu(b)} (x, \bar\theta)  +  m \, {\tilde{\cal F}}^{(b)} (x, \bar\theta) - \frac{1}{4}\, \lambda  (x)\right) \rho    (x),
\label{28} 
\end{eqnarray}
where the superscript $(ac)$  on the super Lagrangian density denotes that we have taken into account the 
anti-chiral superfields in the {\it anti-chiral} super Lagrangian density 
$[\tilde{\cal L}_{(B, {\cal B})}^{(ac)}(x, \bar \theta)]$ {\it which} incorporates  a combination of the  {\it ordinary} fields and {\it anti-chiral} 
superfields with superscript $(b)$ that have been derived earlier in Eq. (34). 
We would like to point out that the ordinary fields are those which are trivially  BRST invariant [cf. Eq. (13)]. Hence, they are {\it independent} 
of the Grassmannian variable.  It is now straightforward to check that we have the following
expression when we apply $\partial_{\bar\theta}$ on (51), namely;
\begin{eqnarray}
\frac {\partial}{\partial \bar\theta}\Big[\tilde{\cal L}^{(ac)}_{(B,\cal B)}\Big] &=& - \partial_\mu \bigg[m\, \varepsilon^{\mu\nu\eta\kappa}\, \tilde \phi_\nu \big(\partial_\eta C_\kappa \big) 
+ B_\nu \big(\partial^\mu C^\nu - \partial^\nu C^\mu  \big)\nonumber\\
& + & \frac{1}{2}\, B^\mu\, \lambda 
- B \big(\partial^\mu C - m C^\mu\big) - \frac{1}{2}\, \big(\partial^\mu \beta \big)\,\rho \bigg] \equiv  
s_b\,{\cal L}_{(B, {\cal B})},
\end{eqnarray}
which demonstrates that the r.h.s. is a total spacetime derivative that has been derived earlier 
in Eq. (15) due to the BRST transformation of the Lagrangian density ${\cal L}_{(B, {\cal B})}$
in the ordinary space.

To capture the anti-BRST invariance of ${\cal L}_{(\bar B, \bar{\cal B})}$ within the framework of ACSA, we
perform the generalization: ${\cal L}_{(\bar B,\bar {\cal B)}}\longrightarrow  \tilde {\cal L}^{(c)}_{(\bar B,\bar {\cal B)}}(x, \theta)$ as
follows      
\begin{eqnarray*}
 \tilde {\cal L}^{(c)}_{(\bar B,\bar {\cal B)}}  (x, \theta)
 &=& \frac{1}{2}\bar {\cal B}_\mu (x) \; \bar{\cal B}^\mu (x) + \bar{\cal B}^\mu (x) \Big(\frac{1}{2}\varepsilon_{\mu\nu\eta\kappa} \, \partial^\nu {\tilde {B}}^{\eta\kappa(ab)} (x, \theta)\nonumber\\
& + & \frac{1}{2}\,\partial_\mu \tilde \varphi (x) + m \;\tilde \phi_\mu (x)\Big)
- \frac{m^2}{4} \,{\tilde {B}}^{{\mu\nu}(ab)}(x, \theta)\,{\tilde {B}}_{\mu\nu}^{(ab)}(x, \theta)\nonumber\\
 &-&\frac{1}{2}\,\partial^\mu {\tilde \Phi}^{\nu (ab)} (x, \theta)
\,\Big(\partial_\mu {{\tilde \Phi}_\nu }^{(ab)} (x, \theta)
  -  \partial_\nu {\tilde \Phi}_\mu ^{(ab)} (x, \theta)\Big)\nonumber\\
 \end{eqnarray*}
\begin{eqnarray} 
 & + & m\, {\tilde B}^{\mu\nu (ab)} (x, \theta)\,\partial_\mu {\tilde\Phi}_\nu^{(ab)} (x, \theta)+ \frac{1}{4}\, {\tilde\Phi}^{\mu\nu} (x) 
{\tilde\Phi}_{\mu\nu} (x)\nonumber\\
& +  & \frac{m}{2}\, \varepsilon^{\mu\nu\eta\kappa} {\tilde B}^{(ab)}_{\mu\nu}(x, \theta) \partial_\eta {{\tilde\phi}}_\kappa (x) - \frac{1}{2}\,\bar B^{\mu} (x) \, \bar B_{\mu}(x)\nonumber\\
& - & \bar B^{\mu} (x)\,\left( \partial^\nu {\tilde B}_{\nu\mu}^{(ab)} (x, \theta) + \frac{1}{2}\, \partial_\mu \tilde\Phi ^{(ab)} (x, \theta)   
+ m \;{\tilde\Phi}_\mu ^{(ab)} (x, \theta) \right) \nonumber\\
& + & \frac{1}{2}\, \bar B (x)\,  \bar B (x))
-  \bar B (x) \,\left(\partial_\mu {\tilde\Phi}^{\mu (ab)} (x, \theta) - \frac{m}{2} \,\tilde\Phi ^{(ab)} (x, \theta) \right)\nonumber\\ 
&-& \frac{1}{2}\,\bar{\cal B} (x)\,\bar{\cal B} (x) + \bar{\cal B}(x)  \left(\partial_\mu \tilde \phi^\mu (x)  - \frac{m}{2} \, 
\tilde\varphi (x)  \right)\nonumber\\
&+&  \Big(\partial_\mu {\tilde{\bar {\cal F}}}^{(ab)} (x, \theta)
- m \,{\tilde {\bar {\cal F}}}_\mu ^{(ab)} (x, \theta)\,\Big)\nonumber\\
&& \Big(\partial^\mu \,{\tilde{\cal F}}^{(ab)} (x, \theta)
- m \,{\tilde {\cal F}}^{\mu (ab)} (x, \theta) \Big)\nonumber\\ 
&-& \Big(\partial_\mu {\tilde {\bar {\cal F}}}_{\nu} ^{(ab)} (x, \theta) - \partial_\nu {\tilde {\bar {\cal F}}}_\mu ^{(ab)} (x, \theta) \Big)\Big(\partial^\mu {\tilde {\cal F}}^{\nu (ab)} (x, \theta)  \Big)\nonumber\\  
&-& \frac{1}{2}\,\partial_\mu {\bar \beta} (x) \,\partial^\mu \beta ^{(ab)} (x, \theta) + \frac{1}{2}\, m^2\, \bar\beta (x)\, \beta ^{(ab)} (x, \theta) \nonumber\\
&-& \frac{1}{2}\left(\partial_\mu {\tilde {\bar {\cal F}}}^{\mu(ab)} (x, \theta) +  m \, {\tilde{\bar {\cal F}}}^{(ab)} (x, \theta) + \frac{1}{4}\, \rho   (x) \right) \lambda  (x) \nonumber\\ 
&-&\frac{1}{2}\left(\partial_\mu {\tilde {\cal F}}^{\mu(ab)} (x, \theta)  +  m \, {\tilde{\cal F}}^{(ab)} (x, \theta) 
- \frac{1}{4}\, \lambda  (x)\right) \rho    (x),
\end{eqnarray}
where we have obtained a {\it chiral} super Lagrangian density from the {\it ordinary} Lagrangian density 
${\cal L}_{(\bar B, \bar{\cal B})}$ [that is characterized by a superscript $(c)$]. This {\it chiral} super
Lagrangian density is made up of the {\it ordinary} fields as well as the {\it chiral} superfields with superscript 
$(ab)$ that have been obtained in Eq. (40). Now we are in the position to apply a derivative (${\partial_\theta}$)
w.r.t. the Grassmannian variable $\theta$ on (53). This operation  leads to the following:
\begin{eqnarray}
\frac {\partial}{\partial\theta}\Big[\tilde{\cal L}^{(c)}_{{(\bar B,\bar {\cal B)}}}\Big] &=& - \partial_\mu \bigg[m\, \varepsilon^{\mu\nu\eta\kappa}\, 
\tilde \phi_\nu \big(\partial_\eta \bar C_\kappa \big) 
- \bar B_\nu \big(\partial^\mu \bar C^\nu - \partial^\nu \bar C^\mu  \big)\nonumber\\ 
& + & \frac{1}{2}\, \bar B^\mu\, \rho 
+ \bar B\; \big(\partial^\mu \bar C - m \;\bar C^\mu\big) - \frac{1}{2}\, \big(\partial^\mu \bar \beta \big)\,\lambda \big]\equiv 
s_{ab}\;{\cal L}_{(\bar B, {\bar {\cal B}})}.
\label{31}
\end{eqnarray}
Keeping in mind the mapping $\partial_\theta \longleftrightarrow s_{ab}$ [4-6], it is straightforward to note
 that we have captured the {\it perfect} anti-BRST invariance  of the Lagrangian density ${\cal L}_{(\bar B, {\bar {\cal B}})}$
 [as we have: $\partial_\theta \tilde {\cal L}^{(c)}_{(\bar B, {\bar {\cal B}})} \equiv s_{ab} {\cal L}_{(\bar B, {\bar {\cal B}})}$].

To establish the existence of the CF-type restrictions  [cf. Eq. (21)] and {\it equivalence} of the Lagrangian 
densities ${\cal L}_{(B, {{\cal B}})}$ and ${\cal L}_{(\bar B, {\bar {\cal B}})}$ w.r.t. the nilpotent
symmetries $s_{(a)b}$, we generalize the Lagrangian density 
${\cal L}_{(B, {{\cal B}})}$ to its {\it chiral} counterpart: ${\cal L}_{(B,\cal B)} \longrightarrow  \tilde {\cal L}^{(c)}_{(B,\cal B)}$ as 
\begin{eqnarray*}
\tilde {\cal L}^{(c)}_{(B,\cal B)}  (x, \theta)
 &=& \frac{1}{2}{\cal B}_\mu (x) \;{\cal B}^\mu (x) - {\cal B}^\mu (x) \Big(\frac{1}{2}\varepsilon_{\mu\nu\eta\kappa} \, \partial^\nu {\tilde {B}}^{\eta\kappa (ab)} (x, \theta)\nonumber\\
& - & \frac{1}{2}\,\partial_\mu \tilde \varphi (x) + m \;\tilde \phi_\mu (x)\Big) 
- \frac{m^2}{4} \,{\tilde {B}}^{\mu\nu (ab)}(x, \theta)\,{\tilde {B}}_{\mu\nu}^{(ab)}(x, \theta)\nonumber\\
 \end{eqnarray*}
\begin{eqnarray}
& - &\frac{1}{2}\,\partial^\mu {\tilde \Phi}^{\nu(ab)} (x, \theta)
\,\Big(\partial_\mu {{\tilde \Phi}_\nu }^{(ab)} (x, \theta)
 -  \partial_\nu {{\tilde\Phi}}_\mu ^{(ab)} (x, \theta)\Big)\nonumber\\ 
& + & {m}\, {\tilde {B}}^{\mu\nu (ab)} (x, \theta)\,\partial_\mu 
{{\tilde \Phi}}_\nu^{(ab)} (x, \theta)+ \frac{1}{4}\, \tilde\Phi^{\mu\nu}(x)\tilde\Phi_{\mu\nu}(x)\nonumber\\
&+& \frac{m}{2}\, \varepsilon^{\mu\nu\eta\kappa} {\tilde B}^{(ab)}_{\mu\nu}(x, \theta) \partial_\eta {{\tilde\phi}}_\kappa (x)  
- \frac{1}{2}\,B^{\mu (ab)} (x, \theta) \, B_{\mu}^{(ab)} (x, \theta)\nonumber\\
& + & B^{\mu (ab)} (x, \theta)\,\Big( \partial^\nu {\tilde B}_{\nu\mu}^{(ab)} (x, \theta) - \frac{1}{2}\, \partial_\mu \tilde\Phi ^{(ab)} (x, \theta)\nonumber\\   
& + & m\; {\tilde\Phi}_\mu ^{(ab)} (x, \theta) \Big) 
 +  \frac{1}{2}\, B^{(ab)} (x, \theta)\,  B^{(ab)} (x, \theta)\nonumber\\
& + &  B^{(ab)} (x, \theta) \,\left(\partial_\mu {\tilde\Phi}^{\mu(ab)} (x, \theta) + \frac{m}{2} \,\tilde\Phi ^{(ab)} (x, \theta) \right)\nonumber\\ 
&-& \frac{1}{2}\,{\cal B} (x)\,{\cal B} (x) - {\cal B}(x)  \left(\partial_\mu \tilde \phi^\mu (x)  + \frac{m}{2} \, \tilde\varphi (x)  \right)\nonumber\\
&+&  \Big(\partial_\mu {\tilde{\bar {\cal F}}}^{(ab)} (x, \theta)
- m \,{\tilde {\bar {\cal F}}}_\mu ^{(ab)} (x, \theta)\,\Big)\nonumber\\
&& \Big(\partial^\mu \,{\tilde{\cal F}}^{(ab)} (x, \theta)
- m \,{\tilde {\cal F}}^{\mu (ab)} (x, \theta) \Big)\nonumber\\ 
&-& \Big(\partial_\mu {\tilde {\bar {\cal F}}}_\nu ^{(ab)} (x, \theta) - \partial_\nu {\tilde {\bar {\cal F}}}_\mu ^{(ab)} (x, \theta) \Big)\Big(\partial^\mu {\tilde {\cal F}}^{\nu (ab)} (x, \theta)  \Big)\nonumber\\  
&-& \frac{1}{2}\,\partial_\mu {\bar \beta} (x) \,\partial^\mu \tilde \beta ^{(ab)} (x, \theta) + \frac{1}{2}\, m^2\, \bar\beta (x)\, \tilde\beta ^{(ab)} (x, \theta) \nonumber\\
&-& \frac{1}{2}\left(\partial_\mu {\tilde {\bar {\cal F}}}^{\mu(ab)} (x, \theta) +  m \, {\tilde{\bar {\cal F}}}^{(ab)} (x, \theta) 
+ \frac{1}{4}\,\rho   (x) \right) \lambda  (x) \nonumber\\ 
&-&\frac{1}{2}\left(\partial_\mu {\tilde {\cal F}}^{\mu (ab)} (x, \theta)  +  
m \, {\tilde{\cal F}}^{(ab)} (x, \theta) - \frac{1}{4}\, \lambda  (x)\right) \rho   (x),
\end{eqnarray}
where the superfields with the superscript $(ab)$ have been obtained earlier [cf. Eq.(40)] after the  applications of 
anti-BRST invariant restrictions and  superscript $(c)$ on ${\cal L}_{(B, {{\cal B}})}$ denotes that we have taken the {\it chiral}
generalization  of  the Lagrangian density ${\cal L}_{(B, {{\cal B}})}$. It will be noted that there are superfields in Eq. (55) as the
 {\it ordinary} fields because they are anti-BRST invariant fields.
It is now elementary to check that we have the following when $\partial_\theta$ operates on $\tilde{\cal L}^{(c)}_{(B,\cal B)}$, namely;
\begin{eqnarray}
\frac {\partial}{\partial \theta}\Big[\tilde{\cal L}^{(c)}_{(B,\cal B)}\Big] &=& - \partial_\mu \bigg[m\, \varepsilon^{\mu\nu\eta\kappa}\, \tilde \phi_\nu \big(\partial_\eta \bar C_\kappa \big)
+ \Big(\partial_\nu B^{\nu\mu} + \frac{1}{2}\, \bar B^\mu + m\; \phi^\mu\Big) \rho \nonumber\\
&+& B_\nu \big(\partial^\mu \bar C^\nu - \partial^\nu \bar C^\mu  \big)  
- B \big(\partial^\mu \bar C - m \;\bar C^\mu\big) - \frac{1}{2}\, \big(\partial^\mu \bar \beta \big)\,\lambda \bigg] \nonumber\\
&+& \frac{1}{2}\, \big[B_\mu + \bar B_\mu + \partial_\mu \varphi \big] \big(\partial^\mu \rho \big) 
+ \partial_\mu\big[B_\nu + \bar B_\nu + \partial_\nu \varphi  \big]\nonumber\\
&& \big(\partial^\mu \bar C^\nu - \partial^\nu \bar C^\mu  \big) 
+ m \big[B_\mu + \bar B_\mu + \partial_\mu \varphi \big]\nonumber\\
 &&\big(\partial^\mu \bar C - m \;\bar C^\mu  \big)
- \frac{m}{2}\, \big[B + \bar B + m \;\varphi \big] \rho \nonumber\\
&-& \partial_\mu \big[B + \bar B + m \;\varphi \big] \big(\partial^\mu \bar C - m\; \bar C^\mu  \big) \equiv s_{ab}\;{\cal L}_{(B,\cal B)}.
\end{eqnarray}
The above equation establishes (keeping in our mind $\partial_\theta \leftrightarrow s_{ab}$) that when the anti-BRST symmetry 
operates on the Lagrangian density ${\cal L}_{(B, {\cal B})}$ in the {\it ordinary} spacetime, we obtain the
variation of ${\cal L}_{(B, {\cal B})}$ such that it transforms to a {\it total} spacetime derivative {\it plus} 
terms that vanish off in the space of fields where the CF-type restrictions [cf. Eq. (21)] are satisfied.
Hence, the Lagrangian density ${\cal L}_{(B, {\cal B})}$ respects {\it both} the BRST and anti-BRST symmetry
transformations {\it together} provided we consider the whole theory on the {\it sub-space} of fields (defined on the flat 4D Minkowskian
spacetime manifold) on which the CF-type restrictions [cf. Eq. (21)] are fully satisfied {\it together}.

Now we capture the BRST invariance of the Lagrangian density ${\cal L}_{(\bar B, \bar{\cal B})}$ within the framework 
of ACSA. Toward this goal in mind, we generalize the {\it ordinary} 4D Lagrangian density ${\cal L}_{(\bar B, \bar{\cal B})}$
onto the (4, 1)-dimensional {\it anti-chiral} supermanifold as the {\it anti-chiral} super Lagrangian density 
 $\tilde{\cal L}^{(ac)}_{(\bar B, \bar{\cal B})} (x, \bar\theta)$. The explicit and lucid form of 
$\tilde{\cal L}^{(ac)}_{(\bar B, \bar{\cal B})} (x, \bar\theta)$ is 
\begin{eqnarray}
\tilde {\cal L}^{(ac)}_{(\bar B,\bar {\cal B)}}  (x, \bar\theta)
 &=& \frac{1}{2}\bar {\cal B}_\mu (x)\; \bar {\cal B}^\mu (x) + \bar {\cal B}^\mu (x) \Big(\frac{1}{2}\varepsilon_{\mu\nu\eta\kappa} \, \partial^\nu {\tilde {B}}^{\eta\kappa (b)} (x, \bar\theta)\nonumber\\
& + & \frac{1}{2}\,\partial_\mu \tilde \varphi (x) + m \;\tilde \phi_\mu (x)\Big)
- \frac{m^2}{4} \,{\tilde {B}}^{\mu\nu (b)}(x, \bar\theta)\,{\tilde {B}}_{\mu\nu}^{(b)}(x, \bar\theta)\nonumber\\
& - & \frac{1}{2}\,\partial^\mu {{\tilde \Phi}}^{\nu (b)} (x, \bar\theta)
\,\Big(\partial_\mu {{\tilde \Phi}}_{\nu} ^{(b)} (x, \bar\theta) - \partial_\nu {{\tilde \Phi}}_\mu ^{(b)} (x, \bar\theta)\Big)  \nonumber\\
&+& {m}\, {\tilde {B}}^{\mu\nu (b)} (x, \bar\theta)\,\partial_\mu {{\tilde \Phi}}_\nu^{(b)} (x, \bar\theta)+ \frac{1}{4}\, \tilde\Phi^{\mu\nu}(x)\tilde\Phi_{\mu\nu}(x)\nonumber\\
& + & \frac{m}{2}\, \varepsilon^{\mu\nu\eta\kappa} {\tilde B}^{(b)}_{\mu\nu}(x, \bar\theta) \;\partial_\eta {{\tilde\phi}}_\kappa (x) 
- \frac{1}{2}\,{\tilde{\bar B}}^{\mu (b)} (x, \bar\theta)\, {\tilde{\bar B}}_{\mu }^{(b)} (x, \bar\theta)\nonumber\\
&-& {\tilde{\bar B}}^{\mu (b)} (x, \bar\theta) \big( \partial^\nu {\tilde B}_{\nu\mu}^{(b)} (x, \bar\theta) + \frac{1}{2}\, \partial_\mu \tilde\Phi ^{(b)} (x, \bar\theta)  \nonumber\\
& + & m \;{\tilde\Phi}_\mu ^{(b)} (x, \bar\theta) \big) + \frac{1}{2}\, {\tilde{\bar B}}^{(b)} (x, \bar\theta)\, {\tilde{\bar B}} ^{(b)} (x, \bar\theta)\nonumber\\ 
& - & {\tilde{\bar B}} ^{(b)} (x, \bar\theta) \,\big(\partial_\mu {\tilde\Phi}^{\mu(b)} (x, \bar\theta)  
-\frac{m}{2} \,\tilde\Phi ^{(b)} (x, \bar\theta) \big)\nonumber\\
& - & \frac{1}{2}\,\bar{\cal B} (x)\,\bar{\cal B} (x) + \bar{\cal B}(x)  \left(\partial_\mu \tilde \phi^\mu (x)  - \frac{m}{2} \, \tilde\varphi (x)  \right)\nonumber\\
&+&  \Big(\partial_\mu {\tilde{\bar {\cal F}}}^{(b)} (x, \bar\theta)
 -  m \,{\tilde {\bar {\cal F}}}_\mu ^{(b)} (x, \bar\theta)\,\Big)\nonumber\\
&& \Big(\partial^\mu \,{\tilde{\cal F}}^{(b)} (x, \bar\theta)
- m \,{\tilde {\cal F}}^{\mu(b)} (x, \bar\theta) \Big)\nonumber\\
&-& \Big(\partial_\mu {\tilde {\bar {\cal F}}}_\nu ^{(b)} (x, \bar\theta) - \partial_\nu {\tilde {\bar {\cal F}}}_\mu ^{(b)} (x, \bar\theta) \Big)\Big(\partial^\mu {\tilde {\cal F}}^{\nu (b)} (x, \bar\theta)  \Big)\nonumber\\  
&-& \frac{1}{2}\,\partial_\mu \tilde{\bar \beta}^{(b)} (x, \bar\theta) \,\partial^\mu \beta (x) + \frac{1}{2}\, m^2\, \tilde{\bar \beta}^{(b)} (x, \bar\theta)\, \beta (x)\nonumber\\
&-& \frac{1}{2}\left(\partial_\mu {\tilde {\bar {\cal F}}}^{\mu(b)} (x, \bar\theta) +  m \, {\tilde{\bar {\cal F}}}^{(b)} (x, \bar\theta) + \frac{1}{4}\,\rho  (x) \right) \lambda (x)\nonumber\\  
&-&\frac{1}{2}\left(\partial_\mu {\tilde {\cal F}}^{\mu (b)} (x, \bar\theta) 
 +  m \, {\tilde{\cal F}}^{(b)} (x, \bar\theta) - \frac{1}{4}\,\lambda  (x)\right) \rho (x), 
\end{eqnarray}
where the superscript $(b)$ on the superfields denotes that these anti-chiral superfields have been obtained  after the 
applications of  BRST-invariant restrictions [cf. Eq. (34)]. In the above {\it super} Lagrangian density 
$\tilde{\cal L}^{(ac)}_{(\bar B, \bar{\cal B})} (x, \bar\theta)$, we have also 4D {\it ordinary} fields 
due to the fact that these fields are BRST-invariant. Keeping in our mind the mapping $\partial_{\bar\theta} \longleftrightarrow s_b$ [4-6],
we operate $\partial_{\bar\theta}$ on the above {\it anti-chiral} super Lagrangian density that leads to:
\begin{eqnarray}
&&\frac {\partial}{\partial \bar\theta}\Big[\tilde{\cal L}^{(ac)}_{{(\bar B,\bar {\cal B)}}}\Big]  =  - \partial_\mu \bigg[m\, \varepsilon^{\mu\nu\eta\kappa}\, \tilde \phi_\nu \big(\partial_\eta C_\kappa \big)
- \Big(\partial_\nu B^{\nu\mu} - \frac{1}{2}\, B^\mu + m\; \phi^\mu\Big) \lambda \nonumber\\
& - & \bar B_\nu \big(\partial^\mu C^\nu - \partial^\nu C^\mu  \big)  
+ \bar B \big(\partial^\mu C - m  \; C^\mu\big) - \frac{1}{2}\, \big(\partial^\mu \beta \big)\,\rho \bigg] \nonumber\\
& + & \frac{1}{2}\, \big[B_\mu + \bar B_\mu + \partial_\mu \varphi \big] \big(\partial^\mu \lambda \big)\nonumber\\ 
& - & \partial_\mu\big[B_\nu + \bar B_\nu + \partial_\nu \varphi  \big] \big(\partial^\mu  C^\nu - \partial^\nu  C^\mu  \big) \nonumber\\
& - & m \big[B_\mu + \bar B_\mu + \partial_\mu \varphi \big] \big(\partial^\mu  C - m \; C^\mu  \big)
- \frac{m}{2}\, \big[B + \bar B + m \;\varphi \big] \lambda \nonumber\\
& + & \partial_\mu \big[B + \bar B + m \;\varphi \big] \big(\partial^\mu  C - m \; C^\mu  \big) \;\equiv\; s_b \,{\cal L}_{(\bar B, \bar{\cal B})}. 
\end{eqnarray} 
Thus, we have captured the BRST symmetry transformation $s_b \,{\cal L}_{(\bar B, \bar{\cal B})}$ [cf. Eq. (25)]
that has been derived [see, the r.h.s. of Eq. (25)] in the {\it ordinary} space  (in the terminology of ACSA to BRST formalism).

We end this sub-section with the remarks that we have expressed the (anti-) BRST invariance
 [cf. Eqs. (56), (58)] within the framework of ACSA. Further, we have derived the CF-type restrictions: 
$B_\mu + \bar B_\mu + \partial_\mu \varphi = 0,\;  B + \bar B + m\;\varphi = 0$ in proving the 
 {\it equivalence} of the Lagrangian densities  ${\cal L}_{(B, {\cal B})}$ and ${\cal L}_{(\bar B, \bar{\cal B})}$
w.r.t. the  nilpotent (anti-)BRST symmetry transformations while expressing the transformations [$(s_b\,{\cal L}_{(\bar B, \bar{\cal B})})$]
and [$(s_{ab}\,{\cal L}_{(\bar B, \bar{\cal B})})$] in the terminology of ACSA. In other words, we note that the r.h.s.
of equations (56) and (58) would become {\it total} spacetime derivatives [cf. Eqs. (24),(25)] if and {\it only}
if we impose the (anti-)BRST invariant CF-type restrictions (21) from {\it outside}.


\subsection{(Anti-)co-BRST Invariance and CF-Type Restrictions: ACSA}

In this sub-section, we prove the (anti-)co-BRST invariance of the Lagrangian densities 
${\cal L}_{(B, {\cal B})}$ and ${\cal L}_{(\bar B, {\bar {\cal B})}}$ and derive the corresponding 
CF-type restrictions: ${\cal B}_\mu + {\bar {\cal B}_\mu} + \partial_\mu {\tilde \phi} = 0$ and
$B + {\cal B} + m\, \tilde\phi = 0$. First of all,
we  concentrate on the co-BRST and anti-co-BRST symmetry invariance of our 4D ordinary  Lagrangian densities 
${\cal L}_{(B, {\cal B})}$ and ${\cal L}_{(\bar B, \bar{\cal B})}$, respectively. In this connection,  
 we generalize the Lagrangian density ${\cal L}_{(B, {\cal B})}$ to its counterpart {\it chiral} 
super Lagrangian density: ${\cal L}_{(B,\cal B)} \longrightarrow  \tilde {\cal L}^{(c, d)}_{(B,\cal B)} (x, \theta)$ as follows
\begin{eqnarray*}
\tilde {\cal L}^{(c, d)}_{(B,\cal B)}  (x, \theta)   & = & \frac{1}{2}{\cal B}_\mu (x)\, {\cal B}^{\mu}(x)\nonumber\\
& - & {\cal B}^{\mu}(x)  \left(\frac{1}{2}\varepsilon_{\mu\nu\eta\kappa} \, \partial^\nu {\tilde {B}}^{\eta\kappa (d)} (x, \theta)
 - \frac{1}{2}\,\partial_\mu \tilde \Phi^{(d)}(x, \theta) + m \;{\it {\tilde\Phi}}_{\mu}^{(d)}(x, \theta)\right)\nonumber\\
&-& \frac{m^2}{4} \,{\tilde {B}}^{\mu\nu (d)}(x, \theta)\,{\tilde {B}}_{\mu\nu}^{(d)}(x, \theta) -
 \frac{1}{2}\,\partial^\mu {{ \phi}}^{\nu } (x)
\,\Big(\partial_\mu {{ \phi}}_{\nu}  (x) - \partial_\nu {{\phi}}_\mu  (x)\Big)  \nonumber\\ 
&+& {m}\, {\tilde {B}}^{\mu\nu (d)} (x, \theta)\,\partial_\mu {{\phi}}_\nu (x)+  \frac{1}{2}\,\partial^\mu {\it {\tilde \Phi}}^{\nu (d)}(x, \theta)
\,\Big(\partial_\mu {\it {\tilde \Phi}}_{\nu }^{(d)}(x, \theta) - \partial_\nu {\it {\tilde \Phi}}_{\mu }^{(d)}(x, \theta)\Big) \nonumber\\
&+& \frac{m}{2}\, {\varepsilon}^{\mu\nu\eta\kappa} {\tilde B}^{(d)}_{\mu\nu}(x, \theta) \partial_\eta {\it {\tilde\Phi}}_{\kappa}^{(d)} 
(x, \theta) - \frac{1}{2}\,B^\mu (x)\, B_\mu (x)\nonumber\\
 \end{eqnarray*}
 \begin{eqnarray}
&+& B^{\mu} (x)\left( \partial^\nu {\tilde B}_{\nu\mu}^{(d)} (x, \theta) - \frac{1}{2}\, \partial_\mu \varphi (x)   
+ m \;{\phi}_\mu (x) \right) + \frac{1}{2}\, B (x)\, B (x)\nonumber\\ 
&+& B (x) \,\left(\partial_\mu {\phi}^\mu (x) + \frac{m}{2} \,\varphi (x) \right) 
- \frac{1}{2}\,{\cal B} (x)\,{\cal B} (x)\nonumber\\
&-&  {\cal B}(x)  \left(\partial^\mu {\it {\tilde \Phi}}_{\mu}^{(d)} (x, \theta)
  + \frac{m}{2} \, {\it {\tilde\Phi}} ^{(d)} (x, \theta)  \right)\nonumber\\
&+&  \Big(\partial_\mu {\tilde{\bar {\cal F}}}^{(d)} (x, \theta)
- m \,{\tilde {\bar {\cal F}}}_\mu ^{(d)} (x, \theta)\,\Big) \Big(\partial^\mu \,{\tilde{\cal F}}^{(d)} (x, \theta)
- m \,{\tilde {\cal F}}^{\mu(d)} (x, \theta) \Big)\nonumber\\ 
&-& \Big(\partial_\mu {\tilde {\bar {\cal F}}}_{\nu}^{(d)} (x, \theta) - \partial_\nu {\tilde {\bar {\cal F}}}_\mu ^{(d)}
 (x, \theta) \Big)\Big(\partial^\mu \tilde {\cal F}^{\nu(d)} (x, \theta)  \Big)\nonumber\\ 
& - & \frac{1}{2}\,\partial_\mu{\bar \beta} (x) \,\partial^\mu \tilde\beta ^{(d)} (x, \theta) 
+ \frac{1}{2}\, m^2\, {\bar \beta}(x)\, \tilde\beta ^{(d)} (x, \theta) \nonumber\\
& - & \frac{1}{2}\left(\partial_\mu {\tilde {\bar {\cal F}}}^{\mu (d)} (x, \theta) +  m \, {\tilde{\bar {\cal F}}}^{(d)} (x, \theta) 
+ \frac{1}{4}\,\rho(x) \right) \lambda (x) \nonumber\\ 
&-& \frac{1}{2}\, \left(\partial_\mu {\tilde {\cal F}}^{\mu(d)} (x, \theta)  +  m \, {\tilde{\cal F}}^{(d)} (x, \theta) 
- \frac{1}{4}\,\lambda (x)\right) \tilde\rho  (x), 
\end{eqnarray}
where the superscript $(c, d)$ on the super Lagrangian density denotes that we have taken the {\it chiral} superfields
 that have been derived after the applications of the co-BRST invariant restrictions [cf. Eq. (41)].
At this stage, keeping in our mind the mapping $s_d \leftrightarrow \partial_\theta$ [4-6], we observe the
operation of $\partial_{\theta}$ on the super Lagrangian density $(\tilde {\cal L}^{(c, d)}_{(B,\cal B)})$ as:
\begin{eqnarray}
\frac{\partial}{\partial\theta} \Big[\tilde{\cal L}^{(c, d)}_{(B, {\cal B})}\Big] &=& - \partial_\mu \, 
\bigg[ m\, \varepsilon^{\mu\nu\eta\kappa} \phi_\nu \big(\partial_\eta \bar C_\kappa \big) 
- {\cal B}_\nu \big(\partial^\mu {\bar C}^\nu-\partial^\nu \bar C^\mu \big) + \frac{1}{2}\, {\cal B}^\mu\, \rho  \nonumber\\
&+&  {\cal B}\, \big(\partial^\mu \bar C - m \,\bar C^\mu \big) - \frac{1}{2}\, \big(\partial^\mu \bar\beta \big)\,\lambda \bigg]
 \equiv s_d \, {\cal L}_{(B, {\cal B})}. 
\end{eqnarray}
The above equation captures the co-BRST invariance of the Lagrangian density ${\cal L}_{(B, {\cal B})}$
in the {\it ordinary} space because the total spacetime derivative is {\it exactly} the same as the {\it one} we have derived in the {\it ordinary}
space [cf. Eq. (17)]. Hence, the action integral $S = \int d^4 x {\cal L}_{(B, {\cal B})}$
would remain invariant for the physically well-defined fields which vanish off at infinity due to the sanctity of Gauss's divergence theorem.

As far as the anti-co-BRST invariance of ${\cal L}_{(\bar B, \bar{\cal B})}$ is concerned [cf. Eq. (16)],
we generalize {\it this} Lagrangian density to its counterpart {\it anti-chiral} super Lagrangian density  
${\cal L}_{(\bar B,\bar {\cal B)}} \longrightarrow  \tilde {\cal L}^{(ac, ad)}_{(\bar B,\bar {\cal B)}} (x, \bar\theta)$ 
on the (4, 1)-dimensional {\it anti-chiral} super sub-manifold  as follows
 \begin{eqnarray*}
\tilde {\cal L}^{(ac, ad)}_{(\bar B,\bar {\cal B)}}  (x, \bar\theta)  & = &  \frac{1}{2}{\bar {\cal B}}_\mu (x)\; \bar{\cal B}^{\mu} (x)\nonumber\\
& + & \bar{\cal B}^{\mu} (x) \big(\frac{1}{2}\varepsilon_{\mu\nu\eta\kappa} \, \partial^\nu {\tilde {B}}^{\eta\kappa (ad)} (x, \bar\theta)\nonumber\\
& + & \frac{1}{2}\,\partial_\mu {\it {\tilde\Phi}} ^{(ad)} (x, \bar\theta) + m \;{\it {\tilde \Phi}}^{(ad)}_\mu (x, \bar\theta) \big)\nonumber\\ 
&-& \frac{m^2}{4} \,{\tilde {B}}^{\mu\nu(ad)}(x, \bar\theta)\,{\tilde {B}}_{\mu\nu}^{(ad)}(x, \bar\theta)
 - \frac{1}{2}\,\partial^\mu {\phi}^\nu(x) \nonumber\\
&&\Big(\partial_\mu {{\phi}_\nu } (x) - \partial_\nu {{\phi}}_\mu (x)\Big)  
+ {m}\, {\tilde {B}}^{\mu\nu (ad)} (x, \bar\theta)\,\partial_\mu {\phi}_\nu (x)\nonumber\\
 \end{eqnarray*}
 \begin{eqnarray}
& + & \frac{1}{2}\, 
\partial^\mu {\it {\tilde\Phi}}^{\nu(ad)}(x, \bar\theta)\Big(\partial_\mu {\it {\tilde\Phi}}_{\nu}^{(ad)}(x, \bar\theta)
 -    \partial_\nu {\it {\tilde\Phi}}_{\mu}^{(ad)}(x, \bar\theta)\Big)\nonumber\\
& + & \frac{m}{2}\, \varepsilon^{\mu\nu\eta\kappa} 
{\tilde B}^{(ad)}_{\mu\nu}(x, \bar\theta) \partial_\eta {\it {\tilde\Phi}}_{\kappa}^{(ad)} (x, \bar\theta)
  - \frac{1}{2}\,\bar B^{\mu} (x) \, \bar B_{\mu}(x) \nonumber\\
&- & \bar B^{\mu} (x)\Big( \partial^\nu {\tilde B}_{\nu\mu}^{(ad)} (x, \bar\theta) 
+ \frac{1}{2}\, \partial_\mu \varphi (x) + m\; {\phi}_\mu (x) \Big)\nonumber\\
& + & \frac{1}{2}\, \bar B (x)\,  \bar B (x))
-  \bar B (x) \,\left(\partial_\mu {\phi}^\mu (x) - \frac{m}{2} \,\varphi (x) \right)\nonumber\\ 
&-& \frac{1}{2}\,\bar{\cal B}  (x)\,\bar{\cal B} (x) + \bar{\cal B} (x) \left(\partial_\mu {\it {\tilde \Phi}}^{\mu (ad)} (x, \bar\theta)  - \frac{m}{2} \, 
{\it {\tilde\Phi}}^{(ad)}  (x, \bar\theta)  \right)\nonumber\\
&+&  \Big(\partial_\mu {\tilde{\bar {\cal F}}}^{(ad)} (x, \bar\theta)
- m \,{\tilde {\bar {\cal F}}}_\mu ^{(ad)} (x, \bar\theta)\,\Big)\nonumber\\
&& \Big(\partial^\mu 
{\tilde{\cal F}}^{(ad)} (x, \bar\theta)
 -  m \,{\tilde {\cal F}}^{\mu(ad)} (x, \bar\theta) \Big)\nonumber\\
&-& \Big(\partial_\mu {\tilde {\bar {\cal F}}}_{\nu}^{(ad)} (x, \bar\theta) - \partial_\nu {\tilde {\bar {\cal F}}}_\mu ^{(ad)} (x, \bar\theta) \Big)\Big(\partial^\mu {\tilde {\cal F}}^{\nu(ad)} (x, \bar\theta)  \Big)\nonumber\\  
&-& \frac{1}{2}\,\partial_\mu \tilde{\bar \beta} ^{(ad)} (x, \bar\theta) \,\partial^\mu \beta (x) + \frac{1}{2}\, m^2\, \tilde{\bar\beta} ^{(ad)} (x, \bar\theta)\, \beta (x)) \nonumber\\
&-& \frac{1}{2}\left(\partial_\mu {\tilde {\bar {\cal F}}}^{\mu(ad)} (x, \bar\theta) +  m \, {\tilde{\bar {\cal F}}}^{(ad)} (x, \bar\theta) + \frac{1}{4}\,\rho (x) \right) \lambda (x) \nonumber\\ 
&-&\frac{1}{2}\left(\partial_\mu {\tilde {\cal F}}^{\mu(ad)} (x, \bar\theta)  +  m \, {\tilde{\cal F}}^{(ad)} (x, \bar\theta) - \frac{1}{4}\,\lambda (x)\right) \rho (x),
\end{eqnarray}
 where the superscript $(ac, ad)$ on the super {\it anti-chiral} Lagrangian density denotes that 
{\it all} the superfields that have been incorporated into the Lagrangian density 
$\tilde{\cal L}^{(ac, ad)}_{(\bar B, \bar{\cal B})}$ are the {\it ones} which have been derived after
the applications of the anti-co-BRST invariant restrictions [cf. Eq. (50)].  
It is crystal clear to note that we have 
\begin{eqnarray}
\frac{\partial}{\partial\bar\theta} \Big[\tilde{\cal L}^{(ac, ad)}_{(\bar B, {\cal{\bar B}})}\Big] & = & - \partial_\mu \, 
\bigg[ m\, \varepsilon^{\mu\nu\eta\kappa} \phi_\nu \big(\partial_\eta C_\kappa \big) 
+ \bar {\cal B}_\nu \big(\partial^\mu C^\nu-\partial^\nu C^\mu \big) + \frac{1}{2}\,\bar {\cal B}^\mu\, \lambda  \nonumber\\
& - & \bar {\cal B} \big(\partial^\mu C - m \,C^\mu \big) + \frac{1}{2}\, \big(\partial^\mu \beta \big)\,\rho \bigg]
\equiv s_{ad}\,{\cal L}_{(\bar B, {\bar{\cal B}})}, 
\end{eqnarray}
where we take  into account the mapping: $\partial_{\bar\theta}\longleftrightarrow s_{ad}.$
Thus, we conclude that the {\it anti-chiral} super Lagrangian density $\tilde{\cal L}^{(ac, ad)}_{(\bar B, \bar{\cal B})}$
is  the sum of  a {\it unique} combination of the {\it anti-chiral} superfields (obtained after the applications of the anti-co-BRST
 invariant restrictions) and {\it ordinary} 4D fields such that its translation along $\bar\theta$-direction of the {\it anti-chiral}
 (4, 1)-dimensional super sub-manifold leads to a total {\it spacetime} derivative  in the {\it ordinary} space.

 The stage is set now to derive the CF-type restrictions connected with the nilpotent (anti-)co-BRST symmetry transformations [cf. Eqs. (18), (19)] within the framework
 of ACSA to BRST formalism. In this connection, we generalize the Lagrangian density ${\cal L}_{(\bar B, \bar{\cal B})}$
 to  its  counterpart {\it chiral} super Lagrangian density $\tilde{\cal L}^{(c, d)}_{(\bar B, \bar{\cal B})}$ [i.e. 
${\cal L}_{(\bar B,\bar {\cal B)}} \longrightarrow  \tilde {\cal L}^{(c, d)}_{(\bar B,\bar {\cal B)}} (x, \theta)$] as follows 
\begin{eqnarray*}
\tilde {\cal L}^{(c, d)}_{(\bar B,\bar {\cal B)}}  (x, \theta) & = &  
\frac{1}{2}{\tilde{\bar{\cal B}}}^{(d)}_\mu (x, \theta)\; {\tilde{\bar{\cal B}}}^{\mu (d)} (x, \theta)
 + {\tilde{\bar{\cal B}}}^{\mu (d)} (x, \theta)\nonumber\\
\end{eqnarray*}
\begin{eqnarray}
&& \Big(\frac{1}{2}\varepsilon_{\mu\nu\eta\kappa} \, \partial^\nu {\tilde {B}}^{\eta\kappa (d)} (x, \theta)
 + 
\frac{1}{2}\,\partial_\mu {\it {\tilde \Phi}} ^{(d)} (x, \theta)\nonumber\\
& + & m \;{\it{\tilde \Phi}}^{(d)}_\mu (x, \theta) \Big)
  -  \frac{m^2}{4} \,{\tilde {B}}^{\mu\nu(d)}(x, \theta)\,{\tilde {B}}_{\mu\nu}^{(d)}(x, \theta)\nonumber\\
& - &\frac{1}{2}\,\partial^\mu {\phi}^\nu(x)
\,\Big(\partial_\mu {{\phi}_\nu } (x) - \partial_\nu {{\phi}}_\mu (x)\Big)\nonumber\\  
 & + & {m}\, {\tilde {B}}^{\mu\nu (d)} (x, \theta)\,\partial_\mu {\phi}_\nu (x)\nonumber\\
& + & \frac{1}{2}\, \partial^\mu{\it {\tilde\Phi}}^{\nu(d)}(x, \theta)\Big(\partial_\mu
{\it {\tilde\Phi}}_{\nu}^{(d)}(x, \theta) -  \partial_\nu {\it {\tilde\Phi}}_{\mu}^{(d)}(x, \theta)\Big)\nonumber\\
& + & \frac{m}{2}\, \varepsilon^{\mu\nu\eta\kappa} {\tilde B}^{(d)}_{\mu\nu}(x, \theta) \partial_\eta {\it {\tilde\Phi}}_{\kappa}^{(d)} (x, \theta)   
- \frac{1}{2}\,\bar B^{\mu} (x) \, \bar B_{\mu}(x)\nonumber\\
&  -  & \bar B ^{\mu} (x)\,\left( \partial^\nu {\tilde B}_{\nu\mu}^{(d)} (x, \theta) + \frac{1}{2}\, \partial_\mu \varphi (x)   
+ m \;{\phi}_\mu (x) \right)\nonumber\\
& + & \frac{1}{2}\, \bar B (x)\,  \bar B (x))
-  \bar B (x) \,\left(\partial_\mu {\phi}^\mu (x) - \frac{m}{2} \,\varphi (x) \right)\nonumber\\ 
&-& \frac{1}{2}\,{\tilde{\bar{\cal B}}}^{(d)} (x, \theta)\,{\tilde{\bar{\cal B}}}^{(d)} (x, \theta)
    +   {\tilde{\bar{\cal B}}}^{(d)}(x, \theta)  \Big(\partial_\mu
 {\it {\tilde \Phi}}^{\mu (d)} (x, \theta)\nonumber\\
  & - & \frac{m}{2} \, {\it {\tilde\Phi}}^{(d)}  (x, \theta)  \Big)
+  \Big(\partial_\mu {\tilde{\bar {\cal F}}}^{(d)} (x, \theta)
- m \,{\tilde {\bar {\cal F}}}_\mu ^{(d)} (x, \theta)\,\Big)\nonumber\\
&& \Big(\partial^\mu \,{\tilde{\cal F}}^{(d)} (x, \theta)
- m \,{\tilde {\cal F}}^{\mu (d)} (x, \theta) \Big)\nonumber\\ 
&-& \Big(\partial_\mu {\tilde {\bar {\cal F}}}_{\nu }^{(d)} (x, \theta) - \partial_\nu {\tilde {\bar {\cal F}}}_\mu ^{(d)} 
(x, \theta) \Big)\Big(\partial^\mu {\tilde {\cal F}}^{\nu(d)} (x, \theta)  \Big)\nonumber\\  
&-& \frac{1}{2}\,\partial_\mu {\bar \beta} (x) \,\partial^\mu \tilde{\beta} ^{(d)} (x, \theta) + 
\frac{1}{2}\, m^2\, \bar\beta (x)\, \tilde{\beta} ^{(d)} (x, \theta) \nonumber\\
&-& \frac{1}{2}\left(\partial_\mu {\tilde {\bar {\cal F}}}^{\mu(d)} (x, \theta) + 
 m \, {\tilde{\bar {\cal F}}}^{(d)} (x, \theta) + \frac{1}{4}\,\rho (x) \right) \lambda (x) \nonumber\\ 
&-&\frac{1}{2}\left(\partial_\mu {\tilde {\cal F}}^{\mu(d)} (x, \theta)  + 
 m \, {\tilde{\cal F}}^{(d)} (x, \theta) - \frac{1}{4}\,\lambda (x)\right) \rho (x),
\end{eqnarray}
where the superscript $(c, d)$ denotes that the {\it chiral} super Lagrangian density 
$\tilde{\cal L}^{(c, d)}_{(\bar B, {\bar {\cal B}})}$ incorporates the {\it chiral} superfields 
[cf. Eq. (45)] derived after the imposition(s) of the  co-BRST invariant restriction(s) [cf. Eq. (42)] and the ordinary 
4D fields. These {\it ordinary} 4D fields are nothing but the {\it trivially} co-BRST invariant fields [cf. Eq. (19].
 Keeping in our mind the mapping $\partial_{\theta} \longleftrightarrow s_d$, we observe the following relationship is true, namely;  
\begin{eqnarray}
\frac{\partial}{\partial\theta} \Big[\tilde{\cal L}^{(c, d)}_{(\bar B, \bar {\cal B})}\Big] &=& - \partial_\mu \bigg[m\, \varepsilon^{\mu\nu\eta\kappa}\,  \phi_\nu \big(\partial_\eta \bar C_\kappa \big)
- \Big(\frac{1}{2}\,\varepsilon^{\mu\nu\eta\kappa}\,\partial_\nu B_{\eta\kappa} - \frac{1}{2}\, {\cal B}^\mu + m \tilde \phi^\mu\Big) \rho \nonumber\\
&+& \bar {\cal B}_\nu \big(\partial^\mu \bar C^\nu - \partial^\nu \bar C^\mu  \big)  
- \bar {\cal B} \big(\partial^\mu \bar C - m \bar C^\mu\big) - \frac{1}{2}\, \big(\partial^\mu \bar \beta \big)\,\lambda \bigg] \nonumber\\
&+& \frac{1}{2}\, \big[{\cal B}_\mu + \bar {\cal B}_\mu + \partial_\mu \tilde \varphi \big] \big(\partial^\mu \rho \big) 
+ \partial_\mu\big[{\cal B}_\nu + \bar {\cal B}_\nu + \partial_\nu \tilde \varphi  \big] \big(\partial^\mu \bar C^\nu 
- \partial^\nu \bar C^\mu  \big) \nonumber\\
&+& m \big[{\cal B}_\mu + \bar {\cal B}_\mu + \partial_\mu \tilde \varphi \big] \big(\partial^\mu \bar C - m \bar C^\mu  \big)
- \frac{m}{2}\, \big[{\cal B} + \bar {\cal B} + m \tilde \varphi \big] \rho \nonumber\\
&-& \partial_\mu \big[{\cal B} + \bar {\cal B} + m \tilde \varphi \big] \big(\partial^\mu \bar C - m \bar C^\mu  \big)\equiv s_d\,
{\cal L}_{(\bar B, \bar {\cal B})}. 
\end{eqnarray}
Thus, we note that we have captured the variation $s_d\,
{\cal L}_{(\bar B, \bar {\cal B})}$ [cf. Eq. (28)] in the terminology of the ACSA to BRST formalism. It is crystal 
clear that if we impose the CF-type restrictions: 
$ {\cal B}_\mu + {\bar{\cal B}_\mu + \partial_\mu\,\tilde \phi = 0,
{\cal B} + \bar {\cal B}} + m\,\tilde \phi = 0$
 from {\it outside}, we obtain $s_d \,{\cal L}_{(\bar B, {\bar{\cal B}})}$
as a total spacetime derivative [cf. Eq. (25)]. It is straightforward to check the invariance $s_{ad} {\cal L}_{(B, {\cal B})}$
on {\it exactly} similar {\it lines} as given in Eqs. (63) and (64). We perform this exercise  {\it concisely} in
   our Appendix A to complement the write-up 
in our present sub-section.

We wrap up this sub-section with the remarks that we have captured the (anti-) co-BRST invariance [cf. Eqs. (18), (19)] as well as 
we have established the existence of the CF-type restrictions: ${\cal B}_\mu + {\bar {\cal B}}_\mu + \partial_\mu\,\tilde \phi = 0$ 
and ${\cal B} + {\bar {\cal B}} + m\, \tilde\phi  = 0$ on our theory. In fact, it is straightforward to note that 
if our whole theory is considered on the space of fields (in the 4D Minkowskian flat spacetime manifold) where the CF-type restrictions:
${\cal B}_\mu + {\bar {\cal B}}_\mu + \partial_\mu\,\tilde \phi = 0$ 
and ${\cal B} + {\bar {\cal B}} + m\, \tilde\phi  = 0$  are satisfied, then, {\it both} the 
Lagrangian densities ${\cal L}_{(B, {\cal B})}$ and ${\cal L}_{(\bar B, {\bar {\cal B}})}$ would respect {\it both}
the nilpotent symmetries (i.e. co-BRST and anti-co-BRST) {\it together}. In other words, we shall have 
$s_d\,{\cal L}_{(B, {\cal B})}$ and  $s_{ad}\, {\cal L}_{(\bar B, \bar {\cal B})}$ as the {\it total} 
spacetime derivatives [cf. Eqs. (16), (17)] as well as the the transformations  
$s_d\, {\cal L}_{(B, {\cal B})}$ and $s_{ad}\, {\cal L}_{(\bar B, {\bar {\cal B})}}$ would {\it also} turn 
out to be the {\it total} spacetime derivatives [cf. Eqs. (26), (27)]. In a subtle manner,
 the observations in Eqs. (56) and (64) establish the existence of the CF-type restrictions which are the 
hallmarks of a BRST {\it quantized} gauge theory.


\section {Conserved Charges and Their Nilpotency and Absolute Anticommutativity Properties: ACSA}

In this section, first of all, we derive the conserved Noether currents, corresponding conserved charges and prove their off-shell 
nilpotency as well as absolute anticommutativity properties (in the {\it ordinary} space) using the BRST formalism. 
 We corroborate the above properties and provide their {\it proof} within the framework of ACSA to BRST formalism, too.
The proof of the absolute anticommutativity property  of the (anti-)BRST as well as (anti-)co-BRST
conserved charges is a {\it novel} as well as {\it surprising}  result in the sense that we have taken into account {\it only} the (anti-)chiral
super expansions of {\it all} the  appropriate superfields. Our present section is divided into {\it three}
sub-sections as illustrated below:


\subsection{Conserved Currents and Charges: Ordinary Space}

According to Noether's theorem, the invariance of the action integrals, corresponding to the off-shell
nilpotent (anti-)BRST and (anti-)co-BRST symmetries, leads to the derivation of the conserved currents. 
We note, in this connection, that the action integral, corresponding to the Lagrangian density 
${\cal L}_{(B, {\cal B})}$, remains {\it perfectly} invariant [cf. Eqs. (15), (17)] under the BRST and co-BRST symmetry 
transformations (without any use of the CF-type restrictions and/or EL-EOMs). Hence, the conserved BRST and co-BRST Noether currents 
(corresponding to the Lagrangian density ${\cal L}_{(B, {\cal B})}$) are: 
 \begin{eqnarray*}
J^\mu_b &=& \varepsilon^{\mu\nu\eta\kappa}\,\big(m\, \tilde \phi_\nu - {\cal B}_\nu \big) \big(\partial_\eta C_\kappa \big)
+ \big(m B^{\mu\nu} - \Phi^{\mu\nu} \big) \big(\partial_\nu C - m \;C_\nu \big) \nonumber\\
&+& B\; \big(\partial^\mu C - m C^\mu \big) + m \;\beta \big(\partial^\mu \bar C - m \bar C^\mu \big) 
- \big(\partial^\mu \bar C^\nu - \partial^\nu \bar C^\mu \big) (\partial_\nu \beta) \nonumber\\
\end{eqnarray*}
\begin{eqnarray}
&-& B_\nu \;\big(\partial^\mu C^\nu - \partial^\nu C^\mu \big) + \frac{1}{2}\, (\partial^\mu \beta)\, \rho
- \frac{1}{2}\, B^\mu\, \lambda, \nonumber\\
J^\mu_d &=& \varepsilon^{\mu\nu\eta\kappa}\,\big(m\;  \phi_\nu -  B_\nu \big) \big(\partial_\eta \bar C_\kappa \big)
+ \Big(\frac{m}{2}\,\varepsilon^{\mu\nu\eta\kappa}\, B_{\eta\kappa} + \tilde \Phi^{\mu\nu} \Big) \big(\partial_\nu \bar C - m \;\bar C_\nu \big) \nonumber\\
&-& {\cal B} \;\big(\partial^\mu \bar C - m \;\bar C^\mu \big) - m \;\bar \beta \,\big(\partial^\mu C - m  C^\mu \big) 
+ \big(\partial^\mu C^\nu - \partial^\nu C^\mu \big) (\partial_\nu \bar \beta)\nonumber\\
&+&  {\cal B}_\nu\; \big(\partial^\mu \bar C^\nu - \partial^\nu \bar C^\mu \big) + \frac{1}{2}\, (\partial^\mu \bar \beta)\, \lambda
- \frac{1}{2}\, {\cal B}^\mu\, \rho.
\end{eqnarray}
The conservation law ($\partial_\mu\,J^\mu _{(r)} = 0, r = b, d $) can be proven by taking 
into account the EL-EOM corresponding to the Lagrangian density ${\cal L}_{(B,  {\cal B})}$
[cf. Eq. (67) below]. These conserved currents lead to the derivation of the conserved and nilpotent charges 
 $(Q_b = \int d^3 x\; J^{0}_b,\; Q_d = \int d^3 x\; J^{0}_d)$ as:
\begin{eqnarray}
Q_b &=& \int d^3x\Big[\epsilon^{ijk}(m\,\tilde{\phi}_i - {\cal B}_i)\,\partial_jC_k + (m\,B^{0i} - \Phi^{0i})(\partial_iC 
- m\,C_i)\nonumber\\
 & - & B_i\;(\partial^0{C}^i - \partial^i{C}^0) + B\;(\partial^0C - m\,C^0) + m\,\beta(\partial^0\bar{C} - m\,\bar{C}^0)\nonumber\\
& - & (\partial_i\beta)(\partial^0\bar{C}^i - \partial^i\bar{C}^0)  + \frac{1}{2}(\partial^0\beta)\rho - \frac{1}{2}B^0\lambda\Big],\nonumber\\
Q_{d} &=& \int d^3x\Big[\epsilon^{ijk}(m\,{\phi}_i - {B}_i)\,\partial_j\bar{C}_k + \big(\frac{m}{2}\,\epsilon^{ijk}B_{jk} + \tilde{\Phi}^{0i}\big)\;\big(\partial_i\bar{C} - m\,\bar{C}_i\big)\nonumber\\
 &+& {\cal B}_i\;(\partial^0\bar{C}^i - \partial^i\bar{C}^0)  
- {\cal B}\;\big(\partial^0\bar{C} - m\,\bar{C}^0\big) - m\,\bar{\beta}\;\big(\partial^0{C} - m\,{C}^0\big)\nonumber\\
&+& (\partial_i\bar{\beta})(\partial^0{C}^i - \partial^i{C}^0)
+  \frac{1}{2}\big(\partial^0\bar{\beta}\big)\lambda - \frac{1}{2}{\cal B}^0\rho\Big]. 
\end{eqnarray}
The above conserved charges can be expressed in different (but equivalent) forms by using the following EL-EOMs that 
are derived from ${\cal L}_{(B, {\cal B})}$, namely; 
\begin{eqnarray*}
&& \varepsilon^{\eta\kappa\mu\nu}\, \partial_\mu {\cal B}_\nu + m^2\,\big(B^{\eta\kappa} - \frac{1}{m}\, \Phi^{\eta\kappa}
- \frac{1}{2m}\, \varepsilon^{\mu\nu\eta\kappa}\, \tilde \Phi_{\mu\nu} \big) + \big(\partial^\eta B^\kappa - \partial^\kappa B^\eta \big)= 0, \nonumber\\
&&  \varepsilon^{\eta\kappa\mu\nu}\, \partial_\mu B_\nu 
+ \frac{m^2}{2}\,\varepsilon^{\eta\kappa\mu\nu}\,\big(B_{\mu\nu} - \frac{1}{m}\, \Phi_{\mu\nu}
- \frac{1}{2m}\, \varepsilon_{\mu\nu\zeta\sigma}\, \tilde \Phi^{\zeta\sigma} \big)\nonumber\\ 
&& -  \big(\partial^\eta {\cal B}^\kappa - \partial^\kappa {\cal B}^\eta \big)= 0, \quad \partial_\mu \Phi^{\mu\nu} - m \big(\partial_\mu B^{\mu\nu} - B^\nu \big) - \partial^\nu B = 0,\nonumber\\
&&\partial_\mu B^\mu + m B = 0, \quad \partial_\mu {\cal B}^\mu + m \;{\cal B} = 0,\quad \Box \bar C - m \Big(\partial_\mu \bar C^\mu + \frac{\rho}{2} \Big) = 0,\nonumber\\
\end{eqnarray*}
\begin{eqnarray}
 &&\partial_\mu \tilde \Phi^{\mu\nu} + m \Big(-\,\frac{1}{2}\,\varepsilon^{\nu\mu\eta\kappa}\partial_\mu B_{\eta\kappa} 
+ {\cal B}^\nu \Big) - \partial^\nu {\cal B} = 0, \quad\big(\Box + m^2 \big) \beta = 0, \nonumber\\
&& \lambda = 2\, \Big(\partial_\mu C^\mu + m \;C\Big), \quad  \rho =  -\,2\, \Big(\partial_\mu \bar C^\mu + m \;\bar C\Big), \nonumber\\
&& \Box C - m \Big(\partial_\mu C^\mu - \frac{\lambda}{2} \Big) = 0, \quad \big(\Box + m^2 \big) \;C_\mu  - \partial_\mu \Big(\partial_\nu C^\nu + m\; C - \frac{\lambda}{2} \Big)= 0, \nonumber\\ 
&& \big(\Box + m^2 \big) \bar\beta = 0, \quad \Big(\Box + m^2 \Big)\; \bar C_\mu  - \partial_\mu\Big(\partial_\nu \bar C^\nu + m \;\bar C 
+ \frac{\rho}{2} \Big)= 0.
\end{eqnarray}
As an additional remark, we mention here that (with $\varepsilon^{0ijk }= \epsilon ^{ijk}$ as 3D Levi-Civita tensor) the above equations (i.e. EL-EOMs) are
{\it useful} in the proof of the conservation ($\partial_\mu\,J^\mu _{(r)} = 0, r = b, d$) law, too.

In exactly similar fashion, we note that the Lagrangian density ${\cal L}_{(\bar B, \bar {\cal B})}$
respects {\it perfect} [cf. Eqs. (14), (16)] anti-BRST and anti-co-BRST symmetries in the sense that
the corresponding action integral remains invariant (without any use of CF-type restrictions). 
As a consequence, we have the following:
\begin{eqnarray}
J^\mu_{ab} &=& \varepsilon^{\mu\nu\eta\kappa}\,\big(m\, \tilde \phi_\nu + \bar {\cal B}_\nu \big) \big(\partial_\eta \bar C_\kappa \big)
+ \big(m \;B^{\mu\nu} - \Phi^{\mu\nu} \big) \big(\partial_\nu \bar C - m \bar C_\nu \big) \nonumber\\
&-& \bar B\, \big(\partial^\mu \bar C - m \bar C^\mu \big) - m \;\bar \beta\, \big(\partial^\mu C - m \, C^\mu \big) 
+ \big(\partial^\mu C^\nu - \partial^\nu C^\mu \big) (\partial_\nu \bar \beta)\nonumber\\
&+& \bar B_\nu \big(\partial^\mu \bar C^\nu - \partial^\nu \bar C^\mu \big) + \frac{1}{2}\, (\partial^\mu \bar \beta)\, \lambda
- \frac{1}{2}\, \bar B^\mu\, \rho,\nonumber\\
J^\mu_{ad} &=& \varepsilon^{\mu\nu\eta\kappa}\,\big(m\;\phi_\nu + \bar B_\nu \big) \big(\partial_\eta C_\kappa \big)
+ \Big(\frac{m}{2}\, \varepsilon^{\mu\nu\eta\kappa}\, B_{\eta\kappa} + \tilde \Phi^{\mu\nu} \Big) \big(\partial_\nu C - m \;C_\nu \big) \nonumber\\
&+& \bar {\cal B} \,\big(\partial^\mu C - m\; C^\mu \big) - m\; \beta \,\big(\partial^\mu \bar C - m \;\bar C^\mu \big) 
+ \big(\partial^\mu \bar C^\nu - \partial^\nu \bar C^\mu \big) (\partial_\nu \beta) \nonumber\\
&-& \bar {\cal B}_\nu \,\big(\partial^\mu C^\nu - \partial^\nu C^\mu \big) - \frac{1}{2}\, (\partial^\mu \beta)\, \rho 
- \frac{1}{2}\, \bar {\cal B}^\mu\, \lambda.
\end{eqnarray}
The conservation law ($\partial_\mu\,J^\mu _{(r)} = 0, r = ab, ad$) can be proven by using 
the following EL-EOMs [besides the onces quoted in Eq.\,(23)], namely;
\begin{eqnarray}
&& \varepsilon^{\eta\kappa\mu\nu}\, \partial_\mu \bar {\cal B}_\nu - m^2\,\Big(B^{\eta\kappa} - \frac{1}{m}\, \Phi^{\eta\kappa}
- \frac{1}{2m}\, \varepsilon^{\eta\kappa\mu\nu}\, \tilde \Phi_{\mu\nu} \Big) 
+ \big(\partial^\eta \bar B^\kappa - \partial^\kappa \bar B^\eta \big)= 0, \nonumber\\
&& \varepsilon^{\eta\kappa\mu\nu}\, \partial_\mu \bar B_\nu 
- \frac{m^2}{2}\,\varepsilon^{\eta\kappa\mu\nu}\,\Big(B_{\mu\nu} - \frac{1}{m}\, \Phi_{\mu\nu}
- \frac{1}{2m}\, \varepsilon_{\mu\nu\zeta\sigma}\, \tilde \Phi^{\zeta\sigma} \Big)\nonumber\\
 && - \big(\partial^\eta \bar {\cal B}^\kappa - \partial^\kappa \bar {\cal B}^\eta \big)= 0, \qquad \partial_\mu \Phi^{\mu\nu} - m \;\big(\partial_\mu B^{\mu\nu} + \bar B^\nu \big) + \partial^\nu \bar B = 0,\nonumber\\
&&  \partial_\mu \tilde \Phi^{\mu\nu} - m \,\Big(\frac{1}{2}\,\varepsilon^{\nu\mu\eta\kappa}\partial_\mu B_{\eta\kappa} 
+ \bar {\cal B}^\nu \Big) + \partial^\nu \bar {\cal B} = 0, \qquad \partial_\mu \bar {\cal B}^\mu + m \;\bar {\cal B} = 0, \nonumber\\
&& \partial_\mu \bar B^\mu + m \;\bar B = 0. 
\end{eqnarray}
The conserved ($\dot Q_{ad}  = \dot Q_{ab} = 0$) and nilpotent ($Q_{ad}^2 = Q_{ab}^2 = 0$) charges $(Q_{ab}, Q_{ad})$ that emerge out from the above conserved currents are: 
\begin{eqnarray}
Q_{ab} &=& \int d^3x\Big[\epsilon^{ijk}\big(m\,\tilde{\phi}_i + \bar{\cal B}_i\big)\,\partial_j\bar{C}_k + \big(m\,B^{0i} - \Phi^{0i})\big(\partial_i\bar{C} 
- m\,\bar{C}_i\big)\nonumber\\
 & + & \bar{B}_i\,\big(\partial^0\bar{C}^i - \partial^i\bar{C}^0\big)  
-\bar{B}\,\big(\partial^0\bar{C} - m\,\bar{C}^0\big) - m\,\bar{\beta}\big(\partial^0{C} - m\,{C}^0\big)\nonumber\\
 &+ &\big(\partial_i\bar{\beta}\big)\big(\partial^0{C}^i - \partial^i{C}^0\big) + \frac{1}{2}\big(\partial^0\bar{\beta}\big)\lambda 
- \frac{1}{2}\bar{B}^0\rho\Big], \nonumber \\
 Q_{ad} &=& \int d^3x\Big[\epsilon^{ijk}\big(m\,{\phi}_i + \bar{B}_i\,\big)\,\partial_j \,C_k + \big(\frac{m}{2}\,\epsilon^{ijk}B_{jk}
 +  \tilde{\Phi}^{0i}\big)\big(\partial_i\,C - m\,C_i\big)\nonumber\\
& - &\bar{\cal B}_i \,\big(\partial^0{C}^i - \partial^i{C}^0\big)  
+ \bar{\cal B}\,\big(\partial^0C - m\,C^0\big) - m\,\beta\big(\partial^0\bar{C} - m\,\bar{C}^0\big)\nonumber\\
& + &(\partial_i\beta)\big(\partial^0\bar{C}^i
 - \partial^i\bar{C}^0\big)  - \frac{1}{2}\big(\partial^0\beta\big)\rho
 - \frac{1}{2}\bar{\cal B}^0\lambda\Big], 
 \end{eqnarray} 
Thus, we have derived the (anti-)BRST $(Q_{(a)b})$ and (anti-)co-BRST  $(Q_{(a)d})$ conserved charges
[cf. Eqs. (66), (70)] from the perfect  invariance of the action integrals corresponding to the Lagrangian densities  
${\cal L}_{(B, {\cal B})}$ and ${\cal L}_{(\bar B, \bar {\cal B})}$. In the above, we have denoted the totally antisymmetric  3D Levi-Civita 
tensor as $\epsilon ^{ijk} \equiv \varepsilon ^{0ijk}$ (with only space indices $i, j, k = 1, 2, 3$).

It is interesting that the above conserved charges $Q_{(a)b}$ and  $Q_{(a)d}$
can be expressed in their equivalent and {\it useful} forms for our further discussions.
For instance, using the EOMs (23) and (67), the BRST charge $(Q_{b})$  can be re-written as:
\begin{eqnarray}
 Q_b^{(1)} &=& \int d^3x\Big[2\,\epsilon^{ijk}\big(m\,\tilde{\phi}_i - {\cal B}_i\big)\,\partial_j C_k + 2\,\big(m\,B^{0i} 
- \Phi^{0i}\big)\big(\partial_i C - m\,C_i\big)\nonumber\\
& - & B_i\big(\partial^0{C}^i - \partial^i{C}^0\big) + B\,\big(\partial^0C - m\,C^0\big) + m\,\beta\,\big(\partial^0\bar{C} - m\,\bar{C}^0 \big)\nonumber\\ 
& + & C\,\big(\partial^0B - m\,B^0\big) - \big(\partial_i\beta \big)\big(\partial^0\bar{C}^i - \partial^i\bar{C}^0\big) - C_i\,\big(\partial^0B^i - \partial^iB^0\big) \nonumber \\ 
&+&    \frac{1}{2}\big(\partial^0\beta \big)\rho - \frac{1}{2}B^0\lambda\Big],\nonumber\\
Q_b^{(2)} &=& \int d^3x\Big[B\,\dot{C} - \dot{B}\,C + \dot{B}_\mu C^\mu - B_\mu\dot{C}^\mu - m\,\big(\beta\,\dot{\bar{C}} - \dot{\beta}\,\bar{C}\big)
 + \big(\partial_{\mu}\beta\big)\,\dot{\bar{C}}^{\mu} \nonumber\\
 &-& \big(\partial_{\mu}\dot{\beta}\big)\,\bar{C}^{\mu}+  \dot{\beta}\,\rho - \beta\,\dot{\rho}\Big]. 
\end{eqnarray}
The above expressions are very interesting for us because they can be expressed in a BRST-{\it exact} form as follows:
\begin{eqnarray}
Q_b^{(1)} &=& \int d^3x\,\,\,s_b\Big[- \epsilon^{ijk}\,\big(m\,\tilde{\phi}_i - {\cal B}_i \big)\,B_{jk} +2\,\big(m\,B^{0i} 
- \Phi^{0i}\big)\,\phi_i + B_i \,B^{0i}  \nonumber \\
&+& B\,\phi_0  +  \big(\partial^0\bar{C} - m\,\bar{C}^0\big)\,C - \big(\partial_0\bar{C}_i - \partial_i\bar{C}_0\big)\,C^i 
- \frac{1}{2}\dot{\beta}\,\bar{\beta} - \frac{1}{2}\bar{C}_0\,\lambda\Big], \nonumber \\ 
Q_b^{(2)} &=& \int d^3x\,\,\,s_b\Big[\bar{C}\,\dot{C} - \dot{\bar{C}}\,C + \dot{\bar{C}}_{\mu}C^{\mu} 
- \bar{C}_{\mu}\dot{C}^{\mu} + \beta\,\dot{\bar{\beta}} - \dot{\beta}\,\bar{\beta}\Big].  
\end{eqnarray}
As a consequence of the above observations, we prove the off-shell nilpotency
 of the above charges in a straightforward fashion because we note that:
\begin{eqnarray}
s_b\,Q_b^{(1)} = - \, i\,\{Q_b^{(1)}, Q_b^{(1)}\} = 0 \quad \Longleftrightarrow \quad s_b^2 = 0 \quad \Longleftrightarrow \quad [Q_b^{(1)}]^2 = 0, \nonumber \\
s_b\,Q_b^{(2)} = - \, i\,\{Q_b^{(2)}, Q_b^{(2)}\} = 0 \quad \Longleftrightarrow \quad s_b^2 = 0 \quad \Longleftrightarrow \quad [Q_b^{(2)}]^2 = 0. 
\end{eqnarray}
Hence, we have proven the off-shell  nilpotency property of the BRST charges.
We note here that the off-shell nilpotency $(s_b ^2 = 0)$ of the BRST symmetry transformations $(s_b)$ 
and off-shell nilpotency $[(Q_b ^{(1, 2)})^2  = 0]$ of the conserved BRST charges ($Q_b ^{(1, 2)}$) are
 deeply related with each-other [cf. Eq. (73)].

We go a step further and, once again, using the EOMs (23) and (67), we obtain {\it another}  equivalent form of the BRST charge  as:
\begin{eqnarray}
Q_b^{(3)} &=& \int d^3x\Big[B\,\dot{C} - \dot{B}\,C + \dot{B}_\mu C^\mu - B_\mu\dot{C}^\mu + \frac{1}{2}(\dot{\beta}\,\rho - \beta\,\dot{\rho})\Big].
\end{eqnarray}
Using the appropriate CF-type restrictions: $B_\mu + {\bar B}_\mu + \partial_\mu \varphi = 0, \;  
B + \bar B + m\,\varphi = 0$ as well as some appropriate EL-EOM (67), we can recast 
the above BRST charge  $Q_b^{(3)}$ in a very interesting form (see, e.g. Appendix B below) as:
\begin{eqnarray}
Q_b^{(3)} &=& \int d^3x \,\big[\dot{\bar{B}}\,{C} - \bar{B}\,\dot{C} + \bar{B}_\mu \dot{C}^\mu - \dot{\bar{B}}_\mu{C}^\mu + \frac{1}{2}(\dot{\beta}\,\rho - \beta\,\dot{\rho}) +  \frac{1}{2}(\dot{\varphi}\,\lambda - \varphi\,\dot{\lambda})\big]. 
\end{eqnarray} 
To be precise, we have used the EL-EOMs: 
$\partial_\mu C^\mu + m\, C = \frac {\lambda}{2}, \; (\Box  + m^2)\,\varphi = (\Box  + m^2)\,  C_\mu = 0$
which emerge out from ${\cal L}_{(B, {\cal B})}$ as well as  ${\cal L}_{(\bar B, \bar{\cal B})}$. This happens because of the fact that 
the ghost part of the coupled (but equivalent) Lagrangian densities [cf. Eqs. (10), (11)] is the {\it same}.
The above expression can be written as {\it an} anti-BRST {\it exact} expression because we have
the following explicit form of the BRST charge, namely; 
\begin{eqnarray}
Q_b^{(3)} &=& \int d^3x \,\,\,s_{ab}\Big[\dot{C}\,C + C_{\mu}\dot{C}^{\mu} + \frac{1}{2}(\dot{\beta}\,\varphi - \beta\,\dot{\varphi})\Big], 
\end{eqnarray}
where $s_{ab}$ stands for the anti-BRST symmetry transformations (12) under which the Lagrangian density 
${\cal L}_{(\bar B, \bar{\cal B})}$ has {\it perfect} invariance [cf. Eq. (14)] {\it without} any use of the
 CF-type restrictions and/or EL-EOMs.

We now concentrate on the anti-BRST charge $Q_{ab}$ [cf. Eq. (70)] which has been derived from the 
{\it perfect} anti-BRST symmetry of ${\cal L}_{(\bar B, \bar{\cal B})}$. Using the EL-EOMs (23) and (69), we observe 
that $Q_{ab}$ can be re-expressed as follows:
\begin{eqnarray}
Q_{ab}^{(1)} &=& \int d^3x\Big[2\,\epsilon^{ijk}\,\big(m\,\tilde{\phi}_i + \bar{\cal B}_i\big)\,\partial_j\bar{C}_k + 2\,\big(m\,B^{0i} - \Phi^{0i}\big)\big(\partial_i\bar{C} - m\,\bar{C}_i \big) \nonumber \\
& +  & \bar{B}_i\,\big(\partial^0\bar{C}^i - \partial^i\bar{C}^0\big) - \bar{B}\;\big(\partial^0\bar{C} - m\,\bar{C}^0\big) - m\,\bar{\beta}\big(\partial^0{C} - m\,{C}^0\big) \nonumber\\
& - & \bar{C}\big(\partial^0\bar{B} - m\,\bar{B}^0\big)  + \big(\partial_i\bar{\beta}\big)\big(\partial^0{C}^i - \partial^i{C}^0\big)
+\bar{C}_i\,\big(\partial^0\bar{B}^i - \partial^i\bar{B}^0\big)\nonumber\\
&  + & \frac{1}{2}\big(\partial^0\bar{\beta}\big)\lambda 
- \frac{1}{2}\bar{B}^0\rho\Big], \nonumber \\
Q_{ab}^{(2)} &=& \int d^3x\Big[\dot{\bar{B}}\,\bar{C} - \bar{B}\,\dot{\bar{C}} + \bar{B}_\mu \dot{\bar{C}}^\mu - \dot{\bar{B}}_\mu\bar{C}^\mu + m(\bar{\beta}\,\dot{{C}} - \dot{\bar{\beta}}\,{C})\nonumber\\
& + & \big(\partial_{\mu}\dot{\bar{\beta}}\big)\,{{C}}^{\mu} - \big(\partial_{\mu}\bar{\beta}\big)\,\dot{C}^{\mu}
 +  \dot{\bar{\beta}}\,\lambda - \bar{\beta}\,\dot{\lambda}\Big]. 
\end{eqnarray}
The above expressions are very interesting because they can be re-written in the following 
{\it exact} form w.r.t. the anti-BRST symmetry $s_{ab}$ as:
\begin{eqnarray}
Q_{ab}^{(1)} &=& \int d^3x\,\,\,s_{ab}\Big[- \epsilon^{ijk}\,\big(m\,\tilde{\phi}_i + \bar{\cal B}_i \big) \,B_{jk} 
+2\,\big(m\,B^{0i}  - \Phi^{0i}\big) \phi_i - \bar{B}_iB^{0i}\nonumber\\
 &-&   \bar{B}\phi_0 + \bar{C}\,\big(\partial^0{C} - m\,{C}^0 \big) 
-\bar{C}_i\,\big(\partial^0{C}^i - \partial^i{C}^0\big) - \frac{1}{2}\dot{\bar{\beta}}\,{\beta} - \frac{1}{2}{C}_0\;\rho\Big], \nonumber \\
Q_{ab}^{(2)} &=& \int d^3x\,\,\,s_{ab}\Big[\dot{\bar{C}}\,{C} - {\bar{C}}\,\dot{C} + {\bar{C}}_{\mu}\dot{C}^{\mu} - \dot{\bar{C}}_{\mu}{C}^{\mu} + \bar{\beta}\,\dot{{\beta}} - \dot{\bar{\beta}}\,{\beta}\Big].
 \end{eqnarray}   
Now it is straightforward to note that we have 
\begin{eqnarray}
s_{ab}\,Q_{ab}^{(1)} = - \, i\,\{Q_{ab}^{(1)}, Q_{ab}^{(1)}\} = 0 \quad \Longleftrightarrow \quad s_{ab}^2 = 0
 \quad \Longleftrightarrow \quad [Q_{ab}^{(1)}]^2 = 0, \nonumber \\
s_{ab}\,Q_{ab}^{(2)} = - \, i\,\{Q_{ab}^{(2)}, Q_{ab}^{(2)}\} = 0 \quad \Longleftrightarrow \quad s_{ab}^2 = 0 
\quad \Longleftrightarrow \quad [Q_{ab}^{(2)}]^2 = 0, 
\end{eqnarray}
where the anti-BRST symmetry transformations $(s_{ab})$ have been quoted in their
full blaze of glory in Eq. (12). Thus, we observe 
that the off-shell nilpotency   $s_{ab}^2 = 0$  of the anti-BRST symmetry transformations 
as well as the anti-BRST charge $Q_{ab}^{(1, 2)}$ are deeply inter connected. We 
further note that, using the CF-type restrictions $ B_\mu + \bar B_\mu + \partial_\mu\phi  = 0, \; B + \bar B + m\,\phi  = 0$, we can {\it also} recast the
anti-BRST charge in an {\it exact } form w.r.t. the BRST symmetry transformations $(s_b)$. For this purpose, first of all, we have an {\it equivalent}  
form of the anti-BRST charge as follows:
\begin{eqnarray}
Q_{ab}^{(3)} &=& \int d^3x\Big[\dot{\bar{B}}\,\bar{C} - \bar{B}\,\dot{\bar{C}} + \bar{B}_\mu \dot{\bar{C}}^\mu - \dot{\bar{B}}_\mu\bar{C}^\mu + \frac{1}{2}(\dot{\bar{\beta}}\,\lambda - \bar{\beta}\,\dot{\lambda})\Big].
\end{eqnarray}
As argued earlier, the above expression can be re-written, using the {\it appropriate} CF-type restrictions
 [cf. Eq. (21)] as well as some appropriate  EL-EOM derived\footnote{ To be precise, we have used the equations of motion:
$\partial_\mu\,\bar C^\mu + m\,\bar C = -\frac{\rho}{2} , (\Box + m^2)\,\bar C_\mu =  (\Box + m^2)\,\phi = 0$
which emerge out from ${\cal L}_{( B, {\cal B}})$     and/or  ${\cal L}_{(\bar B, {\bar{\cal B}})}$  as the EL-EOMs.} 
 from ${\cal L}_{(\bar B, {\bar{\cal B}})}$ and/or ${\cal L}_{( B, {\cal B}})$ [cf. Eqs. (67), (69)]  as
 a BRST-{\it exact} quantity, namely;
\begin{eqnarray}
Q_{ab}^{(3)} &=& \int d^3x\Big[{{B}}\,\dot{\bar{C}} - \dot{B}\,\bar{C} + \dot{B}_\mu \bar{C}^\mu - {{B}}_\mu\dot{\bar{C}}^\mu + m({\varphi}\,\dot{\bar{C}} - \dot{\varphi}\,\bar{C}) \nonumber \\ 
& + & (\partial_\mu\dot{\varphi})\,\bar{C}^\mu - (\partial_\mu{\varphi})\,\dot{\bar{C}}^\mu 
+  \frac{1}{2}(\dot{\bar{\beta}}\,\lambda - \bar{\beta}\,\dot{\lambda})\Big]\nonumber\\
&\equiv & \int d^3 x\, s_b \,\Big[\bar{C}\,\dot{\bar{C}} + \dot{\bar{C}}_{\mu}\bar{C}^{\mu} + \frac{1}{2}(\dot{\bar{\beta}}\,\varphi - \bar{\beta}\,\dot{\varphi})\Big], 
\end{eqnarray}
where the transformations $(s_b)$ are the off-shell nilpotent $(s_b ^2 = 0)$
BRST symmetry transformations quoted in Eq. (13) of our Sec. 3.

At this juncture, we dwell a bit on the equivalent forms of the co-BRST charge $Q_d$ by
exploiting the potential and power of EL-EOMs (23) and (67) that have been  derived from the Lagrangian density 
${\cal L}_{(B, {\cal B})}$. It is straightforward to note that we have the following {\it equivalent}
forms of $Q_d$ due to the  EL-EOM given in (67), namely; 
\begin{eqnarray}
Q_{d}^{(1)} &=& \int d^3x\Big[2\epsilon^{ijk}\big(m\,{\phi}_i - {B}_i\big)\,\partial_j\bar{C}_k + 2\,\big(\frac{m}{2}\,\epsilon^{ijk}B_{jk} + \tilde{\Phi}^{0i}\big)\big(\partial_i\bar{C} - m\,\bar{C}_i\big)  \nonumber \\
&-& {\cal B}\big(\partial^0\bar{C} - m\,\bar{C}^0\big) - m\,\bar{\beta}\big(\partial^0{C} - m\,{C}^0\big) - \bar{C}\big(\partial^0{\cal B} - m\,{\cal B}^0\big)
\nonumber\\
& + & \big(\partial_i\bar{\beta}\big) \big(\partial^0{C}^i - \partial^i{C}^0\big) + \bar{C}_i\big(\partial^0{\cal B}^i - \partial^i{\cal B}^0\big)  + \frac{1}{2}\big(\partial^0\bar{\beta}\big)\lambda\nonumber\\
& - & \frac{1}{2}{\cal B}^0\rho + {\cal B}_i\,\big(\partial^0\bar{C}^i - \partial^i\bar{C}^0\big)\Big],\nonumber\\
Q_{d}^{(2)} &=& \int d^3x\Big[\dot{{\cal B}}\,\bar{C} - {\cal B}\,\dot{\bar{C}} + {\cal B}_\mu \dot{\bar{C}}^\mu - \dot{{\cal B}}_\mu\bar{C}^\mu + m\,\big(\bar{\beta}\,\dot{{C}} - \dot{\bar{\beta}}\,{C}\big)\nonumber \\ 
& + & \big(\partial_{\mu}\dot{\bar{\beta}}\big)\,{{C}}^{\mu} - \big(\partial_{\mu}\bar{\beta}\big)\,\dot{C}^{\mu}
 + \dot{\bar{\beta}}\,\lambda - \bar{\beta}\,\dot{\lambda}\Big]. 
\end{eqnarray}
It is very interesting to point out  that {\it both} the above forms of charges can be precisely 
re-written in the {\it exact} forms w.r.t. the off-shell nilpotent  co-BRST symmetry  transformations $(s_d)$, namely;
\begin{eqnarray}
Q_{d}^{(1)} &=& \int d^3x\,\,\,s_{d}\,\Big[- 2\,\big(m\,{\phi}_i - {B}_i\big)\,B^{0i} +2\,(\frac{m}{2}\,\epsilon^{ijk}\,B_{jk}
 + \tilde{\Phi}^{0i})\,\tilde{\phi}_i   \nonumber\\
& - & \frac{1}{2}\dot{\bar{\beta}}\,{\beta} + \frac{1}{2}\,\epsilon^{ijk}\,{\cal B}_i\,B_{jk} 
 - {\cal B}\,\tilde{\phi}_0 + \bar{C}\,\big(\partial^0{C} - m\,{C}^0\big )\nonumber\\
& - & \bar{C}_i\,\big(\partial^0{C}^i - \partial^i{C}^0\big) - \frac{1}{2}{C}_0\;\rho\Big], \nonumber \\
Q_{d}^{(2)} &=& \int d^3x\,\,\,s_{d}\;\Big[\dot{\bar{C}}\,{C} - {\bar{C}}\,\dot{C} + {\bar{C}}_{\mu}\dot{C}^{\mu} - \dot{\bar{C}}_{\mu}{C}^{\mu} + \bar{\beta}\,\dot{{\beta}} - \dot{\bar{\beta}}\,{\beta}\Big].
\end{eqnarray}
 In other words, we have been able to express $Q_d^{(1, 2)}$ in the co-exact form w.r.t. the
co-BRST symmetry transformation $(s_d)$. As a consequence, we have: 
\begin{eqnarray}
s_d \, Q_d^{(1)} = - \, i \, \{Q_d^{(1)}, Q_d^{(1)}\} = 0 \quad  \Longleftrightarrow \quad s^2_d = 0 \quad \Longleftrightarrow 
\quad [Q_d^{(1)}]^2 = 0, \nonumber \\
s_d \, Q_d^{(2)} = - \, i \, \{Q_d^{(2)}, Q_d^{(2)}\} = 0 \quad \Longleftrightarrow \quad s^2_d = 0 \quad \Longleftrightarrow
\quad [Q_d^{(2)}]^2 = 0.
\end{eqnarray}
Thus, the off-shell nilpotency $(s_d^2 = 0)$ of the co-BRST symmetry $(s_d)$ and corresponding charges 
$Q_d^{(1, 2)}$ are deeply interconnected. Using the EL-EOMs (23) and (67), it can be checked that 
the co-BRST charge $Q_d$ can be further re-expressed in an {\it equivalent} form as:
\begin{eqnarray}
Q_{d}^{(3)} &=& \int d^3x\Big[\dot{{\cal B}}\,\bar{C} - {\cal B}\,\dot{\bar{C}} + {\cal B}_\mu \dot{\bar{C}}^\mu
 - \dot{{\cal B}}_\mu\bar{C}^\mu + \frac{1}{2}(\dot{\bar{\beta}}\,\lambda - \bar{\beta}\,\dot{\lambda})\Big].
\end{eqnarray} 
The above expression, using the CF-type restrictions ${\cal B}_\mu + \bar{\cal B}_\mu + 
\partial_\mu \tilde \varphi = 0$ and ${\cal B} + \bar{\cal B} + m\, \tilde \varphi = 0$ as well as some 
appropriate\footnote{ The El-EOMs that have been exploited for our purpose are:  $\partial_\mu \bar C^\mu + m\, \bar C = 
-\, \frac {\rho}{2}, \; (\Box  + m^2)\,\bar C_\mu = (\Box  + m^2)\,\tilde\varphi = 0$ and these emerge out from 
 ${\cal L}_{(B, {{\cal B}})}$ and/or ${\cal L}_{( \bar B, \bar {\cal B}})$.} and 
useful EL-EOMs from (67) and/or (69), can be re-written in an {\it exact} form w.r.t. the anti-co-BRST symmetry  transformations 
($s_{ad}$) as (see Appendix B for details) 
\begin{eqnarray}
Q_{d}^{(3)} &=& \int d^{3}x\Big[\bar{\cal B}\,\dot{\bar{C}} - {\dot {\bar{\cal{B}}}}\,\bar{C} + 
{\dot{\bar{\cal{B}}}}_\mu \bar{C}^\mu - \bar{\cal B}_{\mu}\dot{\bar{C}}^{\mu} + 
\frac{1}{2}(\dot{\tilde{\varphi}}\,\rho - \tilde{\varphi}\,\dot{\rho})\Big] 
+ \frac{1}{2}(\dot{\bar{\beta}}\,\lambda - \bar{\beta}\,\dot{\lambda})\Big] \nonumber\\
 &\equiv & \int d^3x \,\,\,s_{ad}\;\Big[\bar{C}\,\dot{\bar{C}} + \dot{\bar{C}}_{\mu}\bar{C}^{\mu}
 + \frac{1}{2}({\bar{\beta}}\,\dot{\tilde{\varphi}} - \dot{\bar{\beta}}\,\tilde{\varphi})\Big],
\end{eqnarray}
where the off-shell nilpotent anti-co-BRST symmetry transformations $s_{ad}$ have been explicitly quoted in 
Eq. (18) [cf. Sec. 3].

At our present stage, we now express the anti-co-BRST charge in an appropriate and interesting form by using the 
EL-EOM (23) and (69) which have been derived from ${\cal L}_{(\bar B, \bar {\cal B})}$. It turns out that the following 
couple of {\it equivalent} forms
\begin{eqnarray}
Q_{ad}^{(1)} &=& \int d^3x\Big[2\;\epsilon^{ijk}\big(m\,{\phi}_i + \bar{B}_i\big)\,\partial_jC_k + 2\,\big(\frac{m}{2}\,\epsilon^{ijk}\,B_{jk} + \tilde{\Phi}^{0i}\big) \big(\partial_i C - m\,C_i\big)  \nonumber \\
&+& \bar{\cal B}\big(\partial^0C - m\,C^0\big) - m\,\beta\,\big(\partial^0\bar{C} - m\,\bar{C}^0\big) + C\,\big(\partial^0\bar{\cal B} - m\,\bar{\cal B}^0\big)
\nonumber\\
& + & \big(\partial_i\beta\big)\big(\partial^0\bar{C}^i - \partial^i\bar{C}^0\big)- C_i\,\big(\partial^0\bar{\cal B}^i - \partial^i\bar{\cal B}^0\big)  - \frac{1}{2}(\partial^0\beta)\rho - \frac{1}{2}\bar{\cal B}^0\lambda\nonumber\\
& - & \bar{\cal B}_i\,  \big(\partial^0{C}^i - \partial^i{C}^0\big)\Big], \nonumber \\
Q_{ad}^{(2)} &=& \int d^3 x \Big[\bar{\cal B}\,\dot{C} - { \dot{\bar{\cal B}}}\,C + {\dot {\bar{\cal{B}}}}_\mu C^\mu - \bar{\cal B}_\mu\dot{C}^\mu + m
\big(\beta\,\dot{\bar{C}} - \dot{\beta}\,\bar{C}\big) \nonumber\\
 & + & \big(\partial_{\mu}\dot{\beta}\big)\,{\bar{C}}^{\mu}
 - \big(\partial_{\mu}{\beta}\big)\,\dot{\bar{C}}^{\mu} +  {\beta}\,\dot{\rho} - \dot{\beta}\,{\rho}\Big], 
\end{eqnarray}
can be expressed in the {\it exact} forms w.r.t. the off-shell nilpotent anti-co-BRST symmetry transformations (19) as [cf. Sec. 3 for details]:
\begin{eqnarray}
Q_{ad}^{(1)} &=& \int d^3x\,\,\,s_{ad}\Big[- 2\;\big(m\,{\phi}_i + \bar{B}_i\big)B^{0i} + 2\;\big(\frac{m}{2}\,\epsilon^{ijk}\,B_{jk} + \tilde{\Phi}^{0i}\big)\tilde{\phi}_i + \bar{\cal B}\tilde{\phi}_0 \nonumber \\  
& - & \frac{1}{2}\,\epsilon^{ijk}\,\bar{\cal B}_i\,B_{jk}  + \big(\partial^0\bar{C} - m\,\bar{C}^0\big)\,C - \big(\partial_0\bar{C}_i - \partial_i\bar{C}_0\big)\,C^i - 
\frac{1}{2}\dot{\beta}\,\bar{\beta} - \frac{1}{2}\bar{C}_0\lambda\Big], \nonumber \\
Q_{ad}^{(2)} &=& \int d^3x\,\,\,s_{ad}\Big[\bar{C}\,\dot{C} - \dot{\bar{C}}\,C + \dot{\bar{C}}_{\mu}C^{\mu}
 - \bar{C}_{\mu}\dot{C}^{\mu} + \beta\,\dot{\bar{\beta}} - \dot{\beta}\,\bar{\beta}\Big].  
\end{eqnarray} 
Thus, it is straightforward to point out that the off-shell nilpotency of the equivalent forms of the charges 
$Q_{ad}^{(1)}$ and $Q_{ad}^{(2)}$ can be proven as 
\begin{eqnarray}
s_{ad} \, Q_{ad}^{(1)} = - \, i \, \{Q_{ad}^{(1)}, Q_{ad}^{(1)}\} = 0 \quad  \Longleftrightarrow \quad s^2_{ad} = 0 \quad \Longleftrightarrow 
\quad [Q_{ad}^{(1)}]^2 = 0, \nonumber\\
s_{ad} \, Q_{ad}^{(2)} = - \, i \, \{Q_{ad}^{(2)}, Q_{ad}^{(2)}\} = 0 \quad \Longleftrightarrow \quad s^2_{ad} = 0 \quad \Longleftrightarrow
\quad [Q_{ad}^{(2)}]^2 = 0,
\end{eqnarray}  
where we have used the deep relationship between the continuous symmetry transformations $(s_{ad})$ and their  generators 
($Q_{ad}^{(1, 2)}$).  We note that, using the EL-EOMs (23) and (69), we have yet another interesting and  equivalent  form of the
conserved and off-shell nilpotent 
anti-co-BRST charge, namely;
\begin{eqnarray}
Q_{ad}^{(3)} &=& \int d^3x\Big[\bar{\cal B}\,\dot{C} - {\dot {\bar{\cal{B}}}}\,C + {\dot{ \bar{\cal{B}}}}_\mu C^\mu - \bar{\cal B}_\mu\dot{C}^\mu + \frac{1}{2}({\beta}\,\dot{\rho} - \dot{\beta}\,{\rho})\Big], 
\end{eqnarray} 
which can be recast in a different form by using the CF-type restrictions: ${\cal B}_\mu + \bar{\cal B}_\mu + 
\partial_\mu \tilde \varphi = 0$ and ${\cal B} + \bar{\cal B} + m\, \tilde \varphi = 0$ and some 
appropriate\footnote{The El-EOMs derived from ${\cal L}_{(B, {\cal B})}$ and/or ${\cal L}_{(\bar B, \bar {\cal B})}$
 that have come in handy are:  $\partial_\mu \bar C^\mu + m\, \bar C = \frac {\lambda}{2}, \; (\Box  + m^2)\,\bar C_\mu
 = (\Box  + m^2)\,\tilde\varphi = 0$.} EL-EOMs.  The ensuing  interesting form is:
\begin{eqnarray}
Q_{ad}^{(3)} &=& \int d^3x\Big[\dot{{\cal B}}\,{C} - {\cal B}\,\dot{C} + {\cal B}_\mu \dot{C}^\mu - \dot{{\cal B}}_\mu{C}^\mu  + \frac{1}{2}({\beta}\,\dot{\rho} - \dot{\beta}\,{\rho}) + \frac{1}{2}(\dot{\tilde{\varphi}}\,\lambda - \tilde{\varphi}\,\dot{\lambda})\Big],\nonumber\\
&\equiv & \int d^3x \,\,\,s_{d}\Big[\dot{C}\,C + C_{\mu}\dot{C}^{\mu} + \frac{1}{2}\big(\dot{\beta}\,\tilde{\varphi}
 - \beta\,\dot{\tilde{\varphi}}\big)\Big], 
\end{eqnarray}
where the co-BRST symmetry  transformations $(s_d)$ have been quoted in our Eq. (18) [cf. Sec. 3]. Thus, it is crystal  clear  that
(using the appropriate CF-type restrictions and EL-EOMs), the anti-co-BRST charge can be written in a {\it co-exact} form
w.r.t. the co-BRST symmetry transformations $(s_d)$.

At this crucial juncture, we now comment on the absolute anticommutativity of the (anti-)BRST and (anti-)co-BRST
charges (i.e $\{Q_b, Q_{ab}\} = 0$ and  $\{Q_d, Q_{ad}\} = 0$) which is {\it one} of the {\it decisive} features 
of the conserved and off-shell nilpotent (anti-) BRST and (anti-)co-BRST conserved charges {\it within} the framework of the BRST 
formalism. In this context, we recall that, using the appropriate CF-type restrictions (21), we have been able to express 
(i) the BRST charge ($Q_{b}^{(3)})$ as an {\it exact} form w.r.t. the anti-BRST   transformations $(s_{ab})$ [cf. Eq. (76)],
 (ii)  the anti-BRST charge ($Q_{ab}^{(3)})$ as a {\it BRST-exact} expression [cf. Eq. (81)],
(iii) the co-BRST charge ($Q_{(d)}^{(3)})$ as an {\it exact} form w.r.t. the anti-co-BRST  transformation $s_{ad}$ [cf. Eq. (86)],  and (iv)
 the anti-co-BRST charges ($Q_{(ad)}^{(3)})$ as the  {\it co-exact} form w.r.t. the co-BRST  transformations $s_d$
 [cf. Eq. (91)]. These observations, ultimately, lead to the following proof of the absolute  anticommutativity: 
\begin{eqnarray}
&&s_{b} \, Q_{ab}^{(3)} = - \, i \, \{Q_{ab}^{(3)}, Q_{b}^{(3)}\} = 0 \quad\;  \Longleftrightarrow \quad s^2_{b} = 0, \nonumber \\
&&s_{ab} \, Q_{b}^{(3)} = - \, i \, \{Q_{b}^{(3)}, Q_{ab}^{(3)}\} = 0 \quad \Longleftrightarrow \quad s^2_{ab} = 0, \nonumber \\
&&s_{d} \, Q_{ad}^{(3)} = - \, i \, \{Q_{ad}^{(3)}, Q_{d}^{(3)}\} = 0 \quad \; \Longleftrightarrow \quad s^2_{b} = 0, \nonumber \\
&&s_{ad} \, Q_{d}^{(3)} = - \, i \, \{Q_{d}^{(3)}, Q_{ad}^{(3)}\} = 0 \quad \,\Longleftrightarrow \quad s^2_{ad} = 0.
\end{eqnarray}
We note that the absolute anticommutativity of the BRST charge {\it with} the
anti-BRST charge is deeply connected with the nilpotency $(s_{ab}^2 = 0)$
of the {\it anti-BRST} symmetries $(s_{ab})$. On the other hand, the absolute anticommutativity of the anti-BRST charge {\it with} that 
of the BRST charge is intimately connected with the nilpotency $(s_b^2 =0)$
of the BRST symmetry  $(s_b)$. In exactly similar fashion, the absolute anticommutativity
of the co-BRST charge {\it with} the anti-co-BRST charge  is related with the nilpotency 
$(s_{ad}^2 = 0)$ of the anti-co-BRST  symmetry  $(s_{ad})$. On the other hand, the absolute anticommutativity
of the anti-co-BRST charge {\it with} the
co-BRST charge is deeply related with the off-shell nilpotency $(s_d^2 = 0)$ of the co-BRST 
(dual-BRST)  transformations ($s_d$). These observations should be contrasted with the off-shell 
nilpotency properties where one finds that the nilpotency $(s_r^2 = 0, r = b, ab, d, ad)$ 
 of {\it all} the {\it fermionic} symmetries and corresponding fermionic ($Q_r ^ 2  = 0,\; r = b, ab, d, ad$) charges $Q_r$ (with $r = b, ab, d, ad$)
 are {\it individually} deeply connected [cf. Eqs. (73), (79), (84), (89)] with each-other.


\subsection {Nilpotency and Absolute Anticommutativity Properties of (Anti-)BRST Charges: ACSA}

In this subsection, we capture the off-shell nilpotency and absolute anticommutativity of the 
(anti-)BRST charges within the framework of ACSA to BRST formalism. First of all, we focus on the 
conserved BRST charges $(Q_b^{(1, 2)})$ that have been expressed in (72) as the BRST-{\it exact}
forms. It is straightforward to note that the expression for $Q_b^{(2)}$ is {\it simpler} than
the expression for $Q_b^{(1)}$. Thus, keeping in mind the mapping: $\partial_{\bar\theta} 
\leftrightarrow s_b$ [4-6], it can be seen that we can express $Q_b^{(2)}$ (within the framework of 
ACSA to BRST formalism) as
\begin{eqnarray}
 Q_{b}^{(2)} &=& \frac{\partial}{\partial\bar\theta}\int d^3 x\Big[{\tilde {\bar {\cal F}}}^{(b)} (x,  \bar\theta)\,\dot{\tilde {{\cal F}}}^{(b)} (x,  \bar\theta)
- {\dot{\tilde {\bar {\cal F}}}}^{(b)} (x,  \bar\theta)\,{\tilde{\cal F}}^{(b)} (x,  \bar\theta)\nonumber\\
& + & {\dot{\tilde{\bar{\cal F}}}}_\mu^{(b)} (x, \bar\theta)\,
{\tilde {{\cal F}}}^{\mu {(b)}} (x, \bar\theta) - {\tilde {\bar {\cal F}}}_\mu^{(b)} (x, \bar\theta)\,{\dot {\tilde {{\cal F}}}}^{\mu{(b)}} (x, \bar\theta) + {\tilde\beta}^{(b)} (x, \bar\theta)\,  {\dot{\tilde{\bar \beta}}}^{(b)} (x, \bar\theta)\nonumber\\
& - & {\dot{\tilde{\beta}}}^{(b)} (x, \bar\theta)\,{\tilde{\bar \beta}}^{(b)} (x, \bar\theta)  \Big] \nonumber \\
&\equiv & \int\,d\bar\theta\int d^3 x\Big[{\tilde {\bar {\cal F}}}^{(b)} (x,  \bar\theta)\,{\dot{\tilde {{\cal F}}}}^{(b)} (x,  \bar\theta)
- {\dot{\tilde {\bar {\cal F}}}}^{(b)} (x,  \bar\theta)\,{\tilde{\cal F}}^{(b)} (x,  \bar\theta)\nonumber\\
& + & {\dot{\tilde{\bar{\cal F}}}}_{\mu}^ {(b)} (x, \bar\theta)\,
{\tilde {{\cal F}}}^{\mu{(b)}} (x, \bar\theta)  - {\tilde {\bar {\cal F}}}_{\mu}^{(b)} (x, \bar\theta)\,{\dot {\tilde {{\cal F}}}}^{\mu (b)} (x, \bar\theta) + {\tilde\beta}^{(b)} (x, \bar\theta)\,  {\dot{\tilde{\bar \beta}}}^{(b)} (x, \bar\theta)\nonumber\\
& - & {\dot{\tilde{\beta}}}^{(b)} (x, \bar\theta)\,{\tilde{\bar \beta}}^{(b)} (x, \bar\theta)  \Big],
  \end{eqnarray}
where the superscript $(b)$ on the {\it anti-chiral} superfields denotes the fact that these
superfields have been derived after the applications of  BRST-invariant restrictions. In other words, we have used
the super expansions [that have been already written in (34)] which lead to the derivation  of the 
BRST symmetry transformations (13) as the coefficients of the Grassmannian variable $\bar\theta$
[cf. Eq. (34) for details]. It is now crystal clear that:
\begin{eqnarray}
\partial_{\bar\theta} \, Q_b^{(2)} = 0 \quad \quad \Leftrightarrow \quad \quad \partial_{\bar\theta}^2 =0 \quad \quad
 \Leftrightarrow \quad \quad s_b^2 = 0. 
\end{eqnarray}
In ordinary space, the above equation captures the nilpotency property of Eq. (73).   
Hence, we observe that it is the nilpotency $(\partial_{\bar\theta}^2 = 0)$ of the translational 
generator $(\partial_{\bar\theta})$ along the $\bar\theta$-direction of the
(4, 1)-dimensional {\it anti-chiral} super sub-manifold that is responsible  for the off-shell nilpotency of the
BRST charge $Q_b ^{(2)}$. We would like to emphasize that {\it both} $Q_b^{(1)}$ and/or $Q_b^{(2)}$
can be expressed in terms of the derivative $\partial_{\bar\theta}$ and anti-chiral superfields [cf. Eq. (34)]. However, for the 
sake of brevity, we have chosen $Q_b ^{(2)}$ for our purpose. The same type of exercise can be performed for $Q_b ^{(1)}$, too.

At this stage, we concentrate on capturing the off-shell nilpotency of the anti-BRST charges $Q_{ab} ^{(1, 2)}$ that have 
been written in (78) as an {\it exact} form w.r.t. the anti-BRST symmetry transformations $s_{ab}$ [cf. Eq. (12)].
We capture the expression for $Q_{ab}^{(2)}$, for the shake of brevity, within the framework of ACSA 
as 
\begin{eqnarray}
 Q_{(ab)}^{(2)} &=& \frac{\partial}{\partial\theta}\int d^3 x\Big[{\dot{\tilde {\bar {\cal F}}}}^{(ab)} (x,  \theta)\,{\tilde {{\cal F}}}^{(ab)} (x,  \theta)
- {{\tilde {\bar {\cal F}}}}^{(ab)} (x,  \theta)\,{\dot{\tilde{\cal F}}}^{(ab)} (x,  \theta)\nonumber\\
& + & {{\tilde{\bar{\cal F}}}}_{\mu}^{(ab)} (x, \theta)\,
{\dot{\tilde {{\cal F}}}}^{\mu (ab)} (x, \theta) - {\dot{\tilde {\bar {\cal F}}}} ^{(ab)} _\mu (x, \theta)\,{{\tilde {{\cal F}}}}^{\mu (ab)} (x, \theta)\nonumber\\
& + & \tilde{\bar\beta}^{(ab)} (x, \theta)\,  {\dot{\tilde{ \beta}}} ^{(ab)} (x, \theta) -  {\dot{\tilde{\bar\beta}}}^{(ab)} 
(x, \theta)\,{\tilde{\beta}}^{(ab)} (x, \theta)  \Big] \nonumber \\
&\equiv &  \int d\theta\int d^3 x\Big[{\dot{\tilde {\bar {\cal F}}}}^{(ab)} (x,  \theta)\,{\tilde {{\cal F}}}^{(ab)} (x,  \theta)
- {{\tilde {\bar {\cal F}}}}^{(ab)} (x,  \theta)\,{\dot{\tilde{\cal F}}}^{(ab)} (x,  \theta)\nonumber\\
& + & {{\tilde{\bar{\cal F}}}}_\mu ^{(ab)} (x, \theta)\,
{\dot{\tilde {{\cal F}}}}^{\mu(ab)} (x, \theta)  - {\dot{\tilde {\bar {\cal F}}}}_\mu ^{(ab)} (x, \theta)\,{\tilde {{\cal F}}}^{\mu (ab)} (x, \theta)\nonumber\\
& + & \tilde{\bar\beta}^{(ab)} (x, \theta)\,  \dot{\tilde{ \beta}}^{(ab)} (x, \theta) -  \dot{\tilde{\bar\beta}}^{(ab)} (x, \theta)\,{\tilde{\beta}}^{(ab)} (x, \theta)  \Big],
\end{eqnarray}   
where superscript $(ab)$ denotes that the {\it chiral} superfields have been derived after the applications of the 
anti-BRST invariant restrictions. In other words, we have taken into account the super expansion that have been listed in 
Eq. (40). It is straightforward  to note that we have the following:
\begin{eqnarray}
\partial_{\theta} \, Q_{ab}^{(2)} = 0 \quad \quad \Leftrightarrow \quad \quad \partial_{\theta}^2 =0 \quad \quad
 \Leftrightarrow \quad \quad s_{ab}^2 = 0. 
\end{eqnarray}
In the {\it ordinary} space, the above equation is equivalent to the off-shell nilpotency property $([Q_{(a)b}^{(2)}]^2 = 0)$ of the 
anti-BRST charge $Q_{ab}^{(2)}$ [that has been quoted in Eq. (79)] in view of the mapping: $s_{ab}\leftrightarrow \partial_\theta$ [4-6].

At this crucial juncture, we discuss the absolute anticommutativity of the BRST charge {\it with} anti-BRST charge 
in the terminology of ACSA to BRST formalism. We note that {\it one} of the equivalent forms of the BRST charge is Eq. (76)
where the BRST charge has been expressed in the {\it exact} form w.r.t. the anti-BRST symmetry transformations $(s_{ab})$
of Eq. (12). Keeping in mind the mapping: $\partial_\theta\longleftrightarrow s_{ab}$, it is straightforward to express 
(76) as
\begin{eqnarray}
 Q_{b}^{(3)} &=& \frac{\partial}{\partial\theta}\int d^3 x\Big[{\dot {\tilde {{\cal F}}}}^{(ab)} (x,  \theta)\,{\tilde {{\cal
 F}}}^{(ab)} (x,  \theta)
 + {\tilde{{\cal F}}}_\mu ^{(ab)} (x, \theta)\,{\dot{\tilde {{\cal F}}}}^{\mu (ab)} (x, \theta)\nonumber \\
& + & \frac {1}{2}\big(\dot{\tilde\beta} ^{(ab)}(x,  \theta)\, \tilde \Phi ^{(ab)} (x,  \theta) - {\tilde\beta} ^{(ab)}(x,  \theta)\,\,\dot{\tilde \Phi}^{(ab)} (x,  \theta) \big) \Big],\nonumber \\ 
 &\equiv & \int\,d\theta\int d^3 x\Big[{\dot {\tilde {{\cal F}}}} ^{(ab)}(x,  \theta)\,{\tilde {{\cal F}}} ^{(ab)}(x,  \theta)
 + {\tilde{{\cal F}}}_\mu ^{(ab)} (x, \theta)\,{\dot{\tilde {{\cal F}}}}^{\mu (ab)} (x, \theta)\nonumber \\
& + & \frac {1}{2}\big(\dot{\tilde\beta} ^{(ab)}(x,  \theta)\, \tilde \Phi ^{(ab)} (x,  \theta) - {\tilde\beta} ^{(ab)}(x,  \theta)\,\,\dot{\tilde \Phi} ^{(ab)} (x,  \theta) \big) \Big], 
\end{eqnarray}
where the superscript $(ab)$ stands for the {\it chiral} superfields that have been expanded in Eq. (40).
In exactly similar fashion, to capture the absolute anticommutativity of the anti-BRST charge {\it with} BRST
charge, we focus on  the expression for {\it one} of the equivalent forms of the anti-BRST charge $Q_{ab} ^{(3)}$ [that has been quoted in 
Eq. (81) as a BRST-exact quantity]. This expression can be expressed, keeping in mind the mapping: 
$\partial_{\bar\theta} \longleftrightarrow s_b$ (within the framework of ACSA to BRST formalism)  as follows
 \begin{eqnarray}
 Q_{(ab)}^{(3)} &=& \frac{\partial}{\partial\bar\theta}\int d^3 x\Big[{{\tilde {\bar{\cal F}}}}^{(b)} (x,  \bar\theta)\,{\dot {\tilde {\bar{\cal 
F}}}}^{(b)} (x,  \bar\theta) + {\dot{\tilde{\bar{\cal F}}}}_{\mu }^{(b)} (x, \bar\theta)\,{\tilde {\bar {\cal F}}}^{\mu (b)} (x, \bar\theta)\nonumber \\
& + & \frac {1}{2}\big({\dot{\tilde{\bar\beta}}}^{(b)}(x,  \bar\theta)\, \tilde \Phi ^{(b)} (x,  \bar\theta)
 - {\tilde{\bar \beta}}^{(b)}(x,  \bar\theta)\,\,\dot{\tilde \Phi}^{(b)} (x,  \bar\theta) \big) \Big]\nonumber \\
 &\equiv & \int d\theta \int d^3 x\Big[{{\tilde {\bar{\cal F}}}}^{(b)} (x,  \bar\theta)\,{\dot {\tilde {\bar{\cal F}}}}^{(b)} (x,  \bar\theta)
 + {\dot{\tilde{\bar{\cal F}}}}_{\mu}^{(b)} (x, \bar\theta)\,{\tilde {\bar {\cal F}}}^{\mu (b)} (x, \bar\theta)\nonumber \\
& + & \frac {1}{2}\big({\dot{\tilde{\bar\beta}}} ^{(b)}(x,  \bar\theta)\, \tilde \Phi ^{(b)} (x,  \bar\theta)
 - {\tilde{\bar\beta}} ^{(b)}(x,  \bar\theta)\,\,\dot{\tilde \Phi} ^{(b)} (x,  \bar\theta) \big) \Big], 
\end{eqnarray}
where the superscript $(b)$ denotes the {\it anti-chiral} superfields that have been quoted in Eq. (34). It is
elementary to note {\it now} that we have the following: 
 \begin{eqnarray}
 \partial_\theta Q_{b} ^{(3)}  = \partial_{\bar\theta} Q_{ab} ^{(3)} = 0\quad \Longleftrightarrow \quad\partial_\theta ^2 = 0,\quad \partial_{\bar\theta}^2 = 0. 
\end{eqnarray}
In the {\it ordinary} space, the above relationships are nothing but the absolute anticommutativity of the (anti-)BRST
charges in Eq. (92). It is now very interesting to pinpoint the {\it distinct} differences between $\{Q_b^{(3)}, Q_{ab}^{(3)}\} = 0 $
{\it and} $\{ Q_{ab}^{(3)}, Q_b^{(3)} \} = 0 $ within the framework of ACSA to BRST formalism. It turns out that the absolute anticommutativity of the BRST charge {\it with} anti-BRST charge is closely  connected with the nilpotency $(\partial_\theta^2 = 0)$ of the translational generator
$(\partial_\theta)$ along the  {\it chiral} (i.e.  $\theta$) direction of the (4, 1)-dimensional super sub-manifold. On the other hand, the absolute anticommutativity of the  anti-BRST charge {\it with } BRST charge is deeply related  with the nilpotency  $(\partial_{\bar\theta}^2 = 0)$ of the
translational generator along  $\bar\theta$-direction of the (4, 1)-dimensional {\it anti-chiral} super-submanifold.


\subsection {Nilpotency and Absolute Anticommutativity of the (Anti-)co-BRST Charges: ACSA}

We dwell, in this subsection, on the proof of the off-shell nilpotency
as well as absolute anticommutativity properties of the (anti-)co-BRST charges within the framework of ACSA to BRST
formalism. In this context, we note that we have expressed the conserved (anti-)co-BRST charges (i.e. $Q_d^{(1, 2)}, Q_{ad}^{(1,2)})$  
in the (co-){\it exact} forms in Eqs. (83) and (87) w.r.t. the (anti-)co-BRST symmetries $s_{(a)d}$. Keeping in our mind the mappings:
$\partial_{\theta}\leftrightarrow s_d$ and $\partial_{\bar\theta} \leftrightarrow s_{ad}$, we express
the {\it simpler} versions of the co-BRST and anti-co-BRST charges  $Q_d^{(2)}$ and $Q_{ad}^{(2)}$ [cf. Eqs. (83), (87)] as
\begin{eqnarray} 
Q_{d}^{(2)} &=& \frac{\partial}{\partial\theta}\int d^3 x\Big[{\dot{\tilde {\bar {\cal F}}}}^{(d)} (x,  \theta)\,{\tilde{\cal F}}^{(d)} (x,  \theta)  - {\tilde {
\bar {\cal F}}}^{(d)} (x,  \theta)\,{\dot{\tilde {{\cal F}}}}^{(d)} (x,  \theta)\nonumber\\
& + &  {\tilde {\bar {\cal F}}}_{\mu}^{(d)} (x, \theta)\,{\dot {\tilde {{\cal 
F}}}}^{\mu (d)} (x, \theta) -  {\dot{\tilde{\bar{\cal F}}}}_{\mu}^{(d)} (x, \theta)\,{\tilde {{\cal F}}}^{\mu {(d)}} (x, \theta)  
+ {\dot{\tilde{\beta}}}^{(d)} (x, \theta)\nonumber\\
&&{\tilde{\bar \beta}}^{(d)} (x, \theta) -  {\tilde\beta}^{(d)} (x, \theta)\,  {\dot{\tilde{\bar \beta}}}^{(d)} (x,
 \theta)  \Big], \nonumber \\
 Q_{(ad)}^{(2)} &=& \frac{\partial}{\partial\bar\theta}\int d^3 x\Big[{{\tilde {\bar {\cal F}}}}^{(ad)} (x,  \bar\theta)\,{\dot{\tilde{{\cal F}}}}^{(ad)} (x,  \bar\theta)  - {\dot{\tilde {\bar {\cal F}}}}^{(ad)} (x,  \bar\theta)\,{\tilde {{\cal F}}}^{(ad)} (x,  \bar\theta)\nonumber\\
& + &  {\dot{\tilde{\bar{\cal F}}}}_\mu^
{(ad)} (x, \bar\theta)\,{{\tilde {{\cal F}}}}^{\mu (ad)} (x, \bar\theta) - {{\tilde{\bar{\cal F}}}}_\mu^{(ad)} (x, \bar\theta)\,{\dot{\tilde {{\cal F}}}}^{\mu 
(ad)} (x, \bar\theta)\nonumber\\
 & + & {\tilde{\beta}}^{(ad)} (x, \bar\theta)\,{\dot{\tilde{\bar \beta}}}^{(ad)} (x, \bar\theta) -  {\dot{\tilde\beta}}^{(ad)} (x, \bar\theta)\, {\tilde{\bar \beta
}}^{(ad)} (x, \bar\theta)  \Big], 
\end{eqnarray}
where the superscript $(d)$ and $(ad)$ denote the {\it chiral} and {\it anti-chiral} super expansions that have been 
 quoted in Eqs. (45) and (50). It is now elementary exercise to check  the following:
 \begin{eqnarray}
\partial_{\theta} \, Q_{d}^{(2)} = 0 \quad \quad \Longleftrightarrow \quad \quad \partial_{\theta}^2 =0 \quad \quad
\Longleftrightarrow  \quad \quad s_{d}^2 = 0,\nonumber\\
\partial_{\bar\theta} \, Q_{ad}^{(2)} = 0 \quad \quad \Longleftrightarrow  \quad \quad \partial_{\bar\theta}^2 =0 \quad \quad
\Longleftrightarrow  \quad \quad s_{ad}^2 = 0.
\end{eqnarray}
 Thus, it is crystal clear that the nilpotency property of the co-BRST charge  $Q_{d}^{(2)}$ is deeply connected with
the nilpotency $(\partial_\theta ^2 = 0)$ of the translational generator along $\theta$-direction of the {\it chiral}
super-submanifold. On the other hand, we observe  that the nilpotency $(\partial_{\bar\theta} ^2 = 0)$  of the translational 
generator along  $\bar\theta$-direction of the {\it anti-chiral} super submanifold is responsible for the off-shell nilpotency 
of the anti-co-BRST charge $(Q_{ad} ^{(2)})$.

Within the framework of ACSA to BRST approach, we are now in the position to capture the absolute anticommutativity
(i.e. $\{Q_d, Q_{ad}\} = 0$) of the (anti-)co-BRST charges. In this context, we note that the (anti-)co-BRST charges
$[(Q_{ad}^{(3)})\,Q_d^{(3)}]$ have been written in the {\it exact} forms w.r.t. the co-BRST and anti-co-BRST symmetry
transformations [cf. Eqs. (91), (86)]. Keeping in our mind the mappings: $s_d \leftrightarrow \partial_\theta$, 
$s_{ad} \leftrightarrow \partial_{\bar\theta}$, we can express the charges in (86) and (91), within the framework of
the ACSA, as follows
\begin{eqnarray}
 Q_{d}^{(3)} &=& \frac{\partial}{\partial\bar\theta}\int d^3 x\Big[{{\tilde {\bar{\cal F}}}}^{(ad)}
 (x,  \bar\theta)\,{\dot {\tilde {\bar{\cal F}}}}^{(ad)} (x, 
 \bar\theta) + {\dot{\tilde{\bar{\cal F}}}}_{\mu}^{(ad)} (x, \bar\theta)\,{\tilde {\bar {\cal F}}}^{\mu (ad)} (x, \bar\theta)\nonumber \\
& + & \frac {1}{2}\big({\tilde{\bar\beta}}^{(ad)}(x,  \bar\theta)\, {\it{\dot {\tilde \Phi}}}^{(ad)} (x,  \bar\theta) - {\dot{\tilde{\bar\beta}}}^{(ad)}(x,
  \bar\theta)\,
{\it {\tilde \Phi}} ^{(ad)} (x,  \bar\theta) \big) \Big],\nonumber \\ 
 Q_{(ad)}^{(3)} &=& \frac{\partial}{\partial\theta}\int d^3 x\Big[{\dot{\tilde {{\cal F}}}}^{(d)} (x,  \theta)\,{{\tilde {{\cal F}}}}^{(d)} (x,  \theta)
 - {\dot{\tilde{{\cal F}}}}_{\mu}^{(d)} (x, \theta)\,{{\tilde {{\cal F}}}}^{\mu (d)} (x, \theta)\nonumber \\
& + & \frac {1}{2}\,\big({{\dot{\tilde{\beta}}}}^{(d)}(x,  \theta)\, {\it{\tilde \Phi}}^{(d)} (x,  \theta) - {{\tilde{\beta}}}^{(d)}(x, \theta)\,
{\it{\dot{\tilde \Phi}}} ^{(d)} (x,  \theta) \big) \Big],
\end{eqnarray} 
where the superscripts $(d)$ and $(ad)$ on the superfields stand 
for the superfield expansions (45) and (50). It is now straightforward to note that: 
\begin{eqnarray}
&&\partial_{\bar\theta}\,Q_d^{(3)} = 0\quad\Longleftrightarrow \quad \{Q_d^{(3)}, Q_{ad}^{(3)}\} = 0 \quad\Longleftrightarrow \quad \partial_{\bar\theta}^2 = 0, \nonumber\\
&&\partial_{\theta}\,Q_{ad}^{(3)} = 0\quad\Longleftrightarrow  \quad \{Q_{ad}^{(3)},Q_d^{(3)}\} = 0 \quad \Longleftrightarrow \quad \partial_{\theta}^2 = 0.
\end{eqnarray}
In the above, the anticommutators emerge from the realizations of the $\partial_{\bar\theta}\,Q_d^{(3)}  = 0$
and $\partial_{\theta}\,Q_{ad}^{(3)} = 0$ in the terminology  of the symmetry transformations where $\partial_{\bar\theta}\leftrightarrow s_{ad}$
and $\partial_{\theta}\leftrightarrow s_d$. In other words, we have the following
\begin{eqnarray}
&&\partial_{\bar\theta}\,Q_d^{(3)} = 0 \quad\Longleftrightarrow \quad s_{ad}Q_d^{(3)} = -\,i \,\{Q_d^{(3)}, Q_{ad}^{(3)}\} = 0, \nonumber\\
&&\partial_{\theta}\,Q_{ad}^{(3)} = 0\quad\Longleftrightarrow  \quad s_d\,Q_{ad}^{(3)} =-\,i\,\{Q_{ad}^{(3)},Q_d^{(3)}\} = 0,
\end{eqnarray} 
 in the {\it ordinary} 4D Minkowskian flat space. Thus, we have been able to differentiate and discern  {\it between} the anticommutators 
$ \{Q_d^{(3)}, Q_{ad}^{(3)}\} = 0$  and $\{Q_{ad}^{(3)},Q_d^{(3)}\} = 0$
 within  the framework of ACSA to BRST formalism because of the observations $\partial_{\bar\theta}\, Q_d^{(3)} = 0 = \partial_\theta \,Q_{ad}^{(3)}$.

We end this sub-section with the following remarks. We observe that the off-shell nilpotency ($Q_{(a)d}^2 = 0$) of the (anti-)co-BRST
charge $(Q_{(a)d})$ is connected with the nilpotent ($\partial_{\theta}^2 = \partial_{\bar\theta}^2 = 0$) translational generators along 
$(\bar\theta)\theta$-directions of the (anti-) chiral super submanifolds. This result is very much {\it expected} within the framework of ACSA to 
BRST formalism. However, the interesting and intriguing observations are (i) the absolute 
anticommutativity of the co-BRST charge {\it with} the anti-co-BRST
charge is intimately related with the nilpotency ($\partial_{\bar\theta}^2 = 0$) of the 
translational generator ($\partial_{\bar\theta}$) along  the {\it anti-chiral} $\bar\theta$-direction of the {\it anti-chiral}
super submanifold, and (ii) the absolute anticommutativity of the anti-co-BRST charge {\it with} the 
co-BRST charge, however,  is intimately related with the nilpotency ($\partial_\theta^2 = 0$)  of the translational generator 
($\partial_\theta$) along the {\it chiral} $\theta$-direction  of the (4, 1)-dimensional {\it chiral} super sub-manifold 
[of the {\it general} (4, 2)-dimensional supermanifold on which our present {\it ordinary} 4D massive Abelian 2-form theory is generalized
within the framework of ACSA to BRST formalism].


\section{Conclusions}

In our earlier work [24], we have already established that the 4D {\it massive} Abelian 2-form gauge theory
(without any interaction with matter fields) is a {\it massive} model of Hodge theory in exactly same manner as  the 2D Proca
(i.e. 2D massive Abelian 1-form) theory is (see, e.g. [31-33] for details). In our present endeavor,
 we have corroborated the correctness of the nilpotent (fermionic) symmetries of the {\it former}
theory by exploiting the basic tenets and techniques of ACSA to BRST formalism. We would like to lay emphasis on the fact that the 
{\it existence} of the (anti-)BRST and (anti-)co-BRST symmetries for the massive 4D Abelian 2-form gauge theory\footnote{The 
{\it fundamental} off-shell nilpotent (anti-)BRST and (anti-)co-BRST symmetry transformations ``gauge away`` all the spurious degrees 
of freedom that are connected with the {\it modified} form of St$\ddot{u}$ckelberg's formalism where the polar-vector field (${\varphi}_\mu $),
axial-vector field ($\tilde{\varphi}_\mu$), scalar field ($ \varphi $) and pseudo-scalar field (${\tilde\varphi}$) appear. At the end, we have
{\it only} {\it three} physical degrees of freedom (of the BRST-quantized theory) that are connected with the massive 4D free Abelian 2-form theory.}
is very {\it fundamental} as they provide the physical realizations of the nilpotent  (co-)exterior derivatives of differential geometry
at the {\it algebraic} level. The {\it bosonic} symmetry transformations (i.e. the analogue of the Laplacian operator) are derived from 
the above {\it fundamental} off-shell nilpotent symmetries. Thus, our present work  essentially corroborates the correctness 
of the off-shell nilpotent symmetries that have been discussed in our earlier work [24] for the {\it massive}
4D Abelian 2-form theory that has been turned out to be {\it physically} interesting, too. Hence, our present
 endeavor  is important  in its own right as far as the {\it sanctity} of the nilpotent symmetries and CF-type
 restrictions of our 4D {\it massive}  theory are concerned.

We would like to comment on the combination of fields that appear in a specific  manner in the transformation (2)
where  $B_{\mu\nu}\longrightarrow B_{\mu\nu} - \frac {1}{m}(\partial_\mu\,\varphi_{\nu} - \partial_\nu\varphi_{\mu} + \varepsilon_{\mu\nu\eta\kappa}\,\partial^{\eta}\tilde\phi^{\kappa}$). In our earlier work on the {\it local} duality invariance [34] of the 
source-free Maxwell's equations, we have taken the field strength tensor $F_{\mu\nu}$ for the $U(1)$ Abelian 1-form theory as:
 $F_{\mu\nu} = (\partial_\mu\,V_{\nu} - \partial_\nu V_{\mu} + \varepsilon_{\mu\nu\eta\kappa}\,\partial^{\eta} A^{\kappa})$ where $V_\mu$ and $A_\mu$ are the vector and axial-vector potentials. It is interesting to note that, in the context of  our present 4D {\it massive} Abelian 2-form gauge theory, the {\it above}
specific {\it structure} appears very {\it naturally}. We would like  to point out that the {\it two} potential 
approach to electrodynamics has been considered  by {\it other(s)}, too (see, e.g. [35] and references {\it therein)}.
It is essential to pinpoint that the nature of vector and axial-vector $(\phi_\mu, \tilde\phi_\mu)$ fields in our theory
is quite different from the properties of the vector and axial-vector (i.e. $V_\mu, A_\mu$) fields that are present in  our earlier work [34].

One of the novel results of our present investigation is the observation that the off-shell
nilpotent (anti-)BRST and (anti-)co-BRST charges are found to be absolutely anticommuting {\it despite} the
 fact that we have taken {\it only} the (anti-)chiral superfields within the framework of ACSA to BRST formalism.
 In fact, right in the beginning, we have utilized the CF-type restrictions to recast the off-shell nilpotent charges 
 in a {\it specific} form so that the proof of absolute anticommutativity property  could become straightforward.
 Thus, in a subtle manner, {\it this} proof establishes the existence of the CF-type restrictions on our theory.
 We have {\it also} established {\it their} existence by proving the invariances of the Lagrangian 
densities ${\cal L}_{(B, {\cal B})}$ and ${\cal L}_{(\bar B, \bar {\cal B})}$ within the framework of ACSA to BRST formalism 
in Sec. 5. These observations should be contrasted  with the ACSA approach to the ${\cal N} = 2$ SUSY quantum mechanical 
models where we do {\it not} obtain the absolute anticommutativity of the conserved and nilpotent ${\cal N} = 2$ SUSY charges 
(see, e.g. [36-38] for details). Thus, it is self-evident that the observation of the absolute anticommutativity property 
 between the (anti-)BRST and (anti-)co-BRST charges is  indeed a {\it novel} observation  in our present investigation.

Another interesting observation in the context of ACSA to BRST formalism is the result that it distinguishes (cf. Sec. 6 for details) 
between the absolute anticommutativity property of (i) the BRST charge {\it with} the anti-BRST charge, (ii) the anti-BRST charge {\it with} the  BRST charge,
(iii) the co-BRST charge {\it with} the anti-co-BRST charge, and (iv) the anti-co-BRST charge {\it with} the co-BRST charge.
We observe that the proof of the off-shell nilpotency of {\it all} the conserved charges is as expected within the framework of 
ACSA to BRST formalism. However, the proof of the absolute anticommutativity property, within the framework of ACSA to BRST formalism, 
yields some {\it non-trivial} and {\it novel} results. We have discussed these issues elaborately in
 the terminology of the translational generators ($\partial_\theta, \partial_{\bar\theta} $) along the Grassmannian
 directions $(\theta, \bar\theta)$ of the  (4, 1)-dimensional {\it chiral} and {\it anti-chiral } super sub-manifolds
 of the {\it general}  (4, 2)-dimensional supermanifold (cf. Sec. 6) on which our {\it ordinary} 4D massive Abelian 2-form theory has been generalized.

The model under consideration (i.e. 4D massive Abelian 2-form theory) 
 is {\it physically} interesting because it has led to the existence of fields with {\it negative} 
kinetic terms [24]. These fields have turned out to be pseudo-scalar and axial-vector fields  which are 
invoked in the theory on symmetry grounds. They possess well-defined {\it mass} but their kinetic terms
are forced to be {\it negative} if we wish to have our theory to be  a 4D massive field-theoretic model of Hodge theory [24]. 
Such fields have been found to be one of the possible candidates of dark matter and dark energy [39, 40]. 
Furthermore, in the context of cosmological models of Universe, these kinds of fields have been found to be useful for 
explaining cyclic, self-acceleration and bouncing phenomena of the cohomological models (see, e.g. [41-43]).

We would like to pinpoint the fact that the application of ACSA to BRST formalism enables us to derive the proper
(anti-)BRST and (anti-)co-BRST symmetry transformations {\it without} any application of the {\it formal} mathematical techniques
like: horizontality condition, differential geometry, dual-horizontality condition, etc. However, it would be nice to 
corroborate our observations by the formal mathematical techniques as well. Some of us, at present,  are trying to do that [44]. We have
proven the 6D Abelian 3-form gauge theory (without any interaction with matter fields) to be a tractable
 field-theoretic example  for the Hodge theory in our earlier work [see, e.g. [45] for a brief review] 
where we have discussed  the (anti-)BRST and (anti-)co-BRST symmetry transformations. It would be very 
nice future endeavor to apply the theoretical tricks  of   ACSA to BRST formalism and derive the off-shell nilpotent (anti-)BRST and 
(anti-)co-BRST symmetries for the 6D Abelian 3-form {\it massive} gauge theory by applying the St$\ddot{u}$ckelberg
 formalism. We plan to accomplish this goal in our forthcoming future publications [46].\\




\noindent
{\bf\large Acknowledgments}\vskip 0.7 cm

\noindent
S. Kumar and A. Tripathi acknowledge thankfully the BHU-fellowship awarded to them by Banaras Hindu University (BHU), Varanasi.
The DST-INSPIRE fellowship (Govt. of India) awarded to B. Chauhan is gratefully acknowledged, too.
All these authors express their deep  sense of gratitude  toward  the above {\it local} and {\it national} 
funding agencies for the  financial supports. It is pleasure for {\it all} the authors to express their thanks to the 
Reviewer for a few fruitful comments.\vskip 0.7 cm


\begin{center}
{\bf Appendix A: On the Anti-co-BRST Invariance of the Lagrangian Density ${\cal L}_{(B, {\cal B})}$ Using ACSA}\\
\end{center}

\vspace{0.9 cm}

\noindent
To complement the contents of our sub-section 5.2, we capture the anti-co-BRST symmetry invariance of ${\cal L}_{(B, {\cal B})}$
within the framework of ACSA to BRST formalism. Keeping  this goal in mind, we generalize the {\it ordinary} 4D Lagrangian density
${\cal L}_{(B, {\cal B})}$ to its counterpart {\it anti-chiral} super Lagrangian density ${\tilde{\cal L}}_{(B, {\cal B})}^{(ac, ad)}
(x, \bar\theta)$ as:    
\[
\tilde {\cal L}^{(ac, ad)}_{(B,{\cal B)}}  (x, \bar\theta)  =  \frac{1}{2}{\tilde{\cal B}}^{(ad)}_\mu (x, \bar\theta) {\tilde{\cal B}}^{\mu (ad)} (x,
 \bar\theta) \]
\[- {\tilde{\cal B}}^{\mu (ad)} (x, \bar\theta) \left(\frac{1}{2}\varepsilon_{\mu\nu\eta\kappa} \, \partial^\nu {\tilde {B}}^{\eta\kappa (ad)} (x, \bar\theta) - \frac{1}{2}\,\partial_\mu \tilde \Phi ^{(ad)} (x, \bar\theta) + m \;{\it {\tilde \Phi}}^{(ad)}_\mu (x, \bar\theta) \right) \]
\[- \frac{m^2}{4} \,{\tilde {B}}^{\mu\nu(ad)}(x, \bar\theta)\,{\tilde {B}}_{\mu\nu}^{(ad)}(x, \bar\theta)
 - \frac{1}{2}\,\partial^\mu {\phi}^\nu(x)
\,\Big(\partial_\mu {{\phi}_\nu } (x) - \partial_\nu {{\phi}}_\mu (x)\Big)  \]
\[ + {m}\, {\tilde {B}}^{\mu\nu (ad)} (x, \bar\theta)\,\partial_\mu {\phi}_\nu (x)+ \frac{1}{2}\, \partial^\mu{\it {\tilde\Phi}}^{\nu(ad)}(x,
 \bar\theta)\Big(\partial_\mu{\it {\tilde\Phi}}_{\nu}^{(ad)}(x, \bar\theta) -  \partial_\nu{\it {\tilde\Phi}}_{\mu}^{(ad)}(x, \bar\theta)\Big)  \]
\[ +  \frac{m}{2}\, \varepsilon^{\mu\nu\eta\kappa} {\tilde B}^{(ad)}_{\mu\nu}(x, \bar\theta) \partial_\eta {\it {\tilde\Phi}}_{\kappa}^{(ad)} (x, \bar\theta)   
- \frac{1}{2}\,\bar B^{\mu} (x) \, \bar B_{\mu}(x)  \]
\[ +   \bar B ^{\mu} (x)\,\left( \partial^\nu {\tilde B}_{\nu\mu}^{(ad)} (x, \bar\theta) - \frac{1}{2}\, \partial_\mu \phi (x)   
+ m \;{\phi}_\mu (x) \right)  \]
\[ +  \frac{1}{2}\, \bar B (x)\,  \bar B (x))
+  \bar B (x) \,\left(\partial_\mu {\phi}^\mu (x) + \frac{m}{2} \,\phi (x) \right)\]
\[ - \frac{1}{2}\,{\tilde{\cal B}}^{(ad)} (x, \bar\theta)\,{\tilde{\cal B}}^{(ad)} (x, \bar\theta) - {\tilde{\cal B}}^{(ad)}(x, \bar\theta)  \left(\partial_\mu
{\it  {\tilde \Phi}}^{\mu (ad)} (x, \bar\theta)
  + \frac{m}{2} \, {\it {\tilde\Phi}}^{(ad)}  (x, \bar\theta)  \right)  \]
\[ +  \Big(\partial_\mu {\tilde{\bar {\cal F}}}^{(ad)} (x, \bar\theta)
- m \,{\tilde {\bar {\cal F}}}_\mu ^{(ad)} (x, \bar\theta)\,\Big) \Big(\partial^\mu \,{\tilde{\cal F}}^{(ad)} (x, \bar\theta)
- m \,{\tilde {\cal F}}^{\mu (ad)} (x, \bar\theta) \Big)  \]
\[ - \Big(\partial_\mu {\tilde {\bar {\cal F}}}_{\nu }^{(ad)} (x, \bar\theta) - \partial_\nu {\tilde {\bar {\cal F}}}_\mu ^{(ad)} 
(x, \bar\theta) \Big)\Big(\partial^\mu {\tilde {\cal F}}^{\nu(ad)} (x, \bar\theta)  \Big)  \] 
\[ - \frac{1}{2}\,\partial_\mu {\tilde{\bar \beta}}^{(ad)} (x, \bar\theta) \,\partial^\mu \beta  (x) + 
\frac{1}{2}\, m^2\, {\tilde{\bar \beta}}^{(ad)} (x, \bar\theta)\, \beta  (x)  \]
\[ - \frac{1}{2}\left(\partial_\mu {\tilde {\bar {\cal F}}}^{\mu(ad)} (x, \bar\theta) + 
 m \, {\tilde{\bar {\cal F}}}^{(ad)} (x, \bar\theta) + \frac{1}{4}\,\rho (x) \right) \lambda (x)  \] 
\[ -\frac{1}{2}\left(\partial_\mu {\tilde {\cal F}}^{\mu(ad)} (x, \bar\theta)  + 
 m \, {\tilde{\cal F}}^{(ad)} (x, \bar\theta) - \frac{1}{4}\,\lambda (x)\right) \rho (x), 
\eqno (A.1)\]
where the superscript $(ad)$ on the {\it anti-chiral} superfields denotes the super expansions in Eq. (50).
We note that the above super Lagrangian density is a combination of anti-chiral superfields [cf. Eq. (50)]
{\it and} some {\it ordinary} fields that remain invariant under the anti-co-BRST symmetry transformations
$(s_{ad})$. It is straightforward to note that:
\[
\frac{\partial}{\partial\bar\theta} \Big[\tilde{\cal L}^{(ac, ad)}_{(B, {\cal B})}\Big] = 
- \partial_\mu \bigg[m\, \varepsilon^{\mu\nu\eta\kappa}\, \phi_\nu \big(\partial_\eta C_\kappa \big)
+ \Big(\frac{1}{2}\,\varepsilon^{\mu\nu\eta\kappa}\,\partial_\nu B_{\eta\kappa} + \frac{1}{2}\, \bar {\cal B}^\mu 
+ m \,\tilde \phi^\mu\Big) \lambda \]
\[-   {\cal B}_\nu \big(\partial^\mu C^\nu - \partial^\nu C^\mu  \big)  
+ {\cal B} \big(\partial^\mu C - m \, C^\mu\big) + \frac{1}{2}\, \big(\partial^\mu \beta \big)\,\rho \bigg] \]
  \[ +  \frac{1}{2}\; \big[{\cal B}_\mu + \bar {\cal B}_\mu + \partial_\mu \tilde \varphi \big] \big(\partial^\mu \lambda \big) 
- \partial_\mu\big[{\cal B}_\nu + \bar {\cal B}_\nu + \partial_\nu \tilde \varphi  \big] \big(\partial^\mu  C^\nu - \partial^\nu  C^\mu  \big) \]
\[ -  m \big[{\cal B}_\mu + \bar {\cal B}_\mu + \partial_\mu \tilde \varphi \big] \big(\partial^\mu  C - m \, C^\mu  \big)
- \frac{m}{2}\, \big[{\cal B} + \bar {\cal B} + m \,\tilde \varphi \big] \lambda \]
\[ +  \;\partial_\mu \big[{\cal B} + \bar {\cal B} + m \tilde \varphi \big] \big(\partial^\mu  C 
- m\,  C^\mu  \big) \equiv s_{ad}\,{\cal L}_{(B, {\cal B})}.  
 \eqno (A.2)\]
It is now self-evident that we can have {\it perfect} anti-co-BRST invariance [cf. Eq. (28)] of the Lagrangian
density ${\cal L}_{(B, {\cal B})}$ only when the (anti-)co-BRST invariant CF-type restrictions: ${\cal B}_\mu + {\bar{\cal B}}_\mu + \partial_\mu 
\tilde\varphi = 0$ and ${\cal B} + {\bar{\cal B}} + m \, \tilde\varphi = 0$ would be imposed from {\it outside}.
In a subtle manners, in other words, we have derived the (anti-)co-BRST invariant CF-type restrictions from 
the {\it symmetry} invariance of the Lagrangian densities.

We wrap up  this Appendix with the remark that the requirement of the anti-co-BRST invariance of the Lagrangian 
density ${\cal L}_{(B, {\cal B})}$ [cf. Eq. (28)] leads to the derivation of the CF-type restrictions:
${\cal B}_\mu + {\bar{\cal B}}_\mu + \partial_\mu \tilde\varphi = 0$, ${\cal B} + {\bar{\cal B}} + m \, \tilde\varphi = 0$  
which are responsible for (i) the equivalence of the Lagrangian densities ${\cal L}_{(B, {\cal B})}$ and
${\cal L}_{({\bar B}, \bar{\cal B})}$ w.r.t. the (anti-)co-BRST symmetry transformations $s_{(a)d}$, and
(ii) the absolute anticommutativity (i.e. $\{s_d, s_{ad}\} = 0$) of the (anti-)co-BRST symmetry transformations 
$(s_{(a)d})$  [and corresponding (anti-)co-BRST charges $(Q_{(a)d})$] which can be captured within the framework of  ACSA to BRST formalism.
The derivation of the CF-type restriction is novel result in our present investigation.\vskip  2 cm

\begin{center}
{\bf Appendix B: On the Derivation of the Conserved \\Charges $Q_{(b)d}^{(3)}$ from $Q_{(b)d}^{(2)}$}\\
\end{center}

\vspace{0.5 cm}
\noindent

To supplement the contents of Subsecs. 6.1 and 6.2, we perform here the {\it explicit} algebraic computations 
to show that, with the helps of EL-EOMs and CF-type restrictions, we can derive the expressions for the conserved and nilpotent 
$Q_{(b)d}^{(3)}$ (i.e. the {\it exact} forms w.r.t. the anti-BRST and anti-co-BRST symmetry transformations $s_{ab}$ and $s_{ad}$, respectively)
{\it from} the conserved and nilpotent  charges  $Q_{(b)d}^{(2)}$ (i.e. the {\it exact} forms w.r.t. the BRST and co-BRST symmetry transformations 
$s_b$ and $s_d$, respectively). In this connection, first of all, we begin with $Q_{(b)}^{(2)}$ [cf. Eq. (71)]
which can be re-written as 
\[
 Q_{(b)}^{(2)}  =  \int d^3 x \,\Big [(B\,\dot C - \dot B\,C) + (\dot B_\mu \, C^\mu  - B_\mu\, \dot C^\mu) + 
\partial _\mu (\beta\, \dot {\bar C}_\mu - \dot\beta\, \bar C_\mu)\]
\[ +  \dot\beta \,(\partial_\mu \bar C^\mu + m\,\bar C)
-\beta \,(\partial_\mu \dot{\bar C}^\mu + m\,\dot{\bar C}) + (\dot\beta\,\rho - \beta\,\dot\rho)\Big],
\eqno (B.1)\]
where we have taken the {\it total} derivatives  and re-arranged the {\it rest} of the terms. Using the EL-EOM:
$\partial _\mu \bar C^\mu + m\, \bar C = - \frac {\rho}{2}$ {\it and} throwing away the {\it total} {\it space}
derivative terms, we obtain:  
\[
 Q_{(b)}^{(2)} =   \int d^3 x \,\Big [(B\,\dot C - \dot B\,C) + (\dot B_\mu \, C^\mu  - B_\mu\, \dot C^\mu) + 
(\beta\, \ddot {\bar {C^0}} - \ddot\beta\,\bar C ^0) +  \frac {1}{2}(\dot\beta\,\rho - \beta\,\dot\rho)\Big].
\eqno (B.2)\]
Now taking the helps of the following  EL-EOMs, namely;
\[
 (\Box  + m^2)\beta = 0 \;\;\;\Longrightarrow \;\; \ddot \beta = -\,\partial _i \partial^i \beta  - m^2\,\beta, \]
 \[~~~~~(\Box  + m^2)\bar C^\mu = 0 \;\Longrightarrow \;\; \ddot {\bar C}^0 = -\,\partial _i \partial^i \bar C^0 - m^2 \bar C^0,
\eqno (B.3)\]
the above charge, with the application  of Gauss's divergence theorem,  can be re-expressed as: 
\[
 Q_{(b)}^{(2)} =   \int d^3 x \,\Big [(B\,\dot C - \dot B\,C) + (\dot B_\mu \, C^\mu  - B_\mu\, \dot C^\mu) 
+  \frac {1}{2}\,(\dot\beta\,\rho - \beta\,\dot\rho)\Big].
\eqno (B.4)\]
The stage is set to apply the CF-type restrictions: $B_\mu + \bar B_\mu + \partial_\mu \varphi = 0$ and 
$B  + \bar B  + m\,\varphi = 0$ on the above expression  to obtain 
\[
 Q_{(b)}^{(3)}  =   \int d^3 x \,\Big [(\dot{\bar B}\,C - \bar B\,\dot C) + (\bar B_\mu \, \dot C^\mu  - \dot{\bar B}_\mu\, C^\mu) +  \frac {1}{2}\,(\dot\beta\,\rho - \beta\,\dot\rho)\]
\[ +  (\partial_\mu \varphi) \,\dot C^\mu - (\partial_\mu \dot\varphi)\, C^\mu \Big],
\eqno (B.5)\]
where we have denoted the BRST charge by the symbol  $Q_b ^{(3)}$ because we have {\it already} used the CF-type restrictions.
We take the helps of the total derivatives and re-arrange the terms of the above expression to get:
\[
 Q_{(b)}^{(3)}  =   \int d^3 x \,\Big [(\dot{\bar B}\,C - \bar B\,\dot C) + (\bar B_\mu \, \dot C^\mu  
- \dot{\bar B}_\mu\, C^\mu) +  \frac {1}{2}\,(\dot\beta\,\rho - \beta\,\dot\rho)\]
\[ + \partial_\mu (\varphi \,\dot C^\mu - \dot\varphi \, C^\mu)  + \dot\varphi \, (\partial_\mu C^\mu  
+ m\, C) - \varphi\, (\partial_\mu \dot C^\mu + m\, \dot C)\Big].
\eqno (B.6)\]
At this juncture,  we substitutes the EL-EOM: $\partial _\mu C^\mu  + m\, C =  \frac {\lambda}{2}$ and $\partial _\mu \dot C^\mu  + m\, \dot C =  \frac {\dot\lambda}{2}$ to recast the above charge into the following form:
\[
Q_{(b)}^{(3)}  =   \int d^3 x \,\Big [(\dot{\bar B}\,C - \bar B\,\dot C) + (\bar B_\mu \, \dot C^\mu 
- \dot{\bar B}_\mu\, C^\mu)  + \frac {1}{2}\, (\lambda \,\dot\varphi - \dot\lambda\,\varphi)\]
 \[ +   \frac {1}{2}\,(\dot\beta\,\rho - \beta\,\dot\rho)
+ \partial_\mu (\varphi \,\dot C^\mu - \dot\varphi\, C^\mu)\Big].
\eqno (B.7)\]
Applying the Gauss divergence theorem, we drop the {\it total} space derivative terms which  leads to: 
\[
 Q_{(b)}^{(3)}  =  \int d^3 x \,\Big [(\dot{\bar B}\,C - \bar B\,\dot C) + (\bar B_\mu \, \dot C^\mu
  - \dot{\bar B}_\mu\, C^\mu)  + \frac {1}{2}\, (\lambda \,\dot\varphi - \dot\lambda\,\varphi)\]
\[ +   \frac {1}{2}\,(\dot\beta\,\rho - \beta\,\dot\rho)
 + (\varphi\, \ddot C^0 -\, \ddot \varphi \, C^0)\Big].
\eqno (B.8)\]
As mentioned earlier, we have $(\Box  + m^2)\,C^\mu  = 0$ that implies that $\ddot C^0  = -\, \partial_i \, \partial^i C^0 - m^2 C^0$ and the EL-EOM $\partial_\mu B^\mu  + m\, B  = 0$ leads to $(\Box  + m^2)\,\varphi = 0$ [{\it provided}
we use the expressions for $B$ and $B_\mu$ in terms of $\varphi$ and $\varphi_\mu$ as given in Eq. (23)]. This implies  that  $\ddot\varphi 
 = -\, \partial_i \, \partial^i \varphi - m^2 \varphi$. 
Once again, we use the Gauss divergence theorem to obtain the following  
\[
Q_{(b)}^{(3)}  =   \int d^3 x \,\Big[(\dot{\bar B}\,C - \bar B\,\dot C) + (\bar B_\mu \, \dot C^\mu  - \dot{\bar B}_\mu\, C^\mu)
+ \frac {1}{2}\, (\dot\varphi \, \lambda - \varphi\,\dot\lambda) +   \frac {1}{2}\,(\dot\beta\,\rho - \beta\,\dot\rho)\Big],
\eqno (B.9)\] 
which is quoted in Eq. (75) and re-expressed as an {\it exact} form w.r.t. the anti-BRST symmetry
transformations $(s_{ab})$ that are quoted  in Eq. (12).

We now concentrate on the crucial steps that are useful in the derivation of the conserved co-BRST
charge $Q_d ^{(3)}$ [cf. Eq. (86)] from the earlier expression $Q_d ^{(2)}$ [cf. Eq. (82)]. In this context, 
first of all, we note that $Q_d ^{(2)}$ of Eq. (82) can be expressed in terms of the total spacetime derivatives as:
\[
Q_{(d)}^{(2)}  =    \int d^3 x \,\Big[(\dot {\cal B}\,\bar C - {\cal B}\,\dot {\bar C}) + ({\cal B}_\mu \, {\dot {\bar C}^\mu}
 - \dot {\cal B}_\mu\, \bar C^\mu) + \partial_\mu (\dot {\bar\beta}\,C^\mu - \bar\beta\, \dot C^\mu)\]
\[ +  \bar\beta \, (\partial_\mu \dot C^\mu  + m\,\dot C) - \dot {\bar\beta}\, (\partial_\mu C^\mu + m\, C)
 + (\dot{\bar\beta}\,\lambda - \bar\beta\, \dot \lambda)\Big].
\eqno (B.10)\]
Using the EL-EOMs: $\partial_\mu C^\mu  + m\, C = \frac {\lambda}{2}$ and $\partial_\mu \dot C^\mu  + m\, \dot C = \frac {\dot\lambda}{2}$,
we obtain the following form of the above charge, namely;
\[
Q_{(d)}^{(2)}  =    \int d^3 x \,\Big[(\dot {\cal B}\,\bar C - {\cal B}\,\dot {\bar C}) + ({\cal B}_\mu \, {\dot {\bar C} ^\mu}
 - \dot {\cal B}_\mu\, \bar C^\mu)  +  (\ddot{\bar\beta}\, C^0 - \bar\beta\, \ddot C^0)
  +   \frac {1}{2}\,(\dot{\bar\beta}\,\lambda - \bar\beta\, \dot \lambda)\Big],
\eqno (B.11)\] 
where we have re-arranged the terms and have applied the Gauss divergence theorem to drop the 
total {\it space} derivative terms. At this stage, we use the EL-EOMs: $(\Box  + m^2)\,\bar\beta = 0$ and 
 $(\Box  + m^2)\,C^\mu = 0$ to re-express  (C.11) as 
\[
Q_{(d)}^{(2)}  =    \int d^3 x \,\Big[(\dot {\cal B}\,\bar C - {\cal B}\,\dot {\bar C}) + ({\cal B}_\mu \, {\dot {\bar C}^\mu}
 - \dot {\cal B}_\mu\, \bar C^\mu)  +   \frac {1}{2}\,(\dot{\bar\beta}\,\lambda - \bar\beta\, \dot \lambda)\Big],
\eqno (B.12)\]  
where we have used $\ddot{\bar\beta} =  -\, \partial_i\,\partial^i \bar\beta - m^2\,\bar\beta$ and 
 $\ddot{C^0} =  -\, \partial_i\,\partial^i C^0 - m^2\,C^0$ 
and performed the partial integration in the evaluation of the term $(\ddot{\bar\beta}\, C^0 - \bar\beta\, \ddot C^0)$
which turns out to contribute {\it zero} because of Gauss's divergence theorem. At this stage, we apply the CF-type 
restriction: ${\cal B} + \bar {\cal B} + m\, \tilde\varphi  = 0$ and ${\cal B}_\mu + \bar {\cal B}_\mu +
 \partial_\mu \tilde\varphi  = 0$ to recast (C.12) into the following form 
\[
Q_{(d)}^{(3)}    =     \int d^3 x \,\Big[(\dot {\cal B}\,\bar C - {\cal B}\,\dot {\bar C})
 + \partial_\mu (\dot {\tilde \varphi} \, {\bar C}^\mu - \tilde\varphi\, {\dot {\bar C}_\mu}) 
+ \tilde\varphi \, (\partial _\mu {\dot {\bar C }^\mu} + m\,\dot{\bar C})\]
\[ -  \dot{\tilde\varphi}\, (\partial_\mu {\bar C}^\mu + m\, \bar C)
 +  ({\dot {\bar{\cal B}}}_\mu \, {\bar C}^\mu - \bar {\cal B}_\mu\, {\dot {\bar C}^\mu}) 
 +   \frac {1}{2}\,({\dot{\bar\beta}}\,\lambda - \bar\beta\, \dot \lambda)\Big],
\eqno (B.13)\]  
where we have denoted the charge by the symbol  $Q_d ^{(3)}$ after the applications of the CF-type restrictions.
We exploit the usefulness of EL-EOMs: $\partial_\mu\,{\bar C}^\mu + m\,{\bar C} = - \, \frac{\rho}{2}$ and $\partial_\mu\,{\dot {\bar C}^\mu} 
+ m\,{\dot {\bar C}} = - \, \frac{\dot\rho}{2}$ and Gauss's divergence theorem to express the above form of the conserved  charge as:
\[
Q_d^{(3)}  =  \int d^3 x \,\Big[(\bar{\cal B} \, \dot{\bar C} - {\dot{\bar {\cal B}}}\,{\bar C}) + ({\dot{\bar {\cal B}}}_\mu\,{\bar C}^{\mu}
- {\bar{\cal B}}_\mu\,{\dot{\bar C}^\mu}) + \frac{1}{2}\,(\dot{\bar\beta}\,\lambda - \bar\beta\,\dot \lambda)\]
 \[ +  \frac{1}{2}\,(\dot{\tilde \varphi}\,\rho - \tilde \varphi\, \dot{\rho}) 
 +  (\ddot \varphi\,{\bar C}^0 - \varphi\,\ddot{\bar {C^0}})  \Big].
\eqno (B.14)\]  
At this juncture, we take the helps of EL-EOMs: $(\Box + m^2)\,{\bar C}_\mu = 0$ and $(\Box + m^2)\,\varphi = 0$ where the {\it latter} EOM
has been derived from $\partial_\mu \, {\cal B}^\mu + m\,{\cal B} = 0$ with the {\it inputs} from Eq. (23) where ${\cal B}_\mu$ and
${\cal B}$ have been expressed in terms of ${\tilde\phi}_\mu$ and $\tilde\varphi$. The substitutions of the following 
\[
\ddot{\tilde \varphi} = - \, \partial_i\,\partial^i\, \tilde\varphi - m^2\,\tilde\varphi, 
\qquad \ddot{\bar C}^0 = - \, \partial_i\,\partial^i \,{\bar C}^0 - m^2\,{\bar C}^0, 
\eqno (B.15)\] 
and the use of Gauss's divergence theorem demonstrate that the actual  contribution of the {\it last} 
term in (C.14) is zero. Thus, finally, we obtain the following form of the conserved co-BRST charge 
\[
Q_d^{(3)}  =  \int d^3 x \,\Big[(\bar{\cal B} \, \dot{\bar C} - {\dot{\bar {\cal B}}}\,{\bar C}) + ({\dot{\bar {\cal B}}}_\mu\,{\bar C}^{\mu}
- {\bar{\cal B}}_\mu\,{\dot{\bar C}^\mu}) + \frac{1}{2}\,(\dot{\bar\beta}\,\lambda - \bar\beta\,\dot \lambda) 
+  \frac{1}{2}\,(\dot{\tilde \varphi}\,\rho - \tilde \varphi\, \dot{\rho}),
\eqno (B.16)\] 
which has been mentioned in the main body of our text as Eq. (86) {\it and} has been expressed
 as an {\it exact} form w.r.t. the anti-co-BRST symmetry $(s_{ad})$.

We end this Appendix with the closing   remarks that {\it exactly} similar types of exercises have been performed to obtain $Q_{ab}^{(3)}$ and 
$Q_{ad}^{(3)}$ [cf. Eqs. (80), (91)] from the expressions for the {\it same} charges as $Q_{ab}^{(2)}$ and 
$Q_{ad}^{(2)}$ [cf. Eqs. (78), (88)] which are very interesting and illuminating 
as far as the {\it exact} forms of the conserved charges are concerned. Furthermore, we observe that our present discussion 
in {\it this} Appendix is relevant  to the Subsecs. 6.1 as well as 6.2 where we have discussed the off-shell nilpotency  and absolute 
anticommutativity properties of the conserved charges in the {\it ordinary} space and {\it superspace} 
(by exploiting the ACSA to BRST formalism), respectively.\\


\begin{thebibliography}{99}
\bibitem{TM1}       J. Thierry-Mieg, {J. Math. Phys.} {21} (1980) 2834. 
\bibitem{MQ2}       M. Quiros, F. J. De Urries, J. Hoyos, M. L. Mazon, E. Rodrigues,\\
                   {J. Math. Phys.} {22} (1981) 1767.               
\bibitem{RPM3}    R. Delbourgo, P. D. Jarvis, {J. Phys. A: Math. Gen.} {15} (1981) 611. 
\bibitem{BT4}       L. Bonora, M. Tonin, {Phys. Lett.} B {98} (1981) 48.  
\bibitem{BPT5}      L. Bonora,  P. Pasti,  M. Tonin, {Nuovo Cimento} A {64} (1981) 307. 
\bibitem{BPT6}      L. Bonora,  P. Pasti,  M. Tonin,  {Annals of Physics} {144} (1982) 15. 
\bibitem{RPM7}      L. Baulieu, J. Thierry-Mieg, {Nucl. Phys.} B {197} (1982) 477. 
\bibitem{RPM8}      L. Alvarez-Gaume, L. Baulieu, {Nucl. Phys.} B {212} (1983) 255. 
\bibitem{dirac9}    P. A. M. Dirac, {Lectures on Quantum Mechanics}, 
                    Belfer Graduate School of Science,
                    Yeshiva University Press, New York (1964).
\bibitem{sund10}    K. Sundermeyer, {Constrained Dynamics: Lecture Notes in Physics},\\ 
                    Vol. {169} Springer-Verlag, Berlin (1982).                   
\bibitem{RPM11}     L. Bonora, R. P. Malik, {Phys. Lett.} B {655} (2007) 75. 
\bibitem{SM12}      L. Bonora, R. P. Malik, {J. Phys.} A: {Math. Theor.} {43} (2010) 375403.  
\bibitem{RPM13}     R. P. Malik, {Eur. Phy. J.} C {60} (2009) 457. 
\bibitem{RPM14}     R. P. Malik,  {J. Phy. A: Math. Theor.} {39} (2006) 10575. 
\bibitem{RPM15}     R. P. Malik, {Eur. Phys. J.} C {51} (2007) 169. 
\bibitem{RPM16}     R. P. Malik, {J. Phys. A: Math. Gen.}  {37} (2004) 5261. 
\bibitem{SM17}       L. Bonora, {Nucl. Phys.} B {912} (2016) 103. 
\bibitem{MM18}       L. Bonora, R. P. Malik, Universe {7},  (2021)  280.
\bibitem {RTM19}     A. K. Rao, A. Tripathi, R. P. Malik,  \\{Adv. High Energy Phys.}  {2021} (2021) 5593434.
\bibitem {ABRM20}    A. Tripathi, B. Chauhan, A. K. Rao, R. P. Malik,\\{Adv. High Energy Phys.}  {2021} (2021) 2056629.
\bibitem{NJ21}       B. Chauhan, S. Kumar,  R. P. Malik,\\ {Int. J. Mod. Phys.} A {33} (2018) 1850026. 
\bibitem{MM22}       S. Kumar, B. Chauhan, R. P. Malik,\\  {Int. J. Mod. Phys.} A {33} (2018) 1850133. 
\bibitem{SAM23}      A. Shukla, N. Srinivas, R. P. Malik, {Annals of Physics} {394} (2018) 98.  
\bibitem{KKM24}      S. Krishna, R. Kumar, R. P. Malik, Annals of Physics 414 (2020) 168087. 
 \bibitem{RPM25}     R. Kumar,   S. Krishna,  {Eur. Phys. J.} C {77} (2017) 387.  
\bibitem{RPM26}      T. Eguchi, P. B. Gilkey,  A. Hanson, {Phys. Rep.} {66} (1980) 213.   
\bibitem{SM27}       S. Mukhi, N. Mukanda, {Introduction to Topology, Differential Geometry and Group
                     Theory for Physicists}, Wiley Eastern Private Limited, New Delhi (1990).  
\bibitem{HMS28}      J. W. van Holten, {Phys. Rev. Lett.} {64} (1990) 2863.   
\bibitem{RKSMA29}    S. Deser, A. Gomberoff, M. Henneaux, C. Teitelboim,\\ {Phys. Lett.} B {400} (1997) 80. 
\bibitem{RKSM30}     K. Nishijima, {Prog. Theor. Phys}. {80} (1988) 905. 
\bibitem{RKSM31}     T. Bhanja, D. Shukla, R. P. Malik, {Eur. Phy. J.} C {73} (2013) 2535.  
\bibitem{RKSM32}     A. Tripathi, B. Chauhan, A. K. Rao, R. P. Malik, \\{Adv. High Energy Phys.}  {2020} (2020) 1236518.  
\bibitem {RPm 33}     A. K. Rao, R. P. Malik, { Euro. Phys. Lett.} {135} (2021) 21001.
\bibitem {34}        R. P. Malik, T. Pradhan { Z. Phys. } C 28 (1985) 525.
\bibitem{35}         D. Zwanziger, Phys. Rev. 176 (1968) 1489.
\bibitem {36}        S. Krishna, R. P. Malik,  {Annals of Physics} {355} (2015) 204.
\bibitem {RPm 36}    S. Krishna, R. P. Malik, {Europhys. Lett.} {109} (2015) 31001.
\bibitem {REM 37}    S. Krishna, D. Shukla, R. P. Malik, \\ {Int. J. Mod. Phys.} A {31} (2016) 1650113. 
 \bibitem{SK38}      V. M. Zhuravlev, D. A. Kornilov, E. P. Savelova, \\
                     {General Relativity and Gravitation}, 36 (2004) 1736.
\bibitem{SK39}       Y. Aharonov, S. Popescu, D. Rohrlich, L. Vaidman,\\ {Physical
                     Review A} 48 (1993) 4084. 
\bibitem{SK40}       P. J. Steinhardt, N. Turok, 
                     {Physical Review D} 65 (2002) 126003,.
\bibitem{SK41}       K. Koyama, {Classical and Quantum Gravity} 24 (2007) 231.
\bibitem{SK42}       Y. F. Cai, A. Marciano, D.-G. Wang, E. Wilson-Ewing,
                     {Universe} 3 (2017) 1.                  
\bibitem {REM 43}    S. Krishna, R. Kumar, in preparation. 
\bibitem {RP44}      R. Kumar, S. Krishna, A. Shukla, R. P. Malik,\\
                     {Int. J. Mod. Phys.} A {29} (2014) 1450135.
\bibitem {RP45}      R. P. Malik, {etal}, in preparation.
\end{thebibliography}
\end{document}